\documentclass[10pt,journal,compsoc]{IEEEtran}
\usepackage{cite}
\usepackage[T1]{fontenc}% optional T1 font encoding
\usepackage{bm}
\usepackage{amsmath}
\usepackage[top=0.7in, bottom=0.55in, left=0.65in, right=0.65in]{geometry}

\usepackage[table,xcdraw]{xcolor}

\usepackage{amsthm}
\usepackage{float}
\usepackage{setspace}
\usepackage{amssymb}
\usepackage{stfloats}
\usepackage{cite}
\usepackage{ragged2e}
\usepackage{amsfonts}
\usepackage{mathrsfs}
\usepackage{amsmath,amsthm}
\usepackage{array,booktabs}
\usepackage{subfigure}
\usepackage{multirow}
\usepackage{cuted}
\usepackage{multicol}
\usepackage{graphicx}
\usepackage{subfigure}
\usepackage{graphicx,xcolor,bm}
\usepackage{hyperref}
\usepackage{threeparttable}
\usepackage{dcolumn}
\usepackage{setspace}
\usepackage{makecell}
\usepackage{lipsum}
\usepackage{enumerate}
\usepackage{mathrsfs}
\usepackage{bbm}
\usepackage{booktabs}
\usepackage[table]{xcolor}
\usepackage{enumitem}
\usepackage{subcaption}
\newcolumntype{C}{>{\centering\arraybackslash}X}
\newcolumntype{C}[1]{>{\centering\arraybackslash}m{#1}}
\usepackage[protrusion=true,expansion=true]{microtype}
\pdfoutput=1
\usepackage[ruled,vlined]{algorithm2e}
\SetKwInput{KwIn}{Input}
\SetKwInput{KwOut}{Output}
\DontPrintSemicolon

\usepackage{xfrac}
\usepackage{tabularx}
\usepackage{array}
\newcolumntype{L}{>{\raggedright\arraybackslash}X}
\newcolumntype{C}{>{\centering\arraybackslash}X}
\usepackage{cite,bm}
\graphicspath{{figures/}}
\def\BibTeX{{\rm B\kern-.05em{\sc i\kern-.025em b}\kern-.08em
    T\kern-.1667em\lower.7ex\hbox{E}\kern-.125emX}}
\usepackage{balance}

\begin{document}
\title{Predictive Rolling-Horizon Optimization for Commitment-Aware Model-Parallel Inference under Spatio-Temporal Edge Dynamics}

\author{Minghui Liwang, \IEEEmembership{Senior Member}, \IEEEmembership{IEEE}, Chenxi Xu, Wei Gong, \IEEEmembership{Member}, \IEEEmembership{IEEE}, Li Li, \\ Wenbo Zhu, Xinlei Yi, \IEEEmembership{Senior Member}, \IEEEmembership{IEEE}, Yuhan Su, Xianbin Wang, \IEEEmembership{Fellow}, \IEEEmembership{IEEE}
%Wei Ni, \IEEEmembership{Fellow}, \IEEEmembership{IEEE}
%Zhang Liu, \IEEEmembership{Member}, \IEEEmembership{IEEE}, Seyyedali Hosseinalipour, \IEEEmembership{Senior Member}, 

%Seyyedali Hosseinalipour, \IEEEmembership{Senior Member}, \IEEEmembership{IEEE},
	%\\Liqun Fu, \IEEEmembership{Senior Member}, \IEEEmembership{IEEE}, Sai Zou, \IEEEmembership{Senior Member}, \IEEEmembership{IEEE}, 
	%Xianbin Wang, \IEEEmembership{Fellow}, \IEEEmembership{IEEE},\\ Wei Ni, \IEEEmembership{Fellow}, \IEEEmembership{IEEE},
	%and Yiguang Hong, \IEEEmembership{Fellow}, \IEEEmembership{IEEE}

\thanks{
M. Liwang (minghuiliwang@tongji.edu.cn), C. Xu (kathyxu92911@gmail.com), W. Gong (weigong@tongji.edu.cn), L. Li (lili@tongji.edu.cn), W. Zhu (wbzhu@tongji.edu.cn), and X. Yi (xinleiyi@tongji.edu.cn) are with the Department of Control Science and Engineering, Shanghai Institute of Intelligent Science and Technology, the State Key Laboratory of Autonomous Intelligent Unmanned Systems,  Shanghai Key Laboratory of Intelligent Autonomous Systems, and also with Frontiers Science Center for Intelligent Autonomous Systems, Ministry of Education, Tongji University, Shanghai, China. Y. Su (ysu@xmu.edu.cn) is with the School of Electronic Science and Engineering, Xiamen University, Xiamen, China. X. Wang (xianbin.wang@uwo.ca) is with the Department of Electrical and Computer Engineering, Western University, ON, Canada.
%W. Ni (Wei.Ni@ieee.org) is with School of Engineering, Edith Cowan University, Perth, Australia.
% H. Dai (hdai@nscu.edu) is with the Department of Electrical and Computer Engineering, North Carolina State University, USA. 

% 可能有AI LAB的老师
	%Corresponding author: Minghui Liwang
}
}

\IEEEtitleabstractindextext{
	\begin{abstract}
		\justifying
Model-parallel inference over dynamic edge systems requires scheduling
decisions that account for not only instantaneous resources but also
future resource contention and reliable service commitments. Existing
edge-inference designs, however, predominantly optimize performance
metrics based on current or short-term system states, without explicitly
coupling current assignments with future commitment fulfillment. To
address this issue, we propose \textit{PROMISE}, a predictive rolling-horizon
framework for commitment-aware model-parallel inference under
spatio-temporal edge dynamics. PROMISE jointly models stochastic task
generation, mobility-induced communication variations, privacy-aware
model partitioning, and load-dependent edge computing capability. We
introduce committed completion time (CCT) as an endogenous service
decision and formulate joint SD--ES mapping and CCT determination to
balance commitment fulfillment and service reward. To address
cross-timeslot coupling, PROMISE estimates future task arrivals and
computational workloads over an adaptive horizon and embeds them into
certainty-equivalent ES-state rollout. A progress-aware surrogate then
evaluates feasible current-stage mappings, while the corresponding CCTs
are analytically recovered from predicted completion times. Only the
first-stage decisions are committed, and the optimization is repeated
with newly observed states in a receding-horizon manner. Numerical
experiments demonstrate robust scheduling performance under diverse
system scales and workload dynamics, while Raspberry-Pi-based experiments
validate the practical feasibility and key system characteristics of
model-parallel edge inference.
\end{abstract}

}

\maketitle
\IEEEdisplaynontitleabstractindextext

\IEEEpeerreviewmaketitle

\setlength{\abovedisplayskip}{1.4pt}
\setlength{\belowdisplayskip}{1.4pt}
\setlength{\skip\footins}{6pt}
\setlength{\footnotesep}{0pc}

\section{Introduction}\label{sec:Intro}
\IEEEPARstart{T}{he} rapid advancement of artificial intelligence (AI), embodied intelligence, and next-generation communication systems is driving a paradigm shift toward ubiquitous intelligence, where massive data are continuously generated and processed within large-scale cyber-physical environments \cite{background,background2,background3}. In this context, intelligent services such as perception, decision-making, and control are increasingly required to operate in real time, pushing computation from centralized cloud platforms toward distributed edge infrastructures. To support this transition, distributed inference has emerged as a key enabling paradigm, where inference tasks (e.g., deep neural networks, DNNs) are collaboratively executed across end devices, edge servers, and networked systems, significantly improving responsiveness and scalability in data-intensive applications \cite{Zhou2023acce,Gao2023task,Zou2024scal,background1}. Despite these advances, current edge inference systems are still largely built on simplified assumptions, where system resources are treated as static or weakly dynamic and scheduling decisions are made in a predominantly myopic manner. Moreover, most existing designs optimize surrogate metrics such as latency or energy consumption, without explicitly modeling long-term system evolution or enabling enforceable service guarantees. 

\subsection{Motivations in Form of Q\&A}
The above limitations motivate a principled formulation of edge inference scheduling, spanning spatio-temporal system modeling, forward temporal dependency, and commitment-aware service provisioning. Accordingly, we articulate the following research questions (RQs) to guide our design. 

\noindent
$\bullet$~\textit{RQ1: How can inference scheduling be systematically modeled in a spatio-temporally evolving edge system where workload, communication conditions, and computational capacity co-evolve under uncertainty?} Edge inference scheduling fundamentally requires a tractable representation of a non-stationary system in which workload demand and service capability co-evolve over time. Existing models largely decouple these dynamics or assume static resource characteristics, thereby failing to capture the feedback-driven nature of edge systems. We formulate a unified spatio-temporal abstraction that jointly models stochastic workload generation, communication variability, and load-dependent computational degradation. By converting environmental uncertainty into structured state dynamics through multi-slot workload estimation and resource-state evolution modeling, the resulting framework preserves the essential system behaviors while enabling principled optimization under uncertainty.

\noindent
$\bullet$~\textit{RQ2: How can the inherently forward-coupled temporal dependencies
in dynamic edge inference be transformed into a tractable optimization
framework while accounting for future feasibility?} The fundamental difficulty of dynamic edge inference scheduling stems from the forward propagation of decision impacts across time, whereby present actions continuously reshape future system states and feasible decision regions. Rather than treating such inter-temporal dependencies as an intractable stochastic process, we recast them as a structured predictive optimization problem. By embedding forecasted workload and resource evolution into the decision model, the original sequentially coupled scheduling process is transformed into a tractable finite-horizon optimization framework that explicitly accounts for future system dynamics. This establishes a principled mechanism for balancing long-term foresight and online computational efficiency in evolving edge environments.

\noindent
$\bullet$~\textit{RQ3: How can an edge inference system provide reliable service commitments under load-dependent computational degradation and stochastic future arrivals?} Reliable service provisioning for edge inference requires explicit service commitments under uncertain workload evolution and
load-dependent computational degradation. Existing scheduling frameworks primarily optimize expected performance metrics, leaving service reliability largely implicit and difficult to verify. We address this challenge through a commitment-aware service abstraction based on \textit{committed completion time}, which links service commitments to predicted workload and resource dynamics. By embedding commitment feasibility directly into the scheduling process, reliability becomes a quantifiable and enforceable property rather than a best-effort outcome. This establishes a principled framework for delivering predictable and explicitly verifiable service commitments in spatio-temporally evolving edge inference.

\begin{figure}[t!]
	\centering
	\setlength{\abovecaptionskip}{-0.0 mm}
	\includegraphics[width=1\columnwidth]{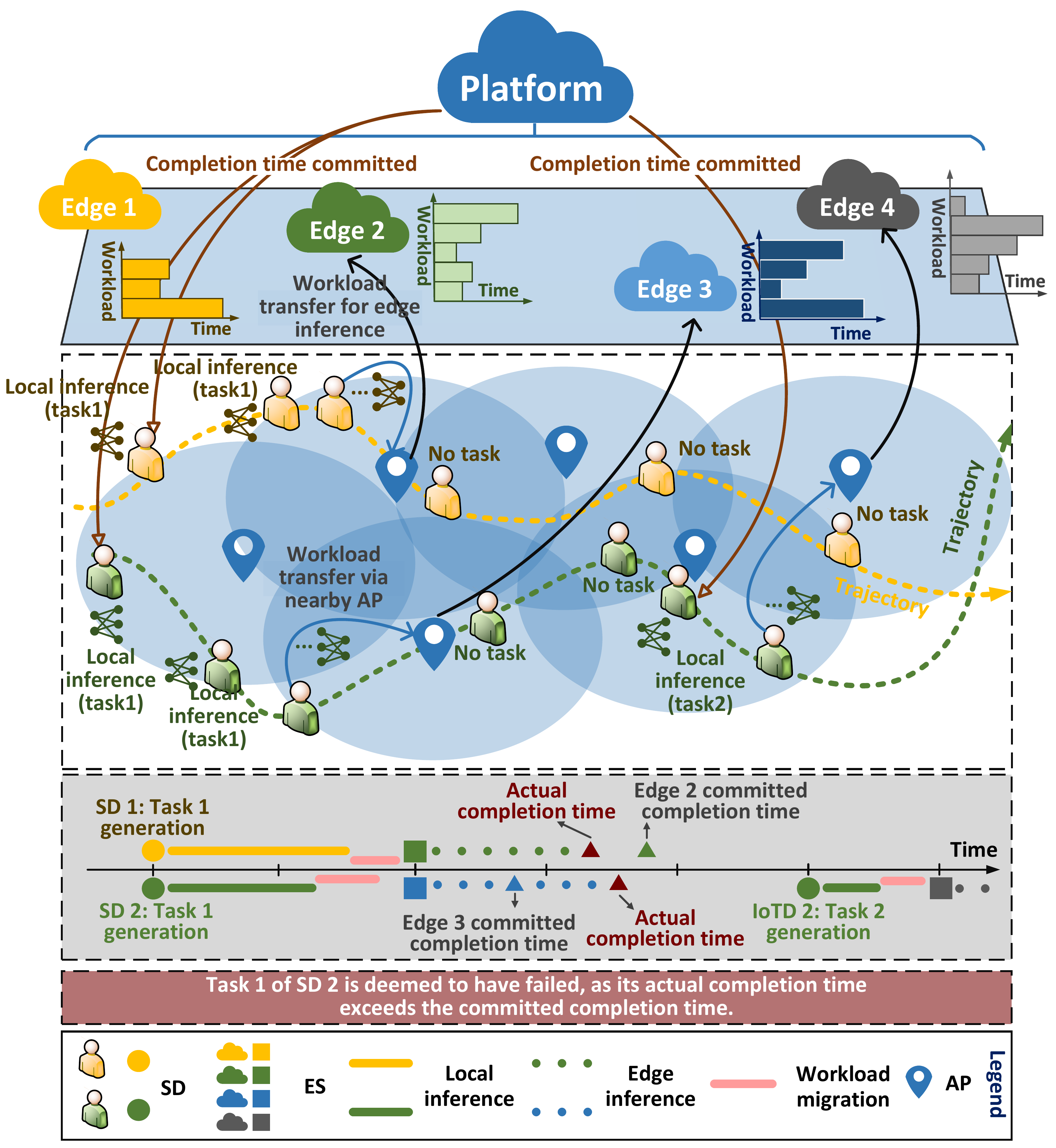}
	\caption{Schematic of our PROMISE.}
	\label{fig:system}
\end{figure}

%------------------------------------------literature review-----------------------------------
%------------------------------------------literature review-----------------------------------
%------------------------------------------literature review-----------------------------------
%------------------------------------------literature review-----------------------------------
\subsection{Literature Review}
We position PROMISE against existing distributed inference studies from three complementary perspectives: system dynamics, temporal decision mechanisms, and QoS/service abstractions ( detailed discussions and individual representative studies can be found in Appx.~\ref{relatedwork}). Table~\ref{tab:taxonomy} summarizes representative studies along these dimensions.

\noindent
$\bullet$~\textit{View 1: partial dynamics vs. coupled spatio-temporal evolution.}
Existing edge-inference studies have progressively evolved from heterogeneous but relatively stable computing environments \cite{Zhou2023acce,Gao2023task,Zou2024scal,Ren2022fine,Zheng2025opt} toward more dynamic settings involving wireless variations, device failures, runtime resource fluctuations, stochastic workloads, and user mobility \cite{iot2026static,Wang2024fail,Han2024s2e,Ye2025resou,Li2024dis,Sun2025ene,Liu2024moei}. Nevertheless, these dynamics are typically modeled individually or treated as currently observed system states. Our PROMISE instead focuses on the coupled spatio-temporal evolution of workload demand, communication conditions, and load-dependent ES capability, thereby explicitly capturing how current task admissions reshape future inference resources and task completion.

\noindent
$\bullet$~\textit{View 2: reactive/adaptive scheduling vs. prediction-aware rolling-horizon optimization.} A large body of distributed inference research determines partitioning, offloading, or scheduling decisions primarily from currently available task and resource states \cite{Zhou2023acce,Gao2023task,Zou2024scal,Ren2022fine,Liu2025ada,Chen2025adap}, while subsequent studies have extended the scope toward multi-task scheduling, dynamic rescheduling, long-term control, and learning-based adaptation \cite{Ye2025resou,Zheng2025opt,Xu2023dis,Shi2023auto,Li2024dis,Sun2025ene,Liu2024moei,Cao2024learn,Samikwa2024disnet,Lin2025top}. Although these approaches substantially enhance temporal adaptability, they generally do not expose an explicit predicted multi-timeslot workload/resource trajectory to the current scheduling decision. PROMISE closes this gap by embedding future demand estimation into ES-state rollout and repeatedly optimizing current decisions through a prediction-aware receding-horizon optimization.

\noindent
$\bullet$~\textit{View 3: metric-/constraint-driven QoS vs. commitment-based service provisioning.}
Existing inference systems predominantly characterize QoS through latency-oriented objectives \cite{Li2023adap,Wang2026scal,Bao2025joint}, multi-objective performance criteria \cite{Xue2022ddpqn,Han2024s2e}, or explicit considerations of fairness, secure resources, energy dynamics, mobility, social welfare, application-specific efficiency, resource utilization, and load balancing \cite{Xu2023dis,Li2024dnn,Wang2023decen,Wang2026mobi,Xu2025cadec,Qiao2025on,Dong2024dnn,Xin2026load}. These formulations provide increasingly sophisticated metric- or constraint-driven QoS, but typically do not treat a platform-announced completion commitment as an endogenous scheduling variable. PROMISE introduces CCT-based service provisioning, where commitment tightness and subsequent fulfillment are jointly coupled with scheduling decisions, thereby enabling explicit commitment-aware inference under uncertain system evolution.

\begin{table*}[t]
\centering
\caption{Taxonomy of representative distributed inference studies along the key dimensions motivating PROMISE}
\label{tab:taxonomy}
\scriptsize
\setlength{\tabcolsep}{2.0pt}
\renewcommand{\arraystretch}{0.72}

\begin{tabularx}{\textwidth}{
@{}
>{\raggedright\arraybackslash}p{1.55cm}
L
L
L
L
>{\centering\arraybackslash}p{1.35cm}
@{}
}
\hline
\textbf{Work} &
\textbf{Dynamics Considered} &
\textbf{Temporal Decision Mechanism} &
\textbf{Future-State Use} &
\textbf{QoS/Service Treatment} &
\textbf{Endogenous Commitment} \\
\hline

\textit{Zhou} \cite{Zhou2023acce}
& Channel variation; heterogeneous devices
& State-adaptive partial offloading
& No explicit multi-slot forecast
& Inference-latency reduction
& No \\

\textit{Gao} \cite{Gao2023task}
& Slot-level task/service contention
& Slot-based game and dynamic pricing
& Execution-delay estimation; no workload forecast
& Delay--energy--price cost
& No \\

\textit{Zou} \cite{Zou2024scal}
& Heterogeneous computing resources
& Heterogeneity-aware model-parallel scheduling
& No explicit multi-slot forecast
& Real-time inference performance
& No \\

\textit{Ren} \cite{Ren2022fine}
& Heterogeneous platforms and runtime conditions
& DRL-based elastic partitioning
& No explicit multi-slot forecast
& QoS-constrained multi-objective optimization
& No \\

\textit{Ye} \cite{Ye2025resou}
& Bandwidth constraints; device-level resource dynamics
& Parallelism planning and fault-tolerant rescheduling
& No explicit workload forecast
& Latency/resource efficiency
& No \\

\textit{Xu} \cite{Xu2023dis}
& Server heterogeneity and load imbalance
& Distributed multiple assignment
& No explicit multi-slot forecast
& Proportional fairness/load balancing
& No \\

\textit{Li} \cite{Li2024dis}
& Heterogeneous ES/device resources
& Learning-based fine-grained partitioning
& No explicit multi-slot forecast
& Delay minimization under delay constraints
& No \\

\textit{Sun} \cite{Sun2025ene}
& Multitask inference dynamics
& Lyapunov-guided reinforcement learning
& No explicit multi-slot workload forecast
& Long-term energy optimization
& No \\

\textit{Wang} \cite{Wang2023decen}
& Workload/energy queues and channel states
& Lyapunov-based online control
& Queue-state feedback
& Cumulative latency/energy optimization
& No \\

\textit{Liu} \cite{Liu2024moei}
& User mobility and service migration
& Mobility-aware partitioning/migration
& No explicit multi-slot workload forecast
& Inference-latency optimization
& No \\

\textit{Cao} \cite{Cao2024learn}
& Multi-tier heterogeneous resources
& DRL with different control cycles
& No explicit multi-slot workload forecast
& Latency/accuracy/energy efficiency
& No \\

\textit{Samikwa} \cite{Samikwa2024disnet}
& Dynamic network/device resources
& Resource-aware micro-split inference
& Current resource-state adaptation
& Latency/energy optimization
& No \\

\textit{Lin} \cite{Lin2025top}
& Dynamic asynchronous task arrivals
& Dynamic operator scheduling
& Task-arrival-pattern awareness
& Throughput/resource utilization
& No \\

\textit{Xu} \cite{Xu2025cadec}
& Dynamic user arrivals; edge-cloud heterogeneity
& Online combinatorial auction
& No explicit multi-slot workload forecast
& Social welfare/resource pricing
& No \\

\textit{Bao} \cite{Bao2025joint}
& Heterogeneous edge-cloud/model resources
& MINLP-based optimization with HSGA
& No explicit multi-slot workload forecast
& Completion time/blocking rate
& No \\

\hline
\rowcolor{gray!10}
\textbf{Our PROMISE}
& \textbf{Stochastic demand, mobility-induced communication, and load-dependent ES capacity}
& \textbf{Prediction-guided rolling-horizon optimization}
& \textbf{Explicit multi-slot workload prediction and state rollout}
& \textbf{TCR + CCT-based commitment reward/fulfillment}
& \textbf{Yes} \\
\hline
\end{tabularx}

\vspace{0.5mm}
\parbox{0.99\textwidth}{\scriptsize
\emph{Note:} ``Future-state use'' refers to explicit information about future workload/resource evolution that is incorporated into the evaluation of a \emph{current} scheduling decision. Runtime state adaptation, queue-state feedback, offline profiling, or per-task execution-time estimation alone is not regarded as explicit multi-slot future-state prediction.}
\end{table*}

%------------------------------------------ Contribution---------------------------------------
%------------------------------------------ Contribution---------------------------------------
%------------------------------------------ Contribution---------------------------------------
%------------------------------------------ Contribution---------------------------------------

\subsection{Spotlight and Contribution}
We investigate a novel framework of \underline{p}redictive \underline{ro}lling-horizon optimization for com\underline{m}itment-aware model-parallel \underline{i}nference under \underline{s}patio-temporal \underline{e}dge dynamics (PROMISE), with main contributions summarized as follows.

\noindent
$\bullet$~\textit{Commitment-aware model-parallel inference beyond conventional latency-centric scheduling.}
We investigate a novel commitment-aware edge inference paradigm in which service quality is governed by both application-specified latency tolerance, and \emph{committed completion time} (CCT) proactively announced by the platform. Unlike conventional deadline-constrained inference, where the deadline is an exogenous requirement, the CCT is an endogenous service decision that explicitly trades off commitment aggressiveness against fulfillment reliability: tighter commitments yield higher service rewards, while violations directly impair the platform's credibility. Building on this abstraction, we formulate a long-term joint optimization of SD-ES mapping and CCT determination, thereby elevating inference scheduling from metric-oriented latency minimization to explicit and verifiable service commitment provisioning under dynamic edge conditions.

\noindent
$\bullet$~\textit{Unified spatio-temporal modeling and predictive characterization of dynamic edge inference.}
We establish a joint model, capturing the major sources of spatio-temporal uncertainty in end-edge model-parallel inference, including stochastic task generation, heterogeneous model and data characteristics, mobility-induced communication variations, privacy-aware DNN partitioning, and load-dependent degradation of ES computing capability. More importantly, rather than treating these dynamics as isolated instantaneous disturbances, we characterize their forward impact on future resource contention and task completion. To this end, PROMISE develops an adaptive multi-timeslot demand-estimation mechanism that predicts future task arrivals and edge-side computational workloads over a dynamically selected horizon, providing structured predictive information for evaluating the long-term consequences of current scheduling decisions.

\noindent
$\bullet$~\textit{Prediction-guided rolling-horizon optimization with implicit temporal decoupling.}
To tackle the strong cross-timeslot coupling between current admissions, future resource availability, and commitment fulfillment, we develop a prediction-guided rolling-horizon methodology that converts uncertain future demand into certainty-equivalent ES-state evolution and repeatedly optimizes the currently executable decisions. For every feasible current-stage SD--ES mapping, PROMISE performs model-based forward rollout under uniformly allocated predictive future loading and evaluates the induced execution trajectory using a normalized workload-progress surrogate with an explicit physical interpretation. The resulting predicted completion evolution enables the continuous CCT variables to be analytically recovered rather than jointly searched with the combinatorial mappings. Only the first-stage mapping and its CCTs are committed, whereas future predictive information is discarded and refreshed once new system states become available, thereby preserving long-term foresight while retaining online adaptability.

\noindent
$\bullet$~\textit{Comprehensive software and hardware validation under heterogeneous dynamics.}
We conduct extensive evaluations across numerical and real-world settings to systematically examine PROMISE under diverse workload intensities, communication conditions, computing capabilities, and service requirements. Beyond software-based evaluation, we further implement model-parallel inference on a Raspberry-Pi-based hardware platform to assess the proposed framework under practical computation and communication behaviors. The combined software/hardware study validates PROMISE from complementary
perspectives: numerical experiments evaluate its predictive
commitment-aware scheduling performance under large-scale dynamics,
while the Raspberry-Pi testbed verifies the practical feasibility and
key system characteristics underlying model-parallel edge inference.

\begin{table}[t]
\centering
\caption{Summary of key notations}
\label{tab:notation}
\scriptsize
\setlength{\tabcolsep}{4pt}
\renewcommand{\arraystretch}{0.7}

\begin{tabular}{@{} l >{\raggedright\arraybackslash}p{5cm} @{}}
\toprule
\textbf{Notation} & \textbf{Description} \\
\midrule

\rowcolor{gray!10}
$f_n^{[\tau],\mathsf{ES}}, f_m^{\mathsf{SD}}$
& Computing capability of ES $\bm{s}_n$ and SD $\bm{u}_m$ \\

$f_n^{\mathsf{min}}, f_n^{\mathsf{max}}$
& Minimum/maximum computing capabilities allocatable by ES $\bm{s}_n$ to a single task \\

\rowcolor{gray!10}
$\omega_n^{\mathsf{max}}, \mathsf{sch}_n^{[\tau]}$
& Maximum simultaneously served SDs and projected task-completion evolution of ES $\bm{s}_n$ \\

$\bm{u}_m^{[\tau],\mathsf{SD}}, \alpha_m^{[\tau]}$
& State information and task-generation indicator of SD $\bm{u}_m$ at $\tau$ \\

\rowcolor{gray!10}
$\mathbbm{t}_m^{[\tau],\mathsf{SD}}$
& Task descriptor characterized by type $l$, maximum tolerable duration $t_m^{[\tau],\mathsf{max}}$, data-batch number $\mathbb{D}_m^{[\tau]}$, and privacy-risk tolerance $\rho_m^{[\tau]}$ \\

$z_{m,l}^{[\tau],\mathsf{part}}$
& DNN partition point selected for the type-$l$ task generated by SD $\bm{u}_m$ \\

\rowcolor{gray!10}
$\mathbbm{c}_m^{[\tau],\mathsf{SD}}, \mathbbm{c}_m^{[\tau],\mathsf{ES}}, \mathbbm{i}_m^{[\tau]}$
& Per-batch local workload, edge workload, and intermediate-feature size \\

$R_m^{[\tau],\mathsf{SD}}$
& Transmission time per unit data \\

\rowcolor{gray!10}
$\hat{\tau}_m^{[\tau],\mathsf{LCompC}},
\hat{\tau}_m^{[\tau],\mathsf{TransC}}$
& Local-completion and intermediate-feature-arrival points on the slot-indexed timeline \\

$\left\lceil \hat{\tau}_m^{[\tau],\mathsf{TransC}} \right\rceil$
& Scheduling-timeslot index corresponding to $\hat{\tau}_m^{[\tau],\mathsf{TransC}}$ \\

\rowcolor{gray!10}
$\hat{\tau}_{m,n}^{[\tau],\mathsf{ECompC}},
t_{m,n}^{[\tau],\mathsf{ECompC}}$
& Edge-completion timeslot and processing duration within its terminal slot \\

$\hat{t}_{m,n}^{[\tau]},
t_{m,n}^{[\tau],\mathsf{end}}$
& CCT (duration) and ACT (slot-indexed time point) \\

\rowcolor{gray!10}
$t_m^{[\tau],\mathsf{SDloc}},
t_m^{[\tau],\mathsf{SDTrans}},
t_m^{[\tau],\mathsf{ESmax}}$
& Local inference, transmission, and maximum allowable edge-processing delays \\

$\mathbb{C}^{[\tau]}, \mathbb{C}, \hat{\mathbb{C}}^{[\tau']}$
& Instantaneous, cumulative, and predictive task completion metrics \\

\rowcolor{gray!10}
$\mathbb{R}_{m,n}^{[\tau]}, \mathbb{R}, \hat{\mathbb{R}}^{[\tau']}$
& Per-task, cumulative, and predictive service rewards \\

$\tau', \tau'^{+}, X^{[\tau']}$
& Scheduling timeslot, rollout timeslot, and adaptive prediction horizon \\

\rowcolor{gray!10}
$\bm{Y}^{[\tau']}, \bm{W}^{[\tau']}$
& Predicted future task-arrival and computational-workload profiles \\

$\mathcal{Y}^{[\tau'],\mathsf{new}},
\mathcal{Y}_n^{[\tau'],\mathsf{old}},
\mathcal{Y}_n^{[\tau'],\mathsf{all}}$
& Newly schedulable, unfinished, and overall active task sets \\

\rowcolor{gray!10}
$\mathcal{Y}_n^{[\tau'],\mathsf{lv}},
\mathcal{Y}^{[\tau'],\mathsf{act}}$
& Leaving tasks and tasks affected by the current decision \\

$\kappa, \pi, \Pi^{[\tau']}$
& Reordered task index, predictive rollout trajectory, and trajectory set \\

\rowcolor{gray!10}
$\mathrm{w}_\kappa^{[\tau'],\mathsf{rem}},
\Delta\mathrm{w}_\kappa^{[\tau']}$
& Remaining and processed workloads of task $\kappa$ \\

$t_\kappa^{\mathsf{arr}},
t_\kappa^{\mathsf{max}},
\widetilde{t}_{\kappa,\pi}^{[\tau'],\mathsf{end}}$
& Arrival point, maximum tolerable edge duration, and predicted completion point \\

\rowcolor{gray!10}
$\mathbbm{1}_\kappa^{[\tau'],\mathsf{pred}}$
& Predictive commitment-satisfaction indicator \\

$\mathbbm{r}_{\kappa,\pi}^{[\tau'^{+}]},
\mathbbm{V}^{[\tau']}(\pi)$
& Normalized processing reward and cumulative rollout value \\

\bottomrule
\end{tabular}
\end{table}

%------------------------------------------------section 2-------------------------------------------%
%------------------------------------------------section 2-------------------------------------------%
%------------------------------------------------section 2-------------------------------------------%
%------------------------------------------------section 2-------------------------------------------%

\section{System Model}
\label{sec: overview and modeling}

We consider an end-edge collaborative inference architecture comprising three entities: \textit{(i)} a set of service demanders (SDs), denoted by $\mathcal{U}$, including mobile devices such as smartphones and autonomous vehicles; \textit{(ii)} a set of geographically distributed edge servers (ESs), denoted by $\mathcal{S}$, which collaboratively execute the offloaded DNN segments; and \textit{(iii)} a management platform (PT) that performs system-level orchestration and service provisioning. Due to local resource limitations and privacy requirements, each SD executes the front part of its DNN locally and offloads the remaining computation to an assigned ES. The PT is connected to all ESs through reliable wired links and maintains their workload and computing states for coordinated scheduling\footnote{Here, spatial heterogeneity arises from the geographically distributed
ES infrastructure and mobility-dependent SD access conditions, while
temporal dynamics arise from the evolving task demand, communication
conditions, and load-dependent computing states. Their interaction
jointly determines the time-varying resource availability experienced by
distributed inference tasks.}. Time is slotted as $\mathcal{T}=\{1,\ldots,\tau,\ldots,|\mathcal{T}|\}$, with $\Delta\tau$ denoting the duration of each timeslot. SDs stochastically generate DNN inference tasks over time. Once a task becomes ready for edge-side processing, the PT selects its serving ES and announces a committed completion time (CCT) according to the observed system state and predicted future workload evolution. The corresponding intermediate features are then forwarded to the selected ES through a nearby access point (AP). A task is regarded as successfully served only if its edge-side inference is completed within the announced CCT; otherwise, a commitment violation occurs and degrades the PT's service credibility. 

%------------------------------------------------section 2.1---------------------------------------
%------------------------------------------------section 2.1---------------------------------------
%------------------------------------------------section 2.1---------------------------------------
%------------------------------------------------section 2.1---------------------------------------

\subsection{Modeling of ESs and Platform}
\label{sec: server modeling}

\begin{figure}[h!t]
\centering
\subfigure[]{\includegraphics[width=.495\linewidth]{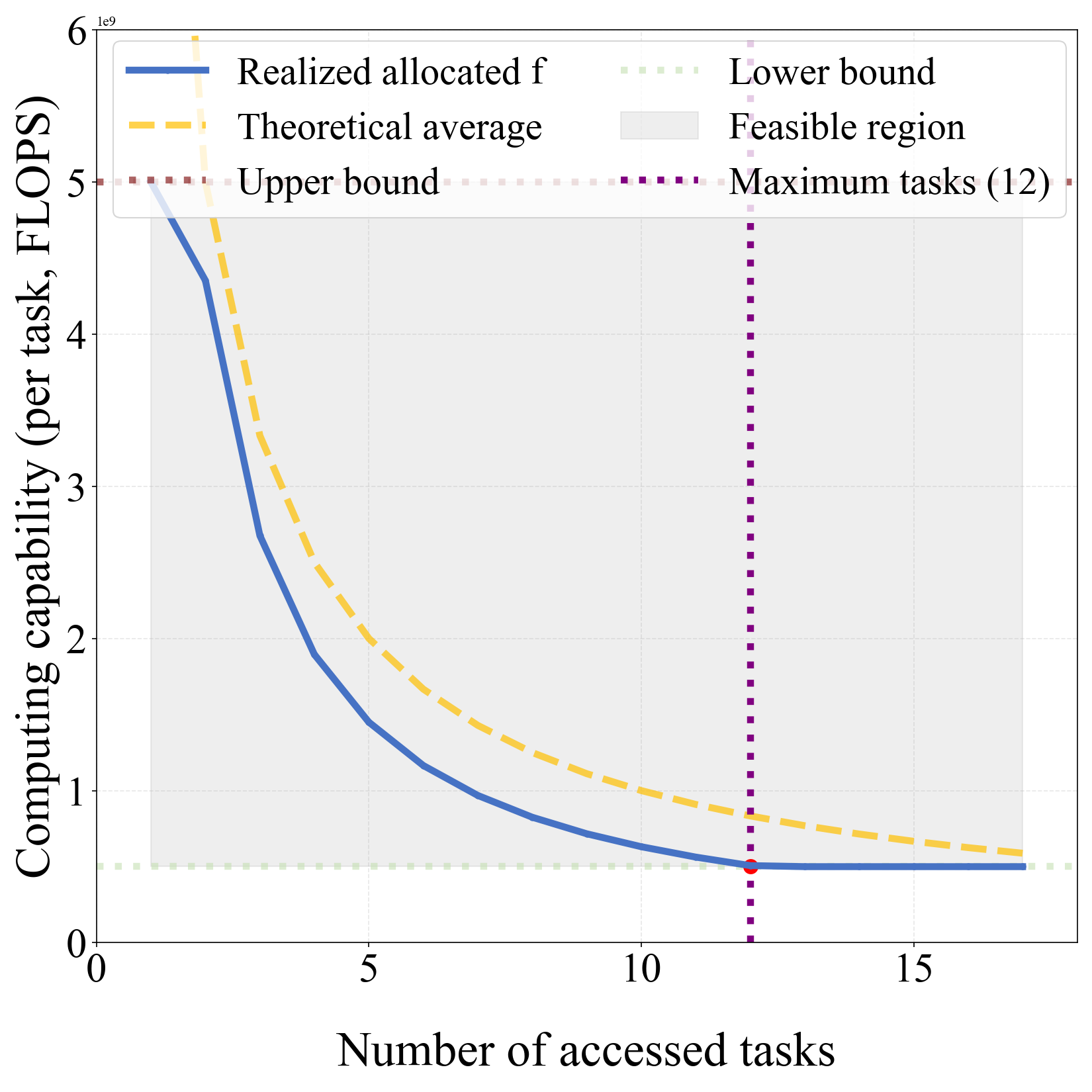}}
\subfigure[]{\includegraphics[width=.495\linewidth]{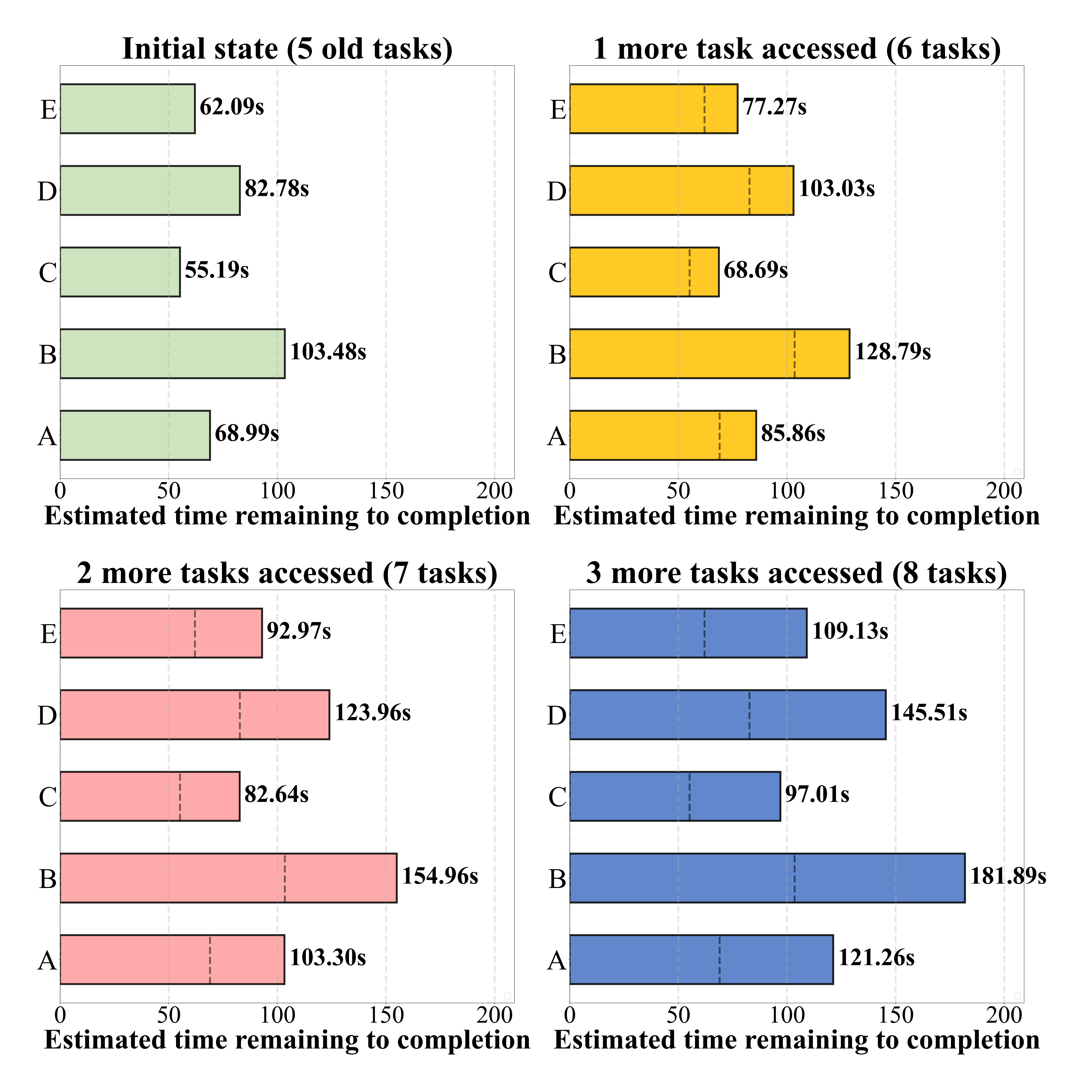}}
\caption{Example of $f_n^{[\tau],\mathsf{ES}}$ (a) and $\mathsf{sch}_n^{[\tau]}$ (b).}
\label{fig:computingspeed}
\end{figure}

The ES set is denoted by
$\mathcal{S}=\{\bm{s}_1,\ldots,\bm{s}_n,\ldots,\bm{s}_{|\mathcal{S}|}\}$.
Each ES $\bm{s}_n$ is characterized by
$\bm{s}_n=\langle f_n^{[\tau],\mathsf{ES}},
\omega_n^{\mathsf{max}},
\mathsf{sch}_n^{[\tau]}\rangle$,
where $f_n^{[\tau],\mathsf{ES}}$ denotes its effective per-task computing capability (e.g., floating-point operations per $\Delta\tau$),
$\omega_n^{\mathsf{max}}$ is the maximum number of simultaneously served SDs, and
$\mathsf{sch}_n^{[\tau]}$ characterizes its projected task-completion evolution at timeslot $\tau$\footnote{Once one or more tasks are admitted to ES $\bm{s}_n$, $\mathsf{sch}_n^{[\tau]}$ records the resulting evolution of the hosted tasks according to workload occupancy and service-capacity allocation. It represents the current ES state when no new task is admitted and is progressively updated after new admissions, thereby capturing their impacts on future resource availability and projected task completion times.}. Unlike static-capacity models, the effective computing capability of an ES depends on its concurrent workload. For an active ES with $\omega_n^{[\tau]}\geq1$, we model $f_n^{[\tau],\mathsf{ES}}
=
\max\left\{
f_n^{\mathsf{min}},
\min\left\{
f_n^{\mathsf{max}},
\frac{f_n^{\mathsf{sum}}}{(\omega_n^{[\tau]})^\beta}
\right\}
\right\}$, where $f_n^{\mathsf{min}}$ and $f_n^{\mathsf{max}}$ denote the minimum and maximum computing capabilities allocatable by $\bm{s}_n$ to a single task, respectively, $f_n^{\mathsf{sum}}$ is its aggregate computing capacity, $\omega_n^{[\tau]}$ is the number of concurrently served SDs, and $\beta$ controls the load-dependent degradation \cite{Liu2024moei,Tang2021joint}. This abstraction captures the reduction in effective per-task computing capability as resource contention increases (also see examples in Fig. \ref{fig:computingspeed}).

%f2

Task metadata are reported to the PT through nearby APs. The PT observes the global ES states and determines the serving ES for each schedulable task, while the corresponding intermediate features are subsequently forwarded to the selected ES after local DNN execution. Scheduling is performed at integer-valued timeslots; hence, a task becoming ready at a fractional time instant is scheduled at the immediately following integer timeslot. Since the PT and ESs communicate through reliable wired links, their control-plane latency is neglected \cite{Xu2023dis}.

%------------------------------------------------section 2.2---------------------------------------
%------------------------------------------------section 2.2---------------------------------------
%------------------------------------------------section 2.2---------------------------------------
%------------------------------------------------section 2.2---------------------------------------

\subsection{Modeling of SDs with Privacy Preservation}
\label{sec: SD modeling}

\noindent
$\bullet$~\textit{SDs and task generation:}
The SD set is denoted by
$\mathcal{U}=\{\bm{u}_1,\ldots,\bm{u}_m,\ldots,\bm{u}_{|\mathcal{U}|}\}$.
At the beginning of timeslot $\tau$, the scheduling-relevant state of SD $\bm{u}_m$ is represented by
$\bm{u}_m^{[\tau],\mathsf{SD}}
=\langle f_m^{\mathsf{SD}},\alpha_m^{[\tau]}\rangle$,
where $f_m^{\mathsf{SD}}$ denotes its local computing capability and
$\alpha_m^{[\tau]}\in\{0,1\}$ is the task-generation indicator. Specifically, every $\mathbbm{n}_m$ timeslots\footnote{To capture heterogeneous temporal demand patterns, task-generation intervals are allowed to differ across SDs.}, SD $\bm{u}_m$ may generate an inference task, where
$\alpha_m^{[\tau]}=1$ indicates that
$\mathbbm{t}_m^{[\tau],\mathsf{SD}}$
is generated at the beginning of $\tau$. We consider heterogeneous DNN types collected by
$\mathbb{L}=\{1,\ldots,l,\ldots,|\mathbb{L}|\}$.
When $\alpha_m^{[\tau]}=1$, the generated task is characterized by $\mathbbm{t}_m^{[\tau],\mathsf{SD}}=\left\langle
l,t_m^{[\tau],\mathsf{max}},\mathbb{D}_m^{[\tau]},\rho_m^{[\tau]}\right\rangle$, where $l\in\mathbb{L}$ denotes the DNN type, $t_m^{[\tau],\mathsf{max}}$ is the maximum tolerable task-completion duration, $\mathbb{D}_m^{[\tau]}$ is the number of input data batches, and $\rho_m^{[\tau]}$ specifies the maximum tolerable privacy leakage.

\begin{figure}[h!t]
\centering
\subfigure[]{\includegraphics[width=\linewidth]{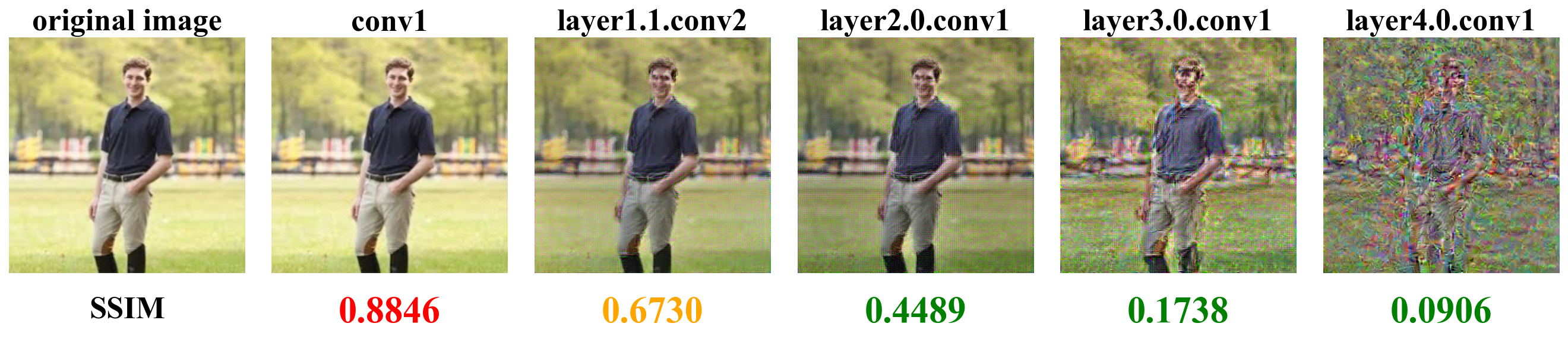}}
\hfill
\subfigure[]{\includegraphics[width=\linewidth]{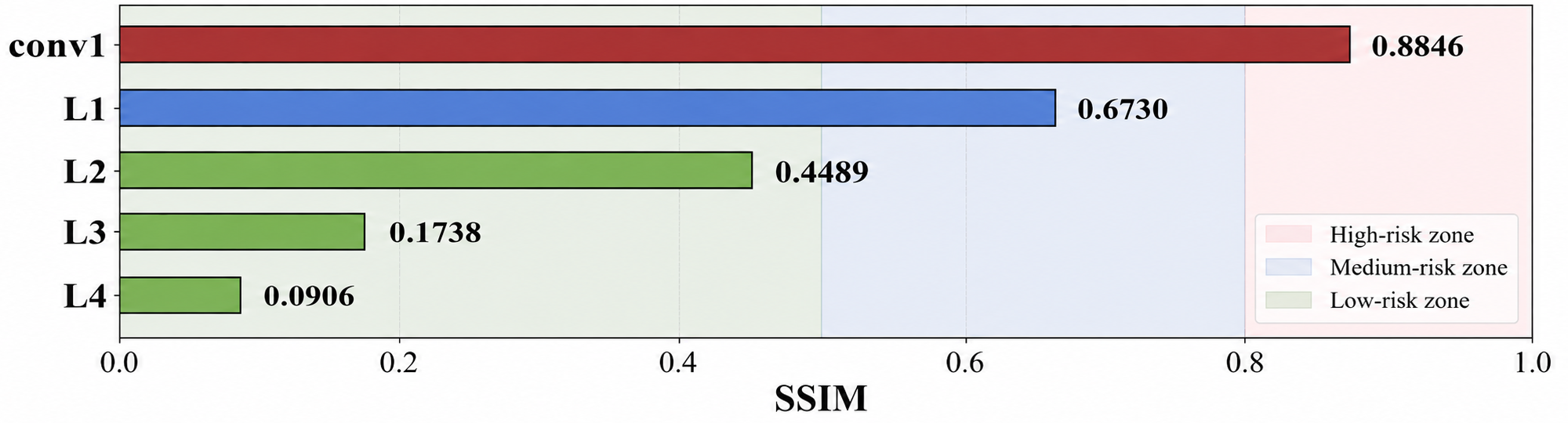}}
\caption{An example of image reconstruction and similarity at different layers of an ImageNet-pretrained ResNet-18 model provided by PyTorch, using a basic mean squared error (MSE)-based inversion method (a), and the variation in reconstruction similarity with increasing model depth (b).}
\label{fig:ssim}
\end{figure}

\noindent
$\bullet$~\textit{Privacy-aware DNN partitioning:}
Due to limited local computing capability and privacy requirements, we adopt model-parallel inference in which the DNN is split between the SD and an ES. Let
$z_{m,l}^{[\tau],\mathsf{part}}\in\{0,1,\ldots,K_l\}$
denote the partition point of the type-$l$ task generated by $\bm{u}_m$ at $\tau$, where $K_l$ is the total number of layers. Layers up to
$z_{m,l}^{[\tau],\mathsf{part}}$
are executed locally, while layers
$z_{m,l}^{[\tau],\mathsf{part}}+1,\ldots,K_l$
are processed by the assigned ES. The partition point is constrained by the task-specific privacy requirement. For image inference, reconstruction leakage can be quantified by reconstructing the input from intermediate activations and measuring the structural similarity (SSIM, see (\ref{SSIM})) between the reconstructed and original images. Prior studies, e.g., \cite{Cheng2025privacy}, show that retaining more front layers locally can reduce reconstruction fidelity of the transmitted representation (see an example Fig.~\ref{fig:ssim}).
\begin{equation}
\label{SSIM}
\mathsf{SSIM}(i,j)=
\frac{(2\mu_i \mu_j + C_1)(2\sigma_{ij} + C_2)}
{(\mu_i^2 + \mu_j^2 + C_1)(\sigma_i^2 + \sigma_j^2 + C_2)},
\end{equation}
where $\mu_i$ and $\mu_j$ are the image means,
$\sigma_i^2$ and $\sigma_j^2$ are their variances,
$\sigma_{ij}$ is the covariance, and $C_1,C_2$ are numerical-stability constants. SSIM lies in $[-1,1]$, with larger values indicating higher structural similarity. For audio and text data, cosine similarity is adopted as the reconstruction-similarity measure \cite{Chang2021neural,Marshall2025diff}. Since our interest lies in reconstruction-based privacy leakage rather than task-specific inference quality, metrics such as word error rate are not considered. We use
$\phi\!\left(z_{m,l}^{[\tau],\mathsf{part}}\right)$
to denote the privacy leakage associated with a partition point, which must satisfy $\phi\!\left(z_{m,l}^{[\tau],\mathsf{part}}\right)
\leq
\rho_m^{[\tau]}$. For a selected partition point,
$\mathbbm{c}_m^{[\tau],\mathsf{SD}}$
and $\mathbbm{c}_m^{[\tau],\mathsf{ES}}$
denote the per-batch computing workloads of the local and edge-side DNN segments, respectively. The former aggregates the computation up to
$z_{m,l}^{[\tau],\mathsf{part}}$,
whereas the latter aggregates the remaining computation from
$z_{m,l}^{[\tau],\mathsf{part}}+1$
to $K_l$. We further denote by
$\mathbbm{i}_m^{[\tau]}$
the intermediate-feature size per batch at the partition point. These quantities are determined by the DNN architecture and selected partition location \cite{Huang2025joint}.

%f3

%------------------------------------------------section 2.3---------------------------------------
%------------------------------------------------section 2.3---------------------------------------
%------------------------------------------------section 2.3---------------------------------------
%------------------------------------------------section 2.3---------------------------------------

\subsection{Uncertain Factors under Spatio-Temporal Evolution}
\label{sec: uncertain modeling}

Inference demand and service capability evolve stochastically across both time and space. We capture the following three major sources of uncertainty.

\noindent
$\bullet$ \textit{Task generation} ($\alpha_m^{[\tau]}$):
Each SD has an individual task-generation interval $\mathbbm{n}_m$ and generates a task probabilistically at the corresponding generation epochs. Specifically, $\Pr\!\left(\alpha_m^{[\tau]}=1\,\middle|\,
\tau=k\mathbbm{n}_m,\;k\in\mathbb{Z}^{+}\right)=p_m$, while $\alpha_m^{[\tau]}=0$ outside these generation epochs. Hence, different SDs exhibit heterogeneous and stochastic inference demand over time.

\noindent
$\bullet$ \textit{Task type and data batches} ($l$ and $\mathbb{D}_m^{[\tau]}$):
Different DNN architectures exhibit heterogeneous computing and intermediate-feature characteristics. Conditioned on task generation, the DNN type is drawn from $\mathbb{L}$; for the uniform setting considered here,
$\Pr(l'=l)=1/|\mathbb{L}|$.
Meanwhile, $\mathbb{D}_m^{[\tau]}$ is modeled as a discrete random variable over a predefined integer range, representing temporal variation in the number of input data batches. These two factors jointly induce time-varying local/edge computation, and transmission workloads.

\noindent
$\bullet$ \textit{Data transmission efficiency} ($R_m^{[\tau],\mathsf{SD}}$):
SD mobility leads to time-varying wireless conditions and hence stochastic transmission delay. We therefore model
$R_m^{[\tau],\mathsf{SD}}$
as a time-varying transmission-time coefficient, i.e., the transmission time per unit batch, which equivalently reflects the inverse of the effective uplink rate. Variations in
$R_m^{[\tau],\mathsf{SD}}$
therefore translate mobility-induced communication dynamics into the inference-transmission delay \cite{Guo2025seamless}.

%------------------------------------------------section 2.4---------------------------------------
%------------------------------------------------section 2.4---------------------------------------
%------------------------------------------------section 2.4---------------------------------------
%------------------------------------------------section 2.4---------------------------------------

\subsection{An Illustrative Example}
\label{sec: example}

We use a task generated by SD $\bm{u}_m$ at timeslot $\tau$ to illustrate the end-to-end operation of PROMISE.

\noindent
$\bullet$~\textit{Step 1: task generation and local inference.}
Suppose
$\mathbbm{t}_m^{[\tau],\mathsf{SD}}$
is generated at $\bm{u}_m$, i.e.,
$\alpha_m^{[\tau]}=1$.
Given its privacy-risk tolerance $\rho_m^{[\tau]}$, a feasible DNN partition point satisfies
$\phi\!\left(z_{m,l}^{[\tau],\mathsf{part}}\right)\leq\rho_m^{[\tau]}$.
The layers up to
$z_{m,l}^{[\tau],\mathsf{part}}$
are then executed locally, after which the intermediate features are transmitted through the nearby AP. Although scheduling decisions are indexed by discrete timeslots, computation and transmission occur on a continuous time axis. We therefore denote the local-inference completion instant and intermediate-feature arrival instant by
${\hat{\tau}}_m^{[\tau],\mathsf{LCompC}}$
and
${\hat{\tau}}_m^{[\tau],\mathsf{TransC}}$,
respectively.

\noindent
$\bullet$~\textit{Step 2: task arrival, CCT determination, and scheduling.}
At
${\hat{\tau}}_m^{[\tau],\mathsf{TransC}}$,
the task becomes ready for edge scheduling. The PT jointly schedules it with other tasks that have become ready since the preceding scheduling epoch at
$\left\lceil{\hat{\tau}}_m^{[\tau],\mathsf{TransC}}\right\rceil$.
If the task is assigned to ES $\bm{s}_n$, the PT announces a CCT
${\hat{t}}_{m,n}^{[\tau]}$,
which should not exceed the residual edge-side latency budget after local inference and intermediate-feature transmission. A tighter feasible CCT yields a higher service reward, but also leaves less tolerance to future resource contention, thereby creating the reward-commitment-fulfillment trade-off central to PROMISE.

\noindent
$\bullet$~\textit{Step 3: edge inference and result feedback.}
The selected ES starts processing the received DNN segment while sharing its computing resources with other active tasks. As tasks arrive/depart, the number of concurrently served SDs changes, causing the effective computing capability and hence task-completion evolution to vary over time. A task completed within its announced CCT
${\hat{t}}_{m,n}^{[\tau]}$
is regarded as successfully served; otherwise, the PT incurs a commitment violation. Upon completion, the inference result is returned to the SD through the nearest AP, with feedback latency neglected.

%------------------------------------------------section 3---------------------------------------------------
%------------------------------------------------section 3---------------------------------------------------
%------------------------------------------------section 3---------------------------------------------------
%------------------------------------------------section 3---------------------------------------------------

\section{Commitment Modeling and Problem Formulation}
\label{sec: core models and problem}

A tighter CCT provides a more responsive service commitment and hence a higher reward to the PT. However, as newly admitted tasks increase future resource contention, the effective ES computing capability may degrade, potentially delaying both new and ongoing tasks and increasing the risk of commitment violations. Such violations reduce the task completion rate and consequently impair the PT's service credibility. We therefore seek to jointly optimize service reliability and commitment reward over the entire scheduling horizon. The corresponding completion-time models, service metrics, and optimization problem are developed below.

%------------------------------------------------section 3.1---------------------------------------------------
%------------------------------------------------section 3.1---------------------------------------------------
%------------------------------------------------section 3.1---------------------------------------------------
%------------------------------------------------section 3.1---------------------------------------------------

\subsection{Actual and Committed Completion Times}
\label{sec: time}

Consider a type-$l$ task $\mathbbm{t}_m^{[\tau],\mathsf{SD}}$ generated by SD $\bm{u}_m$ at timeslot $\tau$. Under partition point $z_{m,l}^{[\tau],\mathsf{part}}$, its local inference delay is $t_m^{[\tau],\mathsf{SDloc}}=\frac{
\mathbb{D}_m^{[\tau]}
\mathbbm{c}_m^{[\tau],\mathsf{SD}}
}{
f_m^{\mathsf{SD}}
}$.
Accordingly, local inference completes at the continuous-time point $\hat{\tau}_m^{[\tau],\mathsf{LCompC}}
=
\tau+
\frac{
t_m^{[\tau],\mathsf{SDloc}}
}{
\Delta\tau
}$.
The resulting intermediate features are then transmitted to the nearby AP, incurring
\begin{equation}
\label{data transmission time}
t_m^{[\tau],\mathsf{SDTrans}}
=
\mathbb{D}_m^{[\tau]}
\mathbbm{i}_m^{[\tau]}
R_m^{[\tau],\mathsf{SD}}.
\end{equation}
Hence, the complete intermediate payload becomes available for edge-side processing at
\[
\hat{\tau}_m^{[\tau],\mathsf{TransC}}
=
\tau+
\frac{
t_m^{[\tau],\mathsf{SDloc}}
+
t_m^{[\tau],\mathsf{SDTrans}}
}{
\Delta\tau
}.
\]
The AP reports the task metadata to the PT, while the payload is buffered until the serving ES is determined. Since scheduling is performed at integer-valued timeslots, the task starts edge-side execution at $\tau^{\mathsf{start}}=\left\lceil
\hat{\tau}_m^{[\tau],\mathsf{TransC}}
\right\rceil$. If the task is assigned to ES $\bm{s}_n$, its actual completion depends on the future effective computing capability of $\bm{s}_n$, which evolves with its concurrent workload. Let
$\hat{\tau}_{m,n}^{[\tau],\mathsf{ECompC}}$
denote the timeslot in which the edge-side computation is completed. It is given by
\begin{equation}
\label{edge completion time point}
\begin{aligned}
\hat{\tau}_{m,n}^{[\tau],\mathsf{ECompC}}
&=
\min
\Bigg\{
T\in\mathbb{Z}^{+}
\,\Bigg|\,
T\geq\tau^{\mathsf{start}},
\\[-1mm]
&\quad
\sum_{\tau''=\tau^{\mathsf{start}}}^{T}
f_n^{[\tau''],\mathsf{ES}}\Delta\tau
\geq
\mathbb{D}_m^{[\tau]}\mathbbm{c}_m^{[\tau],\mathsf{ES}}
\Bigg\}.
\end{aligned}
\end{equation}
That is, $\hat{\tau}_{m,n}^{[\tau],\mathsf{ECompC}}$ is the first timeslot in which the accumulated processing capability is sufficient to complete the required edge-side workload. The completion instant can be further refined within this terminal timeslot. Specifically, the required processing duration inside the terminal slot is
\begin{equation}
\label{edge completion time period}
t_{m,n}^{[\tau],\mathsf{ECompC}}
=
\frac{
\mathbb{D}_m^{[\tau]}
\mathbbm{c}_m^{[\tau],\mathsf{ES}}
-
\displaystyle
\sum_{\tau''=\tau^{\mathsf{start}}}^{
\hat{\tau}_{m,n}^{[\tau],\mathsf{ECompC}}-1
}
f_n^{[\tau''],\mathsf{ES}}\Delta\tau
}{
f_n^{
[\hat{\tau}_{m,n}^{[\tau],\mathsf{ECompC}}],
\mathsf{ES}}
}.
\end{equation}
Accordingly, the actual completion time (ACT) on the continuous slot-indexed timeline is
\begin{equation}
\label{actual completion time point}
t_{m,n}^{[\tau],\mathsf{end}}
=
\hat{\tau}_{m,n}^{[\tau],\mathsf{ECompC}}
-1
+
\frac{
t_{m,n}^{[\tau],\mathsf{ECompC}}
}{
\Delta\tau
}.
\end{equation}

The ACT in (\ref{actual completion time point}) is an \emph{ex-post} quantity because it depends on the ES computing capabilities realized over future timeslots. At the scheduling epoch, the PT instead relies on the current ES state $\mathsf{sch}_n^{[\tau]}$ together with predicted future workload evolution to evaluate the consequence of admitting a task\footnote{Completed tasks may depart from an ES while newly generated tasks are admitted, continuously changing the active workload and hence the effective computing capability. Therefore, a scheduling decision should account not only for the immediate completion feasibility of current tasks but also for the potential impact of subsequent task arrivals.}. Before execution, the PT announces a CCT $\hat{t}_{m,n}^{[\tau]}$, which is a service commitment measured as a duration from the task's edge-side arrival rather than an actual completion time. The maximum admissible edge-side commitment is
\begin{equation}
\label{max edge delay}
t_m^{[\tau],\mathsf{ESmax}}
=
t_m^{[\tau],\mathsf{max}}
-
\left(
t_m^{[\tau],\mathsf{SDloc}}
+
t_m^{[\tau],\mathsf{SDTrans}}
\right),
\end{equation}
and any feasible announced CCT must satisfy
$\hat{t}_{m,n}^{[\tau]}\leq
t_m^{[\tau],\mathsf{ESmax}}$.
Importantly, the CCT is a management-level decision announced before task completion and therefore need not coincide with either the actual completion time or the instantaneous completion estimate contained in $\mathsf{sch}_n^{[\tau]}$.

%------------------------------------------------section 3.2---------------------------------------------------
%------------------------------------------------section 3.2---------------------------------------------------
%------------------------------------------------section 3.2---------------------------------------------------
%------------------------------------------------section 3.2---------------------------------------------------

\subsection{Task Completion and Commitment Reward}
\label{sec: completion and reward}

Unlike conventional deadline-driven inference, PROMISE evaluates service completion against the CCT proactively announced by the PT. For a task generated by $\bm{u}_m$ at $\tau$ and assigned to $\bm{s}_n$, we define
\begin{equation}
\label{commitment indicator}
\mathbbm{1}
\left(
\checkmark\hat{t}_{m,n}^{[\tau]}
\right)
=
\mathbbm{1}
\left(
\left(
t_{m,n}^{[\tau],\mathsf{end}}
-
\hat{\tau}_m^{[\tau],\mathsf{TransC}}
\right)
\Delta\tau
\leq
\hat{t}_{m,n}^{[\tau]}
\right),
\end{equation}
where
$\mathbbm{1}(\checkmark\hat{t}_{m,n}^{[\tau]})=1$
indicates successful fulfillment of the announced CCT and 0 otherwise\footnote{Although the ES index $n$ in $\hat{t}_{m,n}^{[\tau]}$ is not strictly necessary because each task is eventually associated with a single ES, we retain it for notational clarity and consistency with the assignment decision.}. Let $x_{m,n}^{[\tau]}\in\{0,1\}$ indicate whether the task generated by $\bm{u}_m$ at $\tau$ is assigned to $\bm{s}_n$. For a timeslot containing at least one generated task, the corresponding task completion ratio (TCR) is
\begin{equation}
\label{TCR}
\mathbb{C}^{[\tau]}
=
\frac{
\displaystyle
\sum_{m=1}^{|\mathcal{U}|}
\sum_{n=1}^{|\mathcal{S}|}
x_{m,n}^{[\tau]}
\alpha_m^{[\tau]}
\mathbbm{1}
\left(
\checkmark\hat{t}_{m,n}^{[\tau]}
\right)
}{
\displaystyle
\sum_{m=1}^{|\mathcal{U}|}
\alpha_m^{[\tau]}
}.
\end{equation}
When no task is generated at $\tau$, we set $\mathbb{C}^{[\tau]}=0$. PROMISE further rewards tight yet successfully fulfilled service commitments. Specifically, the reward associated with assigning the task of $\bm{u}_m$ to $\bm{s}_n$ is
\begin{equation}
\label{service reward}
\mathbb{R}_{m,n}^{[\tau]}
=
x_{m,n}^{[\tau]}
\alpha_m^{[\tau]}
\mathbbm{1}
\left(
\checkmark\hat{t}_{m,n}^{[\tau]}
\right)
\left(
t_m^{[\tau],\mathsf{ESmax}}
-
\hat{t}_{m,n}^{[\tau]}
\right).
\end{equation}
Hence, among successfully fulfilled commitments, a tighter CCT yields a larger reward, whereas a violated commitment contributes neither completion credit nor service reward. This mechanism discourages unrealistically aggressive announcements while explicitly coupling commitment tightness with fulfillment reliability.

%------------------------------------------------section 3.3---------------------------------------------------
%------------------------------------------------section 3.3---------------------------------------------------
%------------------------------------------------section 3.3---------------------------------------------------
%------------------------------------------------section 3.3---------------------------------------------------

\subsection{Problem Formulation}
\label{sec: problem}

We now formulate commitment-aware inference scheduling over the entire horizon. The PT jointly determines the SD--ES assignments
$x_{m,n}^{[\tau]}$
and CCTs
$\hat{t}_{m,n}^{[\tau]}$
to balance service reliability and commitment reward under stochastic workload, communication, and computing dynamics. Define the cumulative completion metric and service reward as $\mathbb{C}
=
\sum_{\tau=1}^{|\mathcal{T}|}
\mathbb{C}^{[\tau]}$, $
\mathbb{R}=
\sum_{\tau=1}^{|\mathcal{T}|}
\sum_{m=1}^{|\mathcal{U}|}
\sum_{n=1}^{|\mathcal{S}|}
\mathbb{R}_{m,n}^{[\tau]}.$
Accordingly, the resulting long-term optimization\footnote{Here, the superscript $[\tau]$ identifies the task by its generation
timeslot, while its actual scheduling occurs after the corresponding
intermediate features become available for edge-side processing.} is given by $\bm{\mathcal{P}}_0$:
\begin{equation}
\label{P0}
\bm{\mathcal{P}}_0:
~\underset{
x_{m,n}^{[\tau]},
\hat{t}_{m,n}^{[\tau]}
}{
\text{maximize}
}
\quad
\lambda^{\mathsf{cmlp}}\mathbb{C}
+
\lambda^{\mathsf{rwd}}\mathbb{R},
\end{equation}
\begin{center}
\textit{s.t.}
\end{center}
\vspace{-1.2em}
\begin{flalign}
& \text{(C1):}~
\sum_{n=1}^{|\mathcal{S}|}
x_{m,n}^{[\tau]}
=
1,
~\text{if}~
\alpha_m^{[\tau]}=1,
~\forall\,\bm{u}_m\in\mathcal{U},
~\tau\in\mathcal{T},
\notag&\\
& \text{(C2):}~
\omega_n^{[\tau]}
\leq
\omega_n^{\mathsf{max}},
~\forall\,\bm{s}_n\in\mathcal{S},
~\tau\in\mathcal{T},
\notag&\\
& \text{(C3):}~
x_{m,n}^{[\tau]}
\in\{0,1\},
~\forall\,\bm{u}_m\in\mathcal{U},
~\bm{s}_n\in\mathcal{S},
~\tau\in\mathcal{T},
\notag&\\
& \text{(C4):}~
0<
\hat{t}_{m,n}^{[\tau]}
\leq
t_m^{[\tau],\mathsf{ESmax}},
~\text{if}~
x_{m,n}^{[\tau]}=1,
~\forall\,\bm{u}_m\in\mathcal{U},
~\bm{s}_n\in\mathcal{S},
\notag&\\
& \text{(C5):}~
x_{m,n}^{[\tau]}
\leq
\alpha_m^{[\tau]},
~\forall\,\bm{u}_m\in\mathcal{U},
~\bm{s}_n\in\mathcal{S},
~\tau\in\mathcal{T}.
\notag&
\end{flalign}
In $\bm{\mathcal{P}}_0$, $\lambda^{\mathsf{cmlp}}$ and $\lambda^{\mathsf{rwd}}$ control the trade-off between commitment fulfillment and service reward. Constraint (C1) assigns every generated inference task to exactly one ES\footnote{Since we focus on CCT-aware rolling-horizon scheduling for accepted inference requests, all generated tasks are assumed to be admitted for execution. In practice, admission control can be implemented as an independent upper-layer mechanism, while task rejection under server saturation is outside the scope of this work.}; (C2) limits the number of simultaneously served SDs at each ES; (C3) enforces binary assignment; (C4) constrains each announced CCT within the residual edge-side latency budget; and (C5) prevents assignment in the absence of task generation.

\noindent
\textit{Problem structure and challenges:}
Solving $\bm{\mathcal{P}}_0$ online is challenging from three intertwined aspects. First, discrete SD-ES associations are coupled with continuous CCT decisions, yielding a mixed-integer decision space. Second, an assignment made at the current timeslot changes future ES occupancy and hence the effective computing capability available to both newly admitted and ongoing tasks, creating strong cross-timeslot coupling. Third, future task arrivals, task characteristics, communication conditions, and load-dependent service capacities are uncertain and jointly determine whether a current commitment can eventually be fulfilled. Consequently, purely instantaneous optimization cannot explicitly evaluate the downstream feasibility of current commitments. To address these challenges, Sec.~\ref{sec: methodology} develops PROMISE, which combines future-demand estimation with prediction-guided rolling-horizon optimization; further rationale for the adopted modeling assumptions is provided in Appx.~\ref{appxration}.

%------------------------------------------------section 4--------------------------------------------------------
%------------------------------------------------section 4--------------------------------------------------------
%------------------------------------------------section 4--------------------------------------------------------
%------------------------------------------------section 4--------------------------------------------------------

\section{Methodology: Design of PROMISE}
\label{sec: methodology}
Edge inference tasks often span multiple scheduling timeslots, making their completion dependent on future resource availability. Consequently, scheduling decisions exhibit strong temporal coupling: current task assignments consume future computing resources, whereas stochastic future task arrivals and workload variations directly affect the feasibility of current service commitments. Such coupled spatio-temporal dynamics cannot be effectively handled by myopic scheduling, motivating a prediction-aware rolling-horizon optimization framework that explicitly exploits future system evolution. Accordingly, we propose PROMISE, which consists of two tightly coupled modules. \textit{Module A} estimates future workload evolution to provide predictive information for scheduling, enabling Module B to determine SD-ES assignments with future resource availability taken into account. Building upon these predictions, \textit{Module B} then develops an efficient recursive optimization method for commitment-aware inference scheduling under dynamism.

%------------------------------------------------section 4.1
%------------------------------------------------section 4.1
%------------------------------------------------section 4.1
%------------------------------------------------section 4.1
%------------------------------------------------section 4.1

\subsection{Module A. Future $X$-Timeslot Demand Estimation under Spatio-Temporal Coupling}
Inference workloads exhibit pronounced spatio-temporal dynamics. Task generation evolves stochastically over time due to application patterns, user behaviors, and device mobility, while heterogeneous model types, privacy requirements, batch sizes, and wireless conditions induce substantial heterogeneity in resource demand. Consequently, future ES workloads cannot be directly observed at the current scheduling timeslot, making proactive scheduling challenging.

We first distinguish the task-generation time from the scheduling time. Recall that $[\tau]$ indexes the timeslot in which a task is generated. Since the task must first undergo local inference and intermediate-feature transmission before being scheduled, its scheduling time can be later than $\tau$. Accordingly, we use $\tau'$ ($\tau'\in\mathcal{T}$) to denote the scheduling timeslot of newly arrived tasks\footnote{For an active task generated at timeslot $\tau$ by SD $\bm{u}_m$, its scheduling timeslot is given by $\tau'=\left\lceil {\hat{\tau}}^{[\tau],\mathsf{TransC}}_m\right\rceil$.}. At each $\tau'$, Module A predicts the demand evolution over an adaptively selected horizon of $X^{[\tau']}$ future scheduling timeslots. The resulting prediction is subsequently used by Module B to roll forward the ES states and evaluate the downstream impact of the current scheduling decisions. As the scheduling horizon advances, both the prediction horizon and the corresponding demand estimates are updated using newly observed states.

\noindent
$\bullet$~\textit{Determination of $X^{[\tau']}$.}
Under concurrent inference execution, a scheduling decision made at the current timeslot may affect ES states over multiple subsequent timeslots. Specifically, ongoing tasks continuously occupy computing resources, newly admitted tasks introduce additional resource contention and may prolong the execution of existing tasks, while completed tasks release the occupied resources. Therefore, $X^{[\tau']}$ should be sufficiently long to capture the temporal span over which tasks admitted at $\tau'$ can influence future ES states, while avoiding unnecessary prediction and optimization overhead. Specifically, $X^{[\tau']}$ is selected as the smallest positive integer satisfying
$
X^{[\tau']}\Delta\tau
\geq
\max\left\{
t_m^{[\tau],\mathsf{ESmax}}
\,\middle|\,
\alpha_m^{[\tau]}=1,\,
\left\lceil {\hat{\tau}}^{[\tau],\mathsf{TransC}}_m\right\rceil=\tau'
\right\}.
$
Since any feasible CCT announced for a task scheduled at $\tau'$ is upper-bounded by its corresponding $t_m^{[\tau],\mathsf{ESmax}}$, this choice covers the longest admissible commitment window induced by the current scheduling decisions.

\noindent
$\bullet$~\textit{$X^{[\tau']}$-Timeslot demand estimation.}
At scheduling timeslot $\tau'$, the currently arrived tasks and ES states are directly observable. Module A therefore estimates the future inference demand over timeslots $\{\tau'+1,\ldots,\tau'+X^{[\tau']}\}$. By characterizing stochastic task generation together with the associated local-computation and transmission processes, the PT estimates, for each future scheduling timeslot, the expected number of tasks becoming schedulable and their expected aggregate edge-side computing workload. Accordingly, we define $
\bm{Y}^{[\tau']}
=
\left[
\mathbb{E}\!\left(Y^{[\tau'+1]}\right),
\ldots,
\mathbb{E}\!\left(Y^{[\tau'+X^{[\tau']}]}\right)
\right],
$ and $
\bm{W}^{[\tau']}
=
\left[
\mathbb{E}\!\left(W^{[\tau'+1]}\right),
\ldots,
\mathbb{E}\!\left(W^{[\tau'+X^{[\tau']}]}\right)
\right]$, which characterize the predicted task-arrival and workload profiles, respectively. Their analytical derivations are provided in Appx.~\ref{appxdemandestimation}. These prediction profiles serve as the predictive inputs of the rolling-horizon optimization in Module B and are recomputed at every scheduling timeslot as new system states become available.

%------------------------------------------------section 4.2
%------------------------------------------------section 4.2
%------------------------------------------------section 4.2
%------------------------------------------------section 4.2
%------------------------------------------------section 4.2

\subsection{Module B. Forward-Looking Rolling-Horizon Optimization for SD-ES Mapping and CCT Determination with Implicit Time Decoupling}
\label{moduleb}
As established in the previous section, Module A provides the predicted task-arrival and computational-workload profiles over the planning horizon. Since these predictions characterize statistical expectations rather than exact realizations of future system states, the underlying uncertainty cannot be completely eliminated. Instead, by treating the predicted demand as certainty-equivalent inputs, the PT obtains a tractable representation of future workload evolution that can be incorporated into the current scheduling process. Building upon this predictive representation, Module B develops a
receding-horizon optimization framework that approximates the original
time-coupled problem $\bm{\mathcal{P}}_0$ through forward state rollout, while continuously
correcting prediction mismatch as newly observed system states become
available.

%------------------------------------------------section 4.2.1
%------------------------------------------------section 4.2.1
%------------------------------------------------section 4.2.1
%------------------------------------------------section 4.2.1

\noindent
$\bullet$~\textbf{\textit{Transforming $\bm{\mathcal{P}}_0$ to $\bm{\mathcal{P}}_{1}$.}}
Consider an arbitrary scheduling timeslot $\tau'$. At the beginning of $\tau'$, the PT has access to two types of information: \textit{(i) The currently observed ES states}, including the tasks already under processing and the corresponding server-state descriptors $\mathsf{sch}_n^{[\tau']}$, $\forall \bm{s}_n\in\mathcal{S}$. \textit{(ii) The predicted future demand profiles} over the subsequent $X^{[\tau']}$ timeslots, characterized by $\bm{Y}^{[\tau']}$ and $\bm{W}^{[\tau']}$.

Recall that $\bm{\mathcal{P}}_0$ maximizes the cumulative TCR and service reward over the entire time horizon. At scheduling timeslot $\tau'$, however, the eventual completion outcomes of currently admitted tasks are not yet observable, since their execution may span multiple subsequent timeslots and be affected by future workload arrivals. We therefore construct a predictive scheduling-epoch counterpart, denoted by $\bm{\mathcal{P}}_{1}$, in which the completion performance induced by the current decisions is evaluated through the projected ES-state evolution under the demand estimates provided by Module A. Accordingly,
\begin{equation}
\label{P1}
\bm{\mathcal{P}}_{1}:
~\underset{x,\hat{t}}{~\text{maximize}}
\quad
\lambda^{\mathsf{cmlp}}\hat{\mathbb{C}}^{[\tau']}
+
\lambda^{\mathsf{rwd}}\hat{\mathbb{R}}^{[\tau']},
\end{equation}
where $\hat{\mathbb{C}}^{[\tau']}$ and $\hat{\mathbb{R}}^{[\tau']}$ denote the predictive completion ratio and predictive service reward associated with the scheduling decisions made at $\tau'$, respectively. Their explicit forms will be derived after characterizing the rolling evolution of the ES states. Importantly, $\bm{\mathcal{P}}_{1}$ should not be interpreted as a myopic decomposition of $\bm{\mathcal{P}}_0$. Although only the scheduling decisions made at $\tau'$ are ultimately committed, their consequences are evaluated over the subsequent prediction horizon through $\bm{Y}^{[\tau']}$ and $\bm{W}^{[\tau']}$\footnote{At a given scheduling timeslot $\tau'$, newly assigned tasks share the computing resources of an ES with previously admitted tasks that remain unfinished. Consequently, admitting a new task may prolong not only its own execution but also the completion times of existing tasks. This coupling further propagates across subsequent timeslots because future task arrivals alter the number of concurrently served tasks and hence the effective computing capability of the ES (see Sec.~\ref{sec: server modeling}). Therefore, the completion component evaluated at $\tau'$ jointly accounts for newly admitted and previously admitted tasks, while the predicted future arrivals affect their completion feasibility through the projected ES-state evolution. The service reward associated with a previously admitted task, however, is not counted repeatedly, since its CCT was already committed when the task was admitted.}. In this way, the finite-horizon prediction preserves the downstream impact of current scheduling decisions while enabling tractable online optimization.

%------------------------------------------------section 4.2.2
%------------------------------------------------section 4.2.2
%------------------------------------------------section 4.2.2
%------------------------------------------------section 4.2.2

\noindent
$\bullet$~\textbf{\textit{Transforming $\bm{\mathcal{P}}_{1}$ to $\bm{\mathcal{P}}_{1'}$.}}
We first characterize the rolling evolution of tasks hosted by each ES. Let
\begin{equation}
\mathcal{Y}^{[\tau'],\mathsf{new}}
=
\left\{
(m,\tau)
\,\middle|\,
\alpha_m^{[\tau]}=1,\,
\left\lceil
{\hat{\tau}}_m^{[\tau],\mathsf{TransC}}
\right\rceil=\tau'
\right\}
\end{equation}
denote the set of tasks that become ready for scheduling at $\tau'$. The assignment decision partitions $\mathcal{Y}^{[\tau'],\mathsf{new}}$ among the ESs, where
$\mathcal{Y}_n^{[\tau'],\mathsf{new}}\subseteq
\mathcal{Y}^{[\tau'],\mathsf{new}}$
denotes the subset assigned to ES $\bm{s}_n$. Meanwhile,
$\mathcal{Y}_n^{[\tau'],\mathsf{old}}$
collects the previously admitted tasks that remain unfinished at the beginning of $\tau'$. Accordingly, the complete set of tasks requiring processing by $\bm{s}_n$ at the beginning of $\tau'$ is
\begin{equation}
\label{setalltau}
\mathcal{Y}_n^{[\tau'],\mathsf{all}}
=
\mathcal{Y}_n^{[\tau'],\mathsf{new}}
\cup
\mathcal{Y}_n^{[\tau'],\mathsf{old}}.
\end{equation}
Let $\mathcal{Y}_n^{[\tau'],\mathsf{lv}}$ denote the set of tasks completed during timeslot $\tau'$ and hence released from ES $\bm{s}_n$. The unfinished-task set then evolves according to
\begin{equation}
\label{setold}
\mathcal{Y}_n^{[\tau'+1],\mathsf{old}}
=
\mathcal{Y}_n^{[\tau'],\mathsf{all}}
\setminus
\mathcal{Y}_n^{[\tau'],\mathsf{lv}}.
\end{equation}
Therefore, (\ref{setalltau}) and (\ref{setold}) jointly characterize the rolling evolution of the task population at each ES: unfinished tasks are inherited from the preceding timeslot, completed tasks leave the ES, and newly scheduled tasks are admitted.

For notational convenience, each task is mapped to an index $\kappa$ for subsequent state representation, i.e., $\kappa=\phi(m,\tau)$\footnote{At a scheduling timeslot $\tau'$, multiple tasks may originate from different SDs and timeslots. Although scheduling decisions are made at integer-valued timeslots, their actual arrival times may differ. Therefore, the corresponding task indices can be reordered according to a first-come, first-served (FCFS) rule to facilitate subsequent analysis.}. We then revisit the ES-state descriptor $\mathsf{sch}_n^{[\tau']}$ introduced in Sec.~\ref{sec: server modeling}. After admitting $\mathcal{Y}_n^{[\tau'],\mathsf{new}}$, it is represented as
\begin{equation}
\label{schdesign}
\mathsf{sch}_n^{[\tau']}
=
\left\{
\left(
\mathrm{w}_\kappa^{[\tau'],\mathsf{rem}},
t_\kappa^{\mathsf{arr}},
\hat{t}_\kappa,
t_\kappa^{\mathsf{max}}
\right)
\,\middle|\,
\kappa\in\mathcal{Y}_n^{[\tau'],\mathsf{all}}
\right\}.
\end{equation}
Here, $\mathrm{w}_\kappa^{[\tau'],\mathsf{rem}}$ denotes the remaining edge-side computational workload of task $\kappa$ at the beginning of $\tau'$; $t_\kappa^{\mathsf{arr}}$ denotes its actual arrival time for edge-side processing, e.g., $t_\kappa^{\mathsf{arr}}={\hat{\tau}}_\kappa^{\mathsf{TransC}}$; $\hat{t}_\kappa$ is its announced CCT; and $t_\kappa^{\mathsf{max}}$ denotes its maximum tolerable edge-side processing duration. For a newly scheduled task, $t_\kappa^{\mathsf{max}}$ corresponds to the previously defined $t_m^{[\tau],\mathsf{ESmax}}$. The CCTs of previously admitted tasks have already been committed and are therefore retained as fixed state information in $\mathsf{sch}_n^{[\tau']}$, whereas those of newly admitted tasks remain decision variables at $\tau'$.

Following Sec.~\ref{sec: server modeling}, the instantaneous computing capability available to each task depends on the number of concurrently served tasks. Accordingly, the corresponding ES-capacity condition is
\begin{equation}
|\mathcal{Y}_n^{[\tau'],\mathsf{all}}|
\leq
\omega_n^{\mathsf{max}},
\quad
\forall \bm{s}_n\in\mathcal{S}.
\end{equation}
During timeslot $\tau'$, the maximum amount of workload processable for each active task at ES $\bm{s}_n$ is
$\Delta\mathrm{w}_n^{[\tau'],\mathsf{ES}}=\Delta\tau f_n^{[\tau']}$.
Since the remaining workload of a task may be smaller than this amount, its actual processed workload is defined as
\begin{equation}
\Delta\mathrm{w}_\kappa^{[\tau']}
=
\min
\left\{
\mathrm{w}_\kappa^{[\tau'],\mathsf{rem}},
\Delta\mathrm{w}_n^{[\tau'],\mathsf{ES}}
\right\},
\quad
\kappa\in\mathcal{Y}_n^{[\tau'],\mathsf{all}}.
\end{equation}
Accordingly, the remaining workload evolves as
\begin{equation}
\label{workload update}
\mathrm{w}_\kappa^{[\tau'+1],\mathsf{rem}}
=
\mathrm{w}_\kappa^{[\tau'],\mathsf{rem}}
-
\Delta\mathrm{w}_\kappa^{[\tau']}.
\end{equation}
A task completes during timeslot $\tau'$ if
$\mathrm{w}_\kappa^{[\tau'+1],\mathsf{rem}}=0$,
in which case it is included in
$\mathcal{Y}_n^{[\tau'],\mathsf{lv}}$
and no longer occupies the computational resources of $\bm{s}_n$ in the subsequent timeslot.

To distinguish task completion from commitment satisfaction, for any processing timeslot $\tau'^{+}\geq\tau'$, we define
\begin{equation}
\label{events}
\begin{cases}
\mathbb{A}_\kappa^{[\tau'^{+}]}:
~\mathrm{w}_\kappa^{[\tau'^{+}+1],\mathsf{rem}}=0,\\[2mm]
\mathbb{B}_\kappa^{[\tau'^{+}]}:
~(t_\kappa^{\mathsf{end}}-t_\kappa^{\mathsf{arr}})
\Delta\tau
\leq
\hat{t}_\kappa.
\end{cases}
\end{equation}
Here, $t_\kappa^{\mathsf{end}}$ denotes the actual completion time point of task $\kappa$. Event $\mathbb{A}_\kappa^{[\tau'^{+}]}$ indicates that the task completes during timeslot $\tau'^{+}$, whereas $\mathbb{B}_\kappa^{[\tau'^{+}]}$ indicates that its realized edge-side processing duration does not exceed the announced CCT. Therefore, we define the indicator
\begin{equation}
\label{indicatorfunc}
\mathbbm{1}
\left(
\mathbb{A}_\kappa^{[\tau'^{+}]}
~\text{and}~
\mathbb{B}_\kappa^{[\tau'^{+}]}
\right)
\in\{0,1\},
\end{equation}
which characterizes the realized commitment outcome.

The indicator in (\ref{indicatorfunc}), however, is an \emph{ex-post} quantity and is unavailable when the scheduling decision at $\tau'$ is made. To enable forward-looking decision making, the PT instead propagates the current ES states over the prediction horizon according to (\ref{setold}) and (\ref{workload update}), while incorporating the predicted demand profiles $\bm{Y}^{[\tau']}$ and $\bm{W}^{[\tau']}$. Specifically, at each future timeslot $q\in\{\tau'+1,\ldots,\tau'+X^{[\tau']}\}$, the predicted aggregate task arrivals and edge-side workloads are uniformly distributed across the ESs for certainty-equivalent state propagation, such that each ES is associated with an expected load share of $\mathbb{E}(Y^{[q]})/|\mathcal{S}|$ and $\mathbb{E}(W^{[q]})/|\mathcal{S}|$, respectively. These quantities are used only to characterize the anticipated future resource contention during rollout and do not represent actual task-assignment decisions. Let $\widetilde{t}_\kappa^{[\tau'],\mathsf{end}}$
denote the resulting predicted completion time of task $\kappa$ under a candidate scheduling decision made at $\tau'$. If the task is not predicted to complete within the considered horizon, we set
$\widetilde{t}_\kappa^{[\tau'],\mathsf{end}}=+\infty$.
The corresponding predictive commitment indicator is defined as
\begin{equation}
\label{predindicator}
\mathbbm{1}_\kappa^{[\tau'],\mathsf{pred}}
=
\mathbbm{1}
\left(
\left(
\widetilde{t}_\kappa^{[\tau'],\mathsf{end}}
-
t_\kappa^{\mathsf{arr}}
\right)
\Delta\tau
\leq
\hat{t}_\kappa
\right).
\end{equation}
Unlike the realized indicator in (\ref{indicatorfunc}), (\ref{predindicator}) can be evaluated at the current scheduling epoch through predictive state rollout and explicitly depends on the CCT $\hat{t}_\kappa$. Accordingly, the predictive completion ratio evaluated at $\tau'$ is
\begin{equation}
\label{predTCR}
\hat{\mathbb{C}}^{[\tau']}
=
\frac{
\displaystyle
\sum_{\bm{s}_n\in\mathcal{S}}
\sum_{\kappa\in\mathcal{Y}_n^{[\tau'],\mathsf{all}}}
\mathbbm{1}_\kappa^{[\tau'],\mathsf{pred}}
}{
\displaystyle
\sum_{\bm{s}_n\in\mathcal{S}}
\left|
\mathcal{Y}_n^{[\tau'],\mathsf{all}}
\right|
}.
\end{equation}
Both newly admitted and previously admitted tasks are included in (\ref{predTCR}), since the current assignment decisions can affect the future completion feasibility of either type of task. In contrast, the service reward is evaluated only for newly admitted tasks whose CCTs are determined at the current scheduling epoch, thereby avoiding repeated reward accounting for previously committed tasks. Hence,
\begin{equation}
\label{predreward}
\hat{\mathbb{R}}^{[\tau']}
=
\sum_{\bm{s}_n\in\mathcal{S}}
\sum_{\kappa\in\mathcal{Y}_n^{[\tau'],\mathsf{new}}}
\mathbbm{1}_\kappa^{[\tau'],\mathsf{pred}}
\left(
t_\kappa^{\mathsf{max}}
-
\hat{t}_\kappa
\right).
\end{equation}

Substituting (\ref{predTCR}) and (\ref{predreward}) into (\ref{P1}), the scheduling problem at $\tau'$ is explicitly written as
\begin{equation}
%\tag{23}
\label{P1'}
\bm{\mathcal{P}}_{1'}:
~\underset{
\substack{
\left\{\mathcal{Y}_n^{[\tau'],\mathsf{new}}\right\}_{\bm{s}_n\in\mathcal{S}},\\
\left\{\hat{t}_\kappa\right\}_{\kappa\in\mathcal{Y}^{[\tau'],\mathsf{new}}}
}
}{\text{maximize}}
\quad
\lambda^{\mathsf{cmlp}}\hat{\mathbb{C}}^{[\tau']}
+
\lambda^{\mathsf{rwd}}\hat{\mathbb{R}}^{[\tau']}.
\end{equation}
\begin{center}
\textit{s.t.}
\end{center}
\begin{flalign*}
&\text{(C5):}\quad
\bigcup_{\bm{s}_n\in\mathcal{S}}
\mathcal{Y}_n^{[\tau'],\mathsf{new}}
=
\mathcal{Y}^{[\tau'],\mathsf{new}},
&&\\
&\text{(C6):}\quad
\mathcal{Y}_n^{[\tau'],\mathsf{new}}
\cap
\mathcal{Y}_{n'}^{[\tau'],\mathsf{new}}
=
\emptyset,
\quad
\forall n\neq n',
&&\\
&\text{(C7):}\quad
\left|
\mathcal{Y}_n^{[\tau'],\mathsf{all}}
\right|
\leq
\omega_n^{\mathsf{max}},
\quad
\forall \bm{s}_n\in\mathcal{S},
&&\\
&\text{(C8):}\quad
0<\hat{t}_\kappa
\leq
t_\kappa^{\mathsf{max}},
\quad
\forall
\kappa\in\mathcal{Y}^{[\tau'],\mathsf{new}}.
&&
\end{flalign*}
Problem $\bm{\mathcal{P}}_{1'}$ is a certainty-equivalent predictive counterpart of $\bm{\mathcal{P}}_0$ at scheduling timeslot $\tau'$, rather than an exact temporal decomposition. The current task-assignment and CCT decisions are evaluated by rolling the ES states forward over the subsequent $X^{[\tau']}$ timeslots under the predicted demand evolution. Only the decisions associated with the currently schedulable tasks are ultimately committed. Once the system advances to $\tau'+1$, the prediction profiles are recomputed from the newly observed system states and the optimization is solved again. This receding-horizon mechanism preserves the downstream impact of current decisions while continuously correcting prediction mismatch and adapting to newly realized workload/resource dynamics.

\begin{algorithm}[t!]
\footnotesize
\SetAlgoVlined
\LinesNumbered
\caption{Prediction-guided surrogate rollout with uniform future loading for PROMISE}
\label{alg:promise_rollout}

\KwIn{
Current ES states
$\{\mathsf{sch}_n^{[\tau']}\}_{\bm{s}_n\in\mathcal{S}}$;
currently schedulable tasks
$\mathcal{Y}^{[\tau'],\mathsf{new}}$;
predicted demand profiles
$\bm{Y}^{[\tau']}$ and $\bm{W}^{[\tau']}$;
prediction horizon $X^{[\tau']}$
}

\KwOut{
Committed mapping
$\{\mathcal{Y}_n^{[\tau'],\mathsf{new}}\}_{\bm{s}_n\in\mathcal{S}}$
and CCTs
$\{\hat{t}_{\kappa}^{*}\}_{\kappa\in\mathcal{Y}^{[\tau'],\mathsf{new}}}$
}

Construct
$
\mathcal{Y}^{[\tau'],\mathsf{act}}
=
\mathcal{Y}^{[\tau'],\mathsf{new}}
\cup
\left(
\bigcup_{\bm{s}_n\in\mathcal{S}}
\mathcal{Y}_n^{[\tau'],\mathsf{old}}
\right)
$\;

Construct the complete set of feasible current-stage mappings satisfying
(C5)--(C7)\;

Initialize
$\mathbbm{V}^{[\tau'],*}\leftarrow-\infty$\;

\BlankLine
\tcp{Predictively evaluate every feasible current-stage mapping}

\ForEach{feasible current-stage mapping}{
    Initialize a rollout trajectory $\pi$ using the observed ES states
    and the considered mapping\;

    Initialize
    $\mathbbm{V}^{[\tau']}(\pi)\leftarrow0$\;

    \For{$\tau'^{+}=\tau'$ \KwTo $\tau'+X^{[\tau']}$}{

        \If{$\tau'^{+}>\tau'$}{
            Obtain
            $\mathbb{E}(Y^{[\tau'^{+}]})$
            and
            $\mathbb{E}(W^{[\tau'^{+}]})$
            from
            $\bm{Y}^{[\tau']}$ and $\bm{W}^{[\tau']}$\;

            Uniformly distribute the predicted aggregate demand across
            the ESs, such that each $\bm{s}_n$ is associated with
            $\mathbb{E}(Y^{[\tau'^{+}]})/|\mathcal{S}|$
            in expected arrival share and
            $\mathbb{E}(W^{[\tau'^{+}]})/|\mathcal{S}|$
            in expected workload share\;

            Incorporate the resulting certainty-equivalent background
            load into the projected ES states\;
        }

        Compute the projected computing capability of each ES
        according to the resulting rollout state\;

        \ForEach{$\kappa\in\mathcal{Y}^{[\tau'],\mathsf{act}}$}{
            Compute
            $\Delta\mathrm{w}_{\kappa,\pi}^{[\tau'^{+}]}$
            according to the corresponding projected ES state\;

            Compute
            $\mathbbm{r}_{\kappa,\pi}^{[\tau'^{+}]}$
            according to (\ref{singleslotreward})\;

            Update
            $\mathrm{w}_{\kappa,\pi}^{[\tau'^{+}+1],\mathsf{rem}}$
            according to (\ref{workload update})\;

            \If{$\kappa$ completes during $\tau'^{+}$}{
                Record
                $\widetilde{t}_{\kappa,\pi}^{[\tau'],\mathsf{end}}$\;
            }
        }

        Update
        \[
        \mathbbm{V}^{[\tau']}(\pi)
        \leftarrow
        \mathbbm{V}^{[\tau']}(\pi)
        +
        \sum_{\kappa\in\mathcal{Y}^{[\tau'],\mathsf{act}}}
        \mathbbm{r}_{\kappa,\pi}^{[\tau'^{+}]}
        \]

        Update the projected ES task sets according to (\ref{setold})\;

        \If{all tasks in $\mathcal{Y}^{[\tau'],\mathsf{act}}$ have completed}{
            \textbf{break}\;
        }
    }

    Set
    $\widetilde{t}_{\kappa,\pi}^{[\tau'],\mathsf{end}}\leftarrow+\infty$
    for any
    $\kappa\in\mathcal{Y}^{[\tau'],\mathsf{act}}$
    not completed within the rollout horizon\;

    \If{$\mathbbm{V}^{[\tau']}(\pi)>\mathbbm{V}^{[\tau'],*}$}{
        $\mathbbm{V}^{[\tau'],*}\leftarrow
        \mathbbm{V}^{[\tau']}(\pi)$\;

        $\pi^{*}\leftarrow\pi$\;
    }
}

\BlankLine
\tcp{Recover CCTs and implement the first-stage decision}

\ForEach{$\kappa\in\mathcal{Y}^{[\tau'],\mathsf{new}}$}{
    Determine
    $\hat{t}_{\kappa}^{*}$
    from
    $\widetilde{t}_{\kappa,\pi^{*}}^{[\tau'],\mathsf{end}}$
    according to (\ref{optimalCCT})\;
}

Commit only the first-stage SD--ES mapping contained in $\pi^{*}$
and the corresponding CCTs
$\{\hat{t}_{\kappa}^{*}\}$\;

\KwRet
$\{\mathcal{Y}_n^{[\tau'],\mathsf{new}}\}_{\bm{s}_n\in\mathcal{S}}$
and
$\{\hat{t}_{\kappa}^{*}\}_{\kappa\in\mathcal{Y}^{[\tau'],\mathsf{new}}}$\;

\end{algorithm}

%------------------------------------------------section 4.2.3
%------------------------------------------------section 4.2.3
%------------------------------------------------section 4.2.3
%------------------------------------------------section 4.2.3

\noindent
$\bullet$~\textbf{\textit{Prediction-guided surrogate rollout with uniform future loading for tackling $\bm{\mathcal{P}}_{1'}$.}}
Although $\bm{\mathcal{P}}_{1'}$ is formulated over a finite prediction horizon, directly optimizing the multi-timeslot decision sequence remains computationally demanding. In particular, multiple tasks may become simultaneously schedulable at $\tau'$, yielding a combinatorial set of feasible current SD--ES mappings, while further optimizing the assignments of uncertain future arrivals would introduce additional branching across the prediction horizon. PROMISE therefore adopts a prediction-guided rollout architecture. Specifically, all feasible mappings of the currently schedulable tasks are considered subject to (C5)-(C7), whereas the unknown future demand is represented by the certainty-equivalent estimates supplied by Module A rather than by explicitly optimizing its future assignments. For each feasible current-stage mapping, the corresponding ES states are rolled forward over the current timeslot and the subsequent $X^{[\tau']}$ predicted timeslots. The resulting trajectory quantifies how the current decision affects workload processing and task completion under anticipated future resource contention. In this way, PROMISE preserves complete exploration of the decisions that can actually be committed at $\tau'$, while avoiding construction of a multi-stage combinatorial decision tree for future unrealized tasks (pseudocode of the overall procedure is given in Alg.~\ref{alg:promise_rollout}). Rather than exactly solving the predictive objective in \(\bm{\mathcal{P}}_{1'}\), the rollout uses the completion-progress surrogate introduced below to rank all feasible current-stage mappings under a common predicted continuation.

\noindent
\textit{(i) Uniform future loading for predictive state rollout.}
Consider a feasible current-stage mapping and let $\pi$ denote the predictive trajectory induced by this mapping. At the current timeslot $\tau'$, the rollout is initialized using the actually observed ES states and currently schedulable tasks. For each subsequent timeslot $\tau'^{+}\in\{\tau'+1,\ldots,\tau'+X^{[\tau']}\}$, Module A provides the certainty-equivalent aggregate task arrival $\mathbb{E}(Y^{[\tau'^{+}]})$ and edge-side computational workload $\mathbb{E}(W^{[\tau'^{+}]})$. Since the exact future SD--ES associations are unavailable at $\tau'$, PROMISE distributes the predicted aggregate demand uniformly across the ESs. Accordingly, each $\bm{s}_n\in\mathcal{S}$ is associated with an expected arrival share $\mathbb{E}(Y^{[\tau'^{+}]})/|\mathcal{S}|$ and an expected workload share $\mathbb{E}(W^{[\tau'^{+}]})/|\mathcal{S}|$ during predictive state propagation. These quantities characterize certainty-equivalent background loading rather than realized task assignments, and are incorporated into the projected ES states to capture their impacts on future resource contention, effective computing capability, and workload evolution. Importantly, the same predicted future loading is imposed on every feasible current-stage candidate, thereby providing a common continuation model for comparing their downstream consequences without introducing unobserved future scheduling preferences. Once the system advances to the next actual scheduling epoch, these predictive quantities are discarded and replaced by newly observed states and updated predictions.

\noindent
\textit{(ii) Progress-aware surrogate evaluation of predictive trajectories.}
The predictive objective in $\bm{\mathcal{P}}_{1'}$ ultimately depends on whether tasks can satisfy their corresponding CCTs. However, the commitment indicator in (\ref{predindicator}) provides a sparse completion signal and becomes fully determined only after the corresponding predicted completion time is obtained. To continuously quantify the evolution toward task completion during rollout, we exploit the processed workload as a dense completion-oriented surrogate. Specifically, define
\begin{equation}
\label{activeTasks}
\mathcal{Y}^{[\tau'],\mathsf{act}}
=
\mathcal{Y}^{[\tau'],\mathsf{new}}
\cup
\left(
\bigcup_{\bm{s}_n\in\mathcal{S}}
\mathcal{Y}_n^{[\tau'],\mathsf{old}}
\right)
\end{equation}
as the set of tasks whose completion performance can be affected by the scheduling decision at $\tau'$. For each $\kappa\in\mathcal{Y}^{[\tau'],\mathsf{act}}$, its normalized single-timeslot processing reward along trajectory $\pi$ is defined as
\begin{equation}
\label{singleslotreward}
\mathbbm{r}_{\kappa,\pi}^{[\tau'^{+}]}
=
\frac{
\Delta\mathrm{w}_{\kappa,\pi}^{[\tau'^{+}]}
}{
\mathrm{w}_{\kappa}^{[\tau'],\mathsf{rem}}
}.
\end{equation}
According to the workload evolution in (\ref{workload update}), the accumulated reward up to any rollout timeslot $q\geq\tau'$ satisfies
\begin{equation}
\label{progressidentity}
\sum_{\tau'^{+}=\tau'}^{q}
\mathbbm{r}_{\kappa,\pi}^{[\tau'^{+}]}
=
1-
\frac{
\mathrm{w}_{\kappa,\pi}^{[q+1],\mathsf{rem}}
}{
\mathrm{w}_{\kappa}^{[\tau'],\mathsf{rem}}
}.
\end{equation}
Hence, the cumulative single-timeslot reward exactly represents the fraction of the initial remaining workload of task $\kappa$ that has been processed along trajectory $\pi$, increasing from 0 to 1 as the task approaches completion. In particular, the first timeslot at which the right-hand side of (\ref{progressidentity}) reaches 1 identifies the predicted completion timeslot, from which $\widetilde{t}_{\kappa,\pi}^{[\tau'],\mathsf{end}}$ can be obtained and further refined within the terminal timeslot according to the residual workload and the corresponding ES computing capability.

Based on the above characterization, we define the completion-progress return of trajectory $\pi$ as
\begin{equation}
\label{rolloutvalue}
\mathbbm{V}^{[\tau']}(\pi)
=
\sum_{\tau'^{+}=\tau'}^{\tau'+X^{[\tau']}}
\sum_{\kappa\in\mathcal{Y}^{[\tau'],\mathsf{act}}}
\mathbbm{r}_{\kappa,\pi}^{[\tau'^{+}]}.
\end{equation}
%The quantity in (\ref{rolloutvalue}) is not an alternative formulation of $\bm{\mathcal{P}}_{1'}$, but a tractable surrogate used to evaluate the downstream consequences of different current-stage mappings. Under the same predicted future loading, a larger $\mathbbm{V}^{[\tau']}(\pi)$ indicates that a larger fraction of the currently committed workload can be processed within the look-ahead horizon. This generally promotes earlier task completion and faster resource release, thereby improving the prospective feasibility of satisfying the predictive commitments characterized in (\ref{predindicator}). Consequently, (\ref{rolloutvalue}) provides a dense completion-oriented criterion for approximating the impact of the current mapping on the original predictive objective.
The quantity in (\ref{rolloutvalue}) serves as a completion-oriented
surrogate for evaluating the downstream consequences of different
current-stage mappings. Under the same predicted future-loading profile,
the candidate mappings differ in how the currently active workloads are
placed across the ESs and subsequently processed. A larger
$\mathbbm{V}^{[\tau']}(\pi)$ therefore indicates that a larger fraction
of these workloads can be processed within the look-ahead horizon, which
generally promotes earlier predicted completion and earlier resource
release. These effects improve the prospective feasibility of satisfying
the predictive commitments and, for newly admitted tasks, enable tighter
CCTs to be recovered through (\ref{optimalCCT}). Hence,
$\mathbbm{V}^{[\tau']}(\pi)$ provides a tractable mapping-selection
criterion that captures the completion progress underlying the predictive
completion and commitment objectives in $\bm{\mathcal{P}}_{1'}$.
Accordingly, PROMISE uses this progress-based evaluation to efficiently
rank the feasible current-stage mappings, followed by analytical CCT
recovery.

\noindent
\textit{(iii) First-stage mapping selection and analytical CCT recovery.}
Let $\Pi^{[\tau']}$ collect the predictive trajectories induced by all feasible current-stage mappings under the common uniform future-loading model described above. PROMISE selects
\begin{equation}
\label{besttrajectory}
\pi^{*}
=
\arg\max_{\pi\in\Pi^{[\tau']}}
\mathbbm{V}^{[\tau']}(\pi).
\end{equation}
It is worth emphasizing that all feasible mappings of the currently schedulable tasks are retained in this comparison; hence, no heuristic pruning is applied to the current-stage decision space. The approximation instead arises from representing uncertain future demand through the finite-horizon certainty-equivalent prediction and using the progress-oriented return in (\ref{rolloutvalue}) to evaluate its downstream effect. This design avoids recursively optimizing unknown future task assignments while preserving explicit look-ahead evaluation of every feasible decision that may actually be implemented at $\tau'$.

Once $\pi^{*}$ is selected, the predicted completion time
$\widetilde{t}_{\kappa,\pi^{*}}^{[\tau'],\mathsf{end}}$
of each newly admitted task is directly obtained from its rolled-out workload evolution. For a fixed trajectory, the predicted completion time is determined by the projected ES states and is independent of the announced CCT. Accordingly, for each
$\kappa\in\mathcal{Y}^{[\tau'],\mathsf{new}}$,
the CCT is analytically recovered as
\begin{equation}
\label{optimalCCT}
\hat{t}_{\kappa}^{*}
=
\min
\left\{
\left(
\widetilde{t}_{\kappa,\pi^{*}}^{[\tau'],\mathsf{end}}
-
t_{\kappa}^{\mathsf{arr}}
\right)\Delta\tau,
~
t_{\kappa}^{\mathsf{max}}
\right\}.
\end{equation}
When the predicted completion duration does not exceed $t_{\kappa}^{\mathsf{max}}$, any smaller CCT would violate the predictive commitment in (\ref{predindicator}), whereas any larger CCT would reduce the service reward in (\ref{predreward}); thus, the predicted completion duration constitutes the tightest successful commitment. If the task is not predicted to complete within its maximum tolerable duration, no feasible CCT can render the predictive commitment successful, and $t_{\kappa}^{\mathsf{max}}$ is adopted as the least aggressive admissible commitment. Importantly, the analytical recovery in (\ref{optimalCCT}) does not
decouple the CCT from the scheduling decision. The predicted completion
time is itself induced by the selected SD--ES mapping and its projected
ES-state evolution: different mappings result in different resource
contention, workload-processing trajectories, and consequently different
achievable commitment tightness. Therefore, (\ref{optimalCCT}) eliminates
the need for an additional continuous search over CCT only after the
mapping-induced trajectory has been determined. Moreover, the recovered
CCT is committed before future workload realizations become available,
such that prediction mismatch and subsequent stochastic arrivals can
still affect its eventual fulfillment. Finally, only the first-stage SD--ES mapping contained in $\pi^{*}$ and the corresponding CCTs in (\ref{optimalCCT}) are implemented at $\tau'$. Once advancing to $\tau'+1$, the prediction profiles and ES states are updated using newly observed information and the complete procedure is repeated, thereby realizing the receding-horizon operation of PROMISE.

\section{Simulations}
\label{simulation}
All experiments are conducted on a Windows 11 platform equipped with an Intel Core i9-13900H CPU and an NVIDIA GeForce RTX 4060 GPU; implemented in Python 3.11.5 with PyTorch 2.7.0 and CUDA 13.1. Unless otherwise specified, all experiments in this section use the same hardware/software environment for fair and consistent evaluation. 

\subsection{Privacy Evaluation on Real-world Datasets}
To assess the impact of privacy-aware model partitioning on multi-modal inference, we develop a unified evaluation framework spanning vision, text, and audio tasks. We systematically analyze the trade-offs between distributed and full-model inference across accuracy, privacy leakage, and efficiency metrics.

\noindent
$\bullet$~\textit{Datasets and DNN Models}.
To evaluate the differences between distributed and full-model inference across various models and datasets, we conduct experiments on four representative settings: \textit{(i)} ResNet18 on CIFAR-10~\cite{he2016deep}, \textit{(ii)} MobileNetV2 on MNIST~\cite{sandler2018mobilenetv2, lecun1998gradient}, \textit{(iii)} TextCNN on THUCNews~\cite{kim2014convolutional,sun2016thucnews}, and \textit{(iv)} M5 on UrbanSound8K~\cite{dai2017m5,salamon2014dataset}. All models are trained to convergence in a single-machine environment, with the full-model inference accuracy serving as the baseline for subsequent comparisons. To emulate different privacy levels, we select multiple candidate partition points according to network depth: \textit{8 for ResNet18, 8 for MobileNetV2, 5 for TextCNN, and 5 for M5}. We evaluate the impact of model partitioning using multiple metrics, including Top-1 accuracy, Top-5 accuracy, F1-score, and privacy leakage rate, enabling a comprehensive analysis of the privacy–performance trade-off.
%要参考文献

% \begin{table}[htbp]
% \centering
% \caption{Datasets and corresponding models}
% \label{datasetandmodel}
% \begin{tabular}{ccc}
% \toprule
% Modality & Dataset & Model \\
% \midrule
% Image & CIFAR-10 & ResNet18 \\
% Image & MNIST & MobileNetV2 \\
% Text & THUCNews & TextCNN \\
% Audio & UrbanSound8K & M5 \\
% \bottomrule
% \end{tabular}
% \end{table}

% \noindent
% $\bullet$~\textit{Candidate partition points}

\begin{figure}[t!]
	\centering
	\setlength{\abovecaptionskip}{-0.0 mm}
	\includegraphics[width=.9\columnwidth]{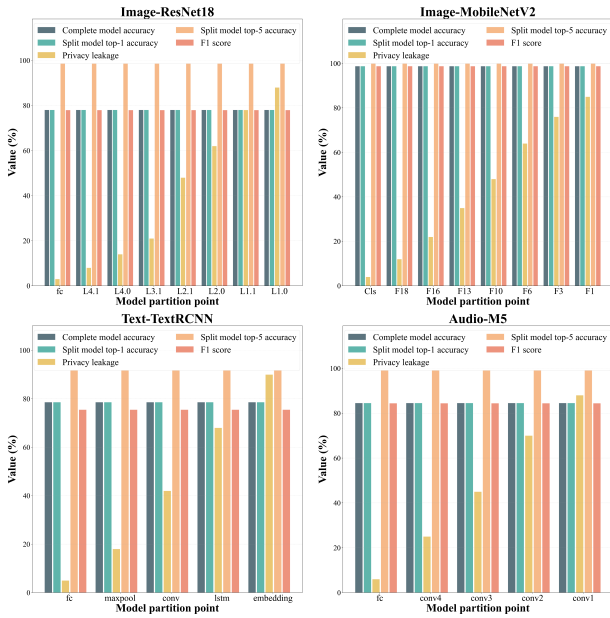}
	\caption{Comparison of inference metrics at different privacy partition points across different datasets and models.}
	\label{fig:realdata}
\end{figure}

\noindent
$\bullet$~\textit{Performance analysis}.As shown in Fig.~\ref{fig:realdata}, distributed inference achieves
identical Top-1 accuracy, Top-5 accuracy, and F1-score compared with
full-model inference. The prediction outputs of individual samples are
also numerically identical within floating-point precision. These results
show that, under identical numerical execution conditions, partitioned
forward propagation preserves the functional behavior of the original
network. Therefore, model partitioning alone does not necessarily
introduce inference degradation. Deviations observed in practical
distributed deployments may instead arise from additional deployment
factors, such as lossy intermediate-feature compression or quantization,
packet loss, heterogeneous numerical implementations, and
platform-specific preprocessing or runtime configurations. This
experiment therefore provides a controlled baseline for separating the
effect of model partitioning from additional deployment-induced factors. In the subsequent hardware evaluation (see Sec. \ref{hardware} for further experiments), we deploy the distributed inference system on real edge devices and investigate practical performance variations by considering communication overhead, computational constraints, and multi-node collaboration. The results presented in this section establish a controlled baseline for analyzing the impact of system-level factors in real-world distributed inference scenarios.

% 四个模型和dataset的推理精度图（共四张折线图）

\subsection{Evaluation on Numerical Datasets}
To evaluate PROMISE under dynamically coupled spatio-temporal inference
scenarios, we establish a discrete time-slot-based simulation environment.

\noindent
$\bullet$~\textit{Parameter and Benchmark Settings}. Key simulation parameter settings are summarized by Table \ref{Simulationparameters}.
\begin{table}[t]
\centering
\caption{Simulation parameters~\cite{computespeedv1,tmc2025}}
\label{Simulationparameters}
\footnotesize
\setlength{\tabcolsep}{1.5pt}  
\begin{tabular}{cccc}
\toprule
\rowcolor{gray!10}
Parameter & Notation & Value/Range & Unit \\
\midrule
\rowcolor{gray!10}
\multicolumn{4}{c}{\textbf{Basic parameters}} \\
Total time slots & $|\mathcal{T}|$  & 400 & slot \\
Slot duration & $\Delta \tau$  & 1.0 & s \\
Number of model types & $|\mathbb{L}|$ & 3 & -- \\
\addlinespace
\rowcolor{gray!10}
\multicolumn{4}{c}{\textbf{Task generation parameters}} \\
Task generation interval & $\mathbbm{n}_m$ & \([5, 20]\) & slot \\
Task generation probability & $p_m$ & \([0.5, 0.9]\) & -- \\
Per-batch workload (SD) & $\mathbbm{c}_m^{[\tau],\mathsf{SD}}$ & \(\{0.08,6.85\}\) & GFLOPs\\
Per-batch workload (ES) & $\mathbbm{c}_m^{[\tau],\mathsf{ES}}$ & \(\{0.085,6.705\}\) & GFLOPs\\
Number of batches & $\mathbb{D}^{[\tau]}_m$ & \([1, 3]\) & batches \\
Privacy requirement & $\rho_m^{[\tau]}$ & \([0.2, 0.8]\) & -- \\
\addlinespace
\rowcolor{gray!10}
\multicolumn{4}{c}{\textbf{Computation/communication parameters}} \\
Computing capability of SD & $f_m^{\mathsf{SD}}$ & \([0.8, 2.0]\) & GFLOPs/s \\
Total computing capability of ES & $f_n^{\mathsf{sum}}$ & \([20, 30]\) & GFLOPs/s \\
Maximum computing capacity & $f^{\mathsf{max}}_n$  & \([5, 10]\) & GFLOPs/s \\
Minimum computing capacity & $f^{\mathsf{min}}_n$ & \([0.5, 2.0]\) & GFLOPs/s \\
%\textcolor{red}{Transmission rate} & \( R_m \) & \([5.0, 7.0]\) & Mbps \\
Resource contention decay factor & \( \beta \) & \([1.1, 1.5]\) & -- \\
\bottomrule
\end{tabular}
\end{table}

To systematically examine the impact of temporal awareness on scheduling
performance, we construct five controlled benchmark strategies under the
same system model and evaluation settings. They span different levels of
future-state information and decision mechanisms, ranging from
information-agnostic assignment and instantaneous optimization to
statistical anticipation, short-horizon prediction, and iterative
optimization. This controlled comparison isolates how different degrees
of temporal information affect scheduling performance under the same
commitment-aware inference setting (see Table~\ref{tab:benchmark} for a
summary).

%\begin{itemize}

\noindent
\textit{(i) Expectation-aware assignment (EAA)} performs service assignment based on statistically characterized workload expectations, exploiting long-term demand patterns while neglecting the explicit evolution of future system states.

\noindent
\textit{(ii) Short-horizon predictive scheduling (SHPS)} leverages one-step-ahead workload forecasting to enhance scheduling decisions, capturing short-term temporal dependencies without performing multi-timeslot future-state exploration.

\noindent
\textit{(iii) Iterative joint optimization (IJO)} designs iterative optimization to jointly refine service scheduling decisions, providing a solution paradigm for handling coupled decision variables without explicit future trajectory prediction.

\noindent
\textit{(iv) Random feasible assignment (RFA)} provides an information-agnostic baseline by randomly exploring feasible service assignment strategies within the admissible decision space.

\noindent
\textit{(v) Instantaneous reward maximization (IRM)} follows a myopic decision paradigm by optimizing the immediate reward at each scheduling epoch, without considering the long-term impact of current decisions on future resource evolution.
%\end{itemize}

\begin{table}[t]
\centering
\caption{Temporal awareness of benchmark methods}
\label{tab:benchmark}
\footnotesize
\setlength{\tabcolsep}{4pt}
\begin{tabular}{lcc}
\toprule
\rowcolor{gray!10}
Method & Future Information & Horizon \\
\midrule
RFA & None & --\\
IRM & Current state & 1 slot\\
EAA & Expected demand & Average \\
SHPS & One-step prediction & Short \\
IJO & Implicit state & Iterative \\
\rowcolor{gray!10}
PROMISE & Multi-slot evaluation & Rolling\\
\bottomrule
\end{tabular}
\end{table}

% \noindent
% $\bullet$~\textit{Evaluation on task generation}.
% %这个地方的任务数量差的挺多 合理吗？但反正我们之后的性能还是比较好的
% %统计量和实际量
% %两个图

\noindent
$\bullet$~\textit{Evaluations on $\mathbb{C}$ and $\mathbb{R}$}. To further validate the scalability of PROMISE, we conduct experiments under varying problem scales by configuring different system sizes and workload settings (see Table \ref{tab:scenarios}), characterizing system dynamism. 

\begin{table}[htbp]
\centering
\caption{Scenario settings}
\label{tab:scenarios}
\footnotesize
\setlength{\tabcolsep}{0.3pt}  
\begin{tabular}{ccccc}
\toprule
\rowcolor{gray!10}
Scenario & \(|\mathcal{U}|\) & \(|\mathcal{S}|\) & \(\mathbbm{n}_m\) & \(\mathbb{D}_m\) \\
\midrule
S\#1 & 10, 15, \ldots, 30 & 5 & [5, 20] & [1, 3] \\
S\#2 & 60 & 3, 4, \ldots, 7 & [5, 20] & [1, 3] \\
S\#3 & 20 & 5 & [5,10], [5,15], \ldots, [5,25] & [1, 3] \\
S\#4 & 30 & 5 & [5, 20] & [1,10], \ldots, [1,50] \\
\bottomrule
\end{tabular}
\end{table}

\begin{figure}[t!]
	\centering
	\setlength{\abovecaptionskip}{-0.0 mm}
	\includegraphics[width=1\columnwidth]{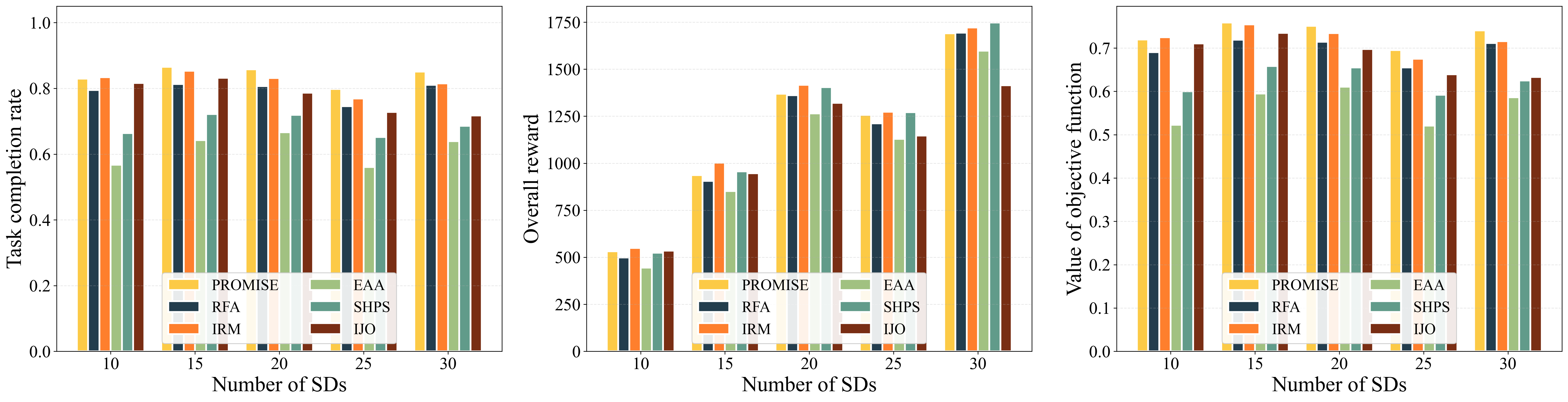}
	\caption{Evaluation on S\#1.}
	\label{fig:set1}
\end{figure}

\begin{figure}[t!]
	\centering
	\setlength{\abovecaptionskip}{-0.0 mm}
	\includegraphics[width=1\columnwidth]{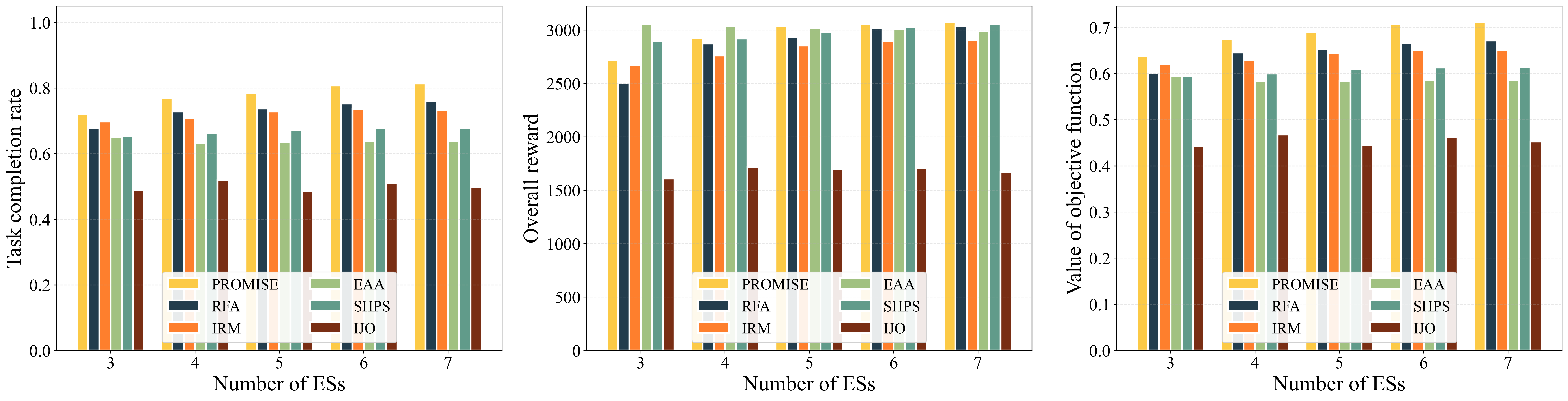}
	\caption{Evaluation on S\#2.}
	\label{fig:set2}
\end{figure}

\begin{figure}[t!]
	\centering
	\setlength{\abovecaptionskip}{-0.0 mm}
	\includegraphics[width=1\columnwidth]{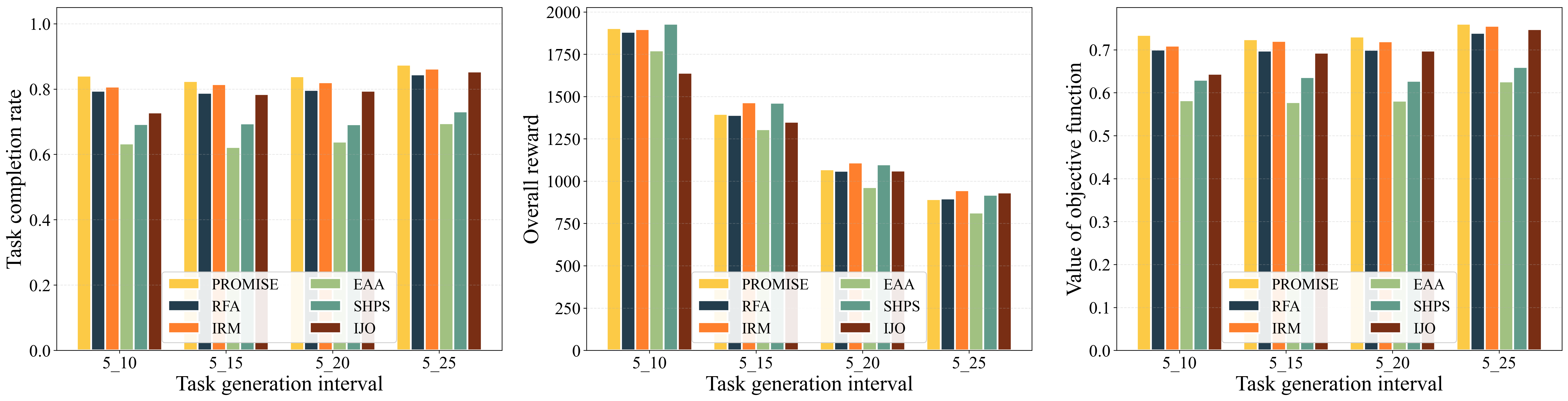}
	\caption{Evaluation on S\#3.}
	\label{fig:set3}
\end{figure}

\begin{figure}[t!]
	\centering
	\setlength{\abovecaptionskip}{-0.0 mm}
	\includegraphics[width=1\columnwidth]{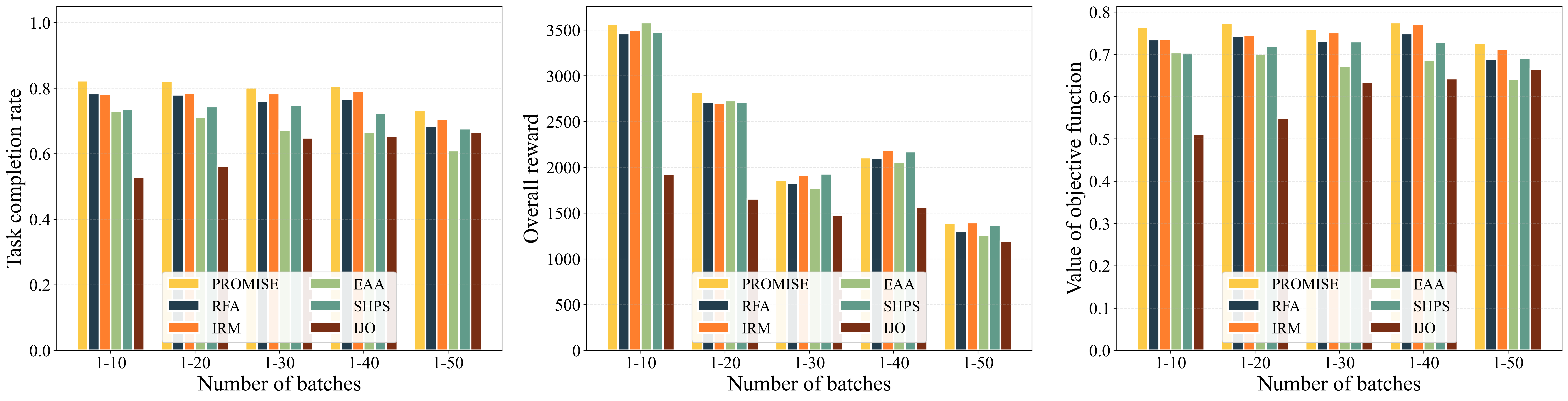}
	\caption{Evaluation on S\#4.}
	\label{fig:set4}
\end{figure}

As shown in Fig.~\ref{fig:set1}, increasing the number of SDs intensifies resource contention among concurrent tasks, leading to an overall decrease
in the task completion rate. Meanwhile, the overall reward increases with
the number of SDs because more generated tasks contribute to the cumulative
service reward over the scheduling horizon. Conversely, Fig.~\ref{fig:set2}
shows that increasing the number of ESs generally improves the task
completion rate by providing more computing resources for concurrent
inference. In Fig.~\ref{fig:set3}, enlarging the task-generation interval
reduces the task-arrival intensity and hence alleviates resource contention,
resulting in generally improved completion rates and objective values,
although the cumulative reward decreases due to fewer generated tasks.
Fig.~\ref{fig:set4} presents the opposite trend: as the number of data
batches increases, the resulting computation and transmission workloads
become heavier, causing all three performance metrics to generally
decrease. Across these heterogeneous operating conditions, PROMISE
consistently maintains a high task completion rate and objective value,
demonstrating its robustness to variations in both system scale and
workload intensity. The performance differences among the considered methods further reveal
the importance of temporal awareness. RFA lacks state-aware decision
guidance, while IRM optimizes only the instantaneous reward. EAA exploits
statistical workload expectations but does not explicitly track future
state evolution; SHPS incorporates one-step workload prediction but is
limited to short-term temporal information; and IJO iteratively refines
the scheduling decisions without explicit future-trajectory prediction.
In contrast, PROMISE predicts multi-timeslot task arrivals/workloads
and embeds them into forward ES-state rollout to evaluate the downstream
impact of each current mapping. By committing only the first-stage
decision and re-optimizing with newly observed states, PROMISE continuously
adapts to prediction mismatch and evolving service contention. This
explains its sustained performance across Figs.~\ref{fig:set1}--\ref{fig:set4},
particularly when temporal resource coupling becomes pronounced.

\begin{figure}[t!]
	\centering
	\setlength{\abovecaptionskip}{-0.0 mm}
	\includegraphics[width=1\columnwidth]{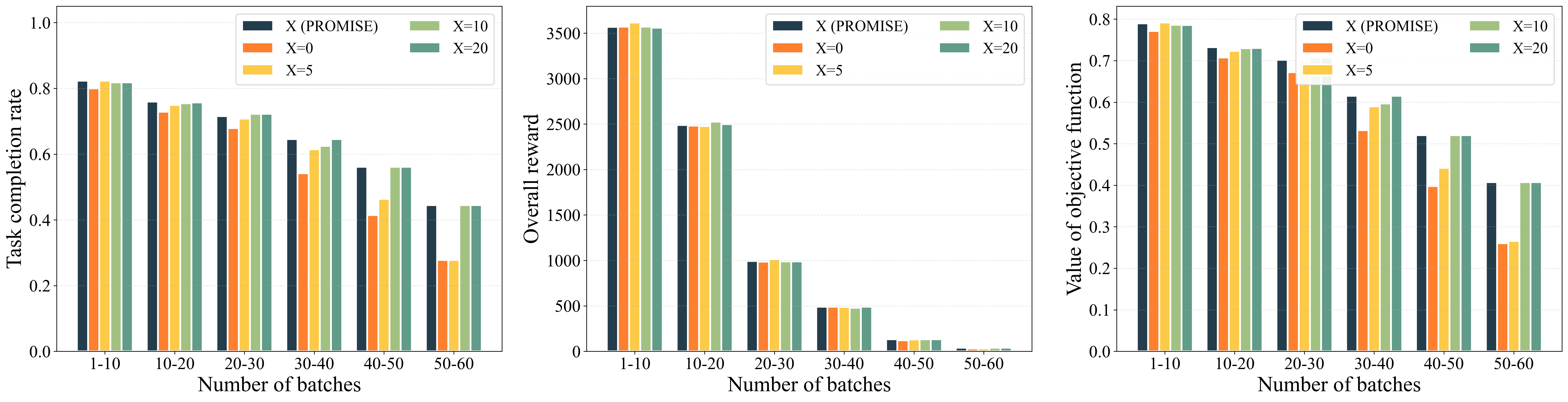}
	\caption{Evaluation on impact of $X^{[\tau']}$.}
	\label{fig:estimationX}
\end{figure}

\noindent
$\bullet$~\textit{Evaluations on $X^{[\tau']}$.} To investigate the impact of the prediction horizon \(X^{[\tau']}\), Fig.~\ref{fig:estimationX}
compares the adaptive horizon strategy of PROMISE, denoted as \(X(\mathrm{PROMISE})\),
with fixed horizons \(X=0,5,10,20\) under different batch workloads. As batches increase, the computation workload and resource occupation duration
of each task increase, leading to an overall degradation in task completion rate,
overall reward, and objective value. Meanwhile, the performance gap among different
horizon settings becomes increasingly evident under heavier workloads. Specifically, \(X=0\) degenerates into a purely instantaneous decision mechanism
without future-state evaluation, resulting in more significant performance
degradation under medium and high workloads. Introducing a non-zero prediction
horizon improves the scheduling performance by enabling forward evaluation of
future resource evolution. However, enlarging a fixed horizon does not yield
monotonic performance gains, since different horizon lengths exhibit distinct
advantages under different workload conditions. This indicates that the look-ahead
horizon should balance sufficient coverage of the temporal impact induced by
current decisions and unnecessary future-state exploration, making fixed
horizons inadequate for dynamic edge environments.

In contrast, \(X(\mathrm{PROMISE})\) adaptively determines the prediction horizon
according to the effective temporal span of current scheduling decisions. It
ensures that the predicted evolution covers the potential impact duration of
committed tasks while avoiding excessive prediction and rollout overhead. The
results demonstrate that the adaptive horizon strategy achieves more stable
performance across varying batch workloads, particularly under heavier loads.
Therefore, the effectiveness of adaptive \(X^{[\tau']}\) lies not in simply
extending the prediction horizon, but in selecting an appropriate look-ahead
depth according to the temporal characteristics of current scheduling decisions.

%----------------------------------------------section 6----------------------------------------------
%----------------------------------------------section 6----------------------------------------------
%----------------------------------------------section 6----------------------------------------------
%----------------------------------------------section 6----------------------------------------------

\section{Hardware Experiments}
\label{hardware}
This section complements the numerical evaluation by validating the
practical deployment and key system characteristics underlying PROMISE
on a Raspberry-Pi-based edge computing testbed, with a focus on system
implementation, performance, and privacy. Main experimental studies are summarized as follows: \textit{(i)} We develop a lightweight edge computing testbed based on a Raspberry Pi cluster and implement the complete pipeline from model export and environment deployment to multi-node collaborative inference, providing a reproducible platform for system-level evaluation. \textit{(ii)} We convert DNN models to the ONNX
format~\cite{onnxformat} and deploy them on Raspberry Pi devices,
verifying their practical deployability and quantifying the accuracy
variation after model conversion. \textit{(iii)} We compare model-parallel distributed inference on inference latency, throughput, and communication overhead. \textit{(iv)} We evaluate the local computing capability of embedded devices through single-node multi-core parallelism and analyze the achievable speedup and performance limits using Amdahl's law~\cite{amdahllaw}. \textit{(v)} We assess the privacy leakage associated with different model partition points through feature inversion attacks, and investigate how the increasing abstraction of intermediate representations affects the feasibility of reconstructing the original input.

% \noindent \textit{(i)} We develop a lightweight edge computing testbed based on a Raspberry Pi cluster and implement the complete pipeline from model export and environment deployment to multi-node collaborative inference, providing a reproducible platform for system-level evaluation.

% \noindent 
% \textit{(ii)} We convert DNN models to the ONNX
% format~\cite{onnxformat} and deploy them on Raspberry Pi devices,
% verifying their practical deployability and quantifying the accuracy
% variation after model conversion.

% \noindent 
% \textit{(iii)} We compare model-parallel distributed inference on inference latency, throughput, and communication overhead.

% \noindent 
% \textit{(iv)} We evaluate the local computing capability of embedded devices through single-node multi-core parallelism and analyze the achievable speedup and performance limits using Amdahl's law~\cite{amdahllaw}.

% \noindent 
% \textit{(v)} We assess the privacy leakage associated with different model partition points through feature inversion attacks, and investigate how the increasing abstraction of intermediate representations affects the feasibility of reconstructing the original input.

Although stochastic task arrivals and time-varying resource states are not explicitly modeled, the real testbed inherently captures practical dynamics arising from task contention, network fluctuations, and device heterogeneity. Given the scale and controllability of the testbed, this section focuses on system implementation and empirical performance, while large-scale stochastic workloads and complex resource dynamics are evaluated in Sec.~\ref{simulation}.

\subsection{System Deployment and Experimental Setup}

\begin{figure}[t!]
	\centering
	\setlength{\abovecaptionskip}{-0.0 mm}
	\includegraphics[width=1\columnwidth]{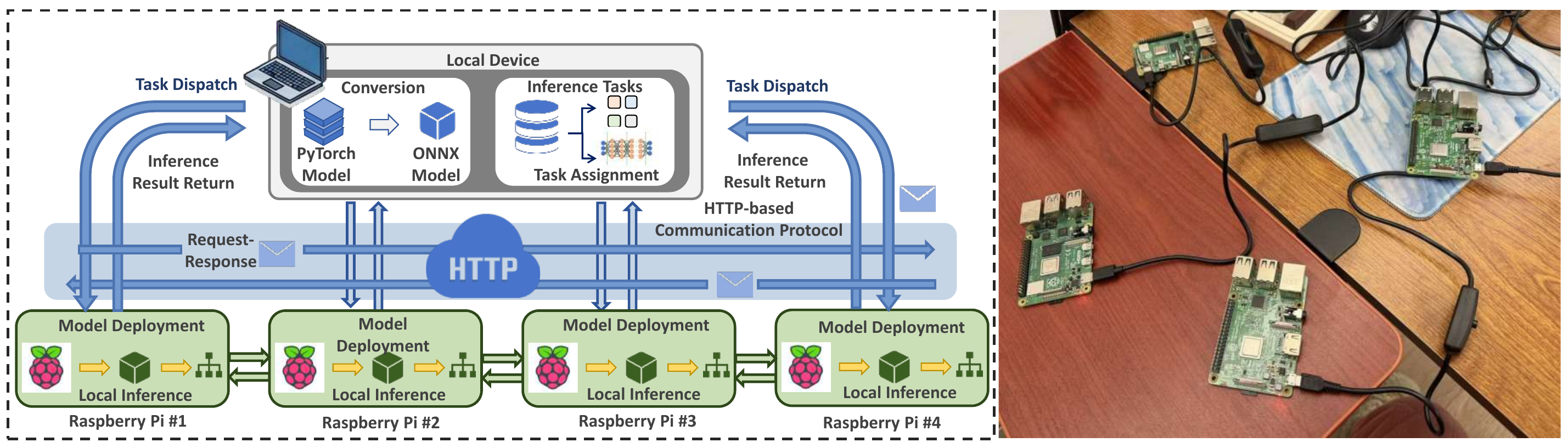}
	\caption{Framework of hardware experiments and Raspberry Pis in real-world test settings.}
	\label{fig:RaspberryPi}
\end{figure}

We construct a real edge computing testbed based on a Raspberry Pi cluster to evaluate the practical deployability and system performance of the proposed distributed inference methods. As illustrated in Fig.~\ref{fig:RaspberryPi}, the testbed consists of 4 heterogeneous edge nodes, including 2 Raspberry Pi 4B and 2 Raspberry Pi 3B, with main hardware specifications summarized in Table~\ref{tab:rpi_config}. On the host, the pretrained PyTorch models are converted to ONNX format for cross-platform deployment. The host serves as the coordinator, partitioning the input data or model according to the specified inference strategy and dispatching the resulting tasks to the edge nodes. Each Raspberry Pi runs ONNX Runtime for local inference and returns the inference results together with the corresponding performance measurements. Communication between the host and edge nodes follows an HTTP-based request-response mechanism through RESTful APIs. Input data or intermediate features are transmitted to the designated nodes, while inference results and runtime statistics are returned in JSON format. The host then aggregates the results and records the computation latency and communication overhead for system-level performance evaluation. All Raspberry Pi nodes are initialized using the official Raspberry Pi Imager with a unified OS image and preconfigured network and user settings, ensuring a consistent runtime environment across the cluster. 

\begin{table}[h]
\centering
\caption{Hardware specifications of Raspberry Pi nodes}
\label{tab:rpi_config}
\footnotesize
\setlength{\tabcolsep}{3pt}  
\begin{tabular}{lcc}
\hline
\rowcolor{gray!10}
\textbf{Component} & \textbf{Raspberry Pi 3B} & \textbf{Raspberry Pi 4B} \\
\hline
CPU & BCM2837, 4$\times$A53@1.2 GHz & BCM2711, 4$\times$A72@1.5 GHz \\
Memory & 1 GB LPDDR2 & 4/8 GB LPDDR4 \\
Storage & microSD & microSD/USB 3.0 SSD \\
Ethernet & 100 Mbps & Gigabit Ethernet \\
Wireless & 2.4 GHz Wi-Fi, BT 4.1 & Dual-band Wi-Fi 5, BT 5.0 \\
\hline
\end{tabular}
\end{table}

%(see Fig. \ref{xxxx}).
%Fig. 树莓派成功按成烧录连上wifi的照片
%\ Image of the experiment in which the Raspberry Pi was successfully flashed and connected to Wi-Fi.

\subsection{Inference Accuracy Validation}

To validate the model conversion and deployment pipeline on embedded devices, we evaluate four representative models covering image classification, speech recognition, and text classification. The pretrained models are converted to the ONNX format and deployed on Raspberry Pi devices using ONNX Runtime. The results are summarized in Table~\ref{tab:onnx_accuracy}, where all models can be successfully deployed and executed on the Raspberry Pi devices, with model-dependent accuracy variations after conversion. The maximum absolute accuracy difference is 3.44\%, observed for M5, while the other models exhibit smaller deviations. Such variations may arise from differences in numerical precision, operator implementations, and pre-processing or runtime configurations across platforms. Overall, the results verify the practical feasibility of the ONNX-based
deployment pipeline and characterize the deployed models used in the
subsequent system experiments.

\begin{table}[h]
\centering
\caption{Inference accuracy before and after ONNX-based deployment}
\label{tab:onnx_accuracy}
\footnotesize
\setlength{\tabcolsep}{4pt} 
\begin{tabular}{lcccc}
\hline
\rowcolor{gray!10}
\textbf{Model} & \textbf{Dataset} & \textbf{Original} & \textbf{Deployed} & $\Delta$ \textbf{(pp)} \\
\hline
M5          & UrbanSound8K & 84.54\% & 81.10\% & -3.44 \\
MobileNetV2 & MNIST        & 98.78\% & 97.90\% & -0.88 \\
ResNet18    & CIFAR-10     & 78.02\% & 78.02\% & +0.00 \\
TextCNN    & THUCNews     & 78.60\% & 76.90\% & -1.70 \\
\hline
\end{tabular}
\end{table}

\subsection{Experiments on Model Parallel Inference}
\label{sec: modelparallel}

We first conduct model-parallel inference experiments on the Raspberry Pi cluster (experimental settings are moved to Appx. \ref{appx: modelparallel} due to space limitation).
Results show that the partition point has a negligible impact on inference accuracy but significantly affects end-to-end latency, primarily due to the trade-off between computation and communication overhead. Partitioning at an early layer reduces the computation on the preceding node but produces larger intermediate features, resulting in higher communication overhead. As the partition point moves deeper, the intermediate features become more compact, reducing communication overhead while increasing the computational load on the preceding node. The optimal partition point varies across model architectures. For ResNet18, partitioning at intermediate layers provides a favorable balance between computation and communication. MobileNetV2 exhibits smaller latency variations across partition points due to the relatively gradual change in intermediate feature sizes. For TextCNN, partitioning near the pooling layer reduces intermediate data transmission and thus improves inference efficiency. Overall, an appropriate partition point can effectively balance computation and communication overhead, thereby reducing the end-to-end latency of model-parallel inference.

\begin{figure}[t!]
	\centering
	\setlength{\abovecaptionskip}{-0.0 mm}
	\includegraphics[width=1\columnwidth]{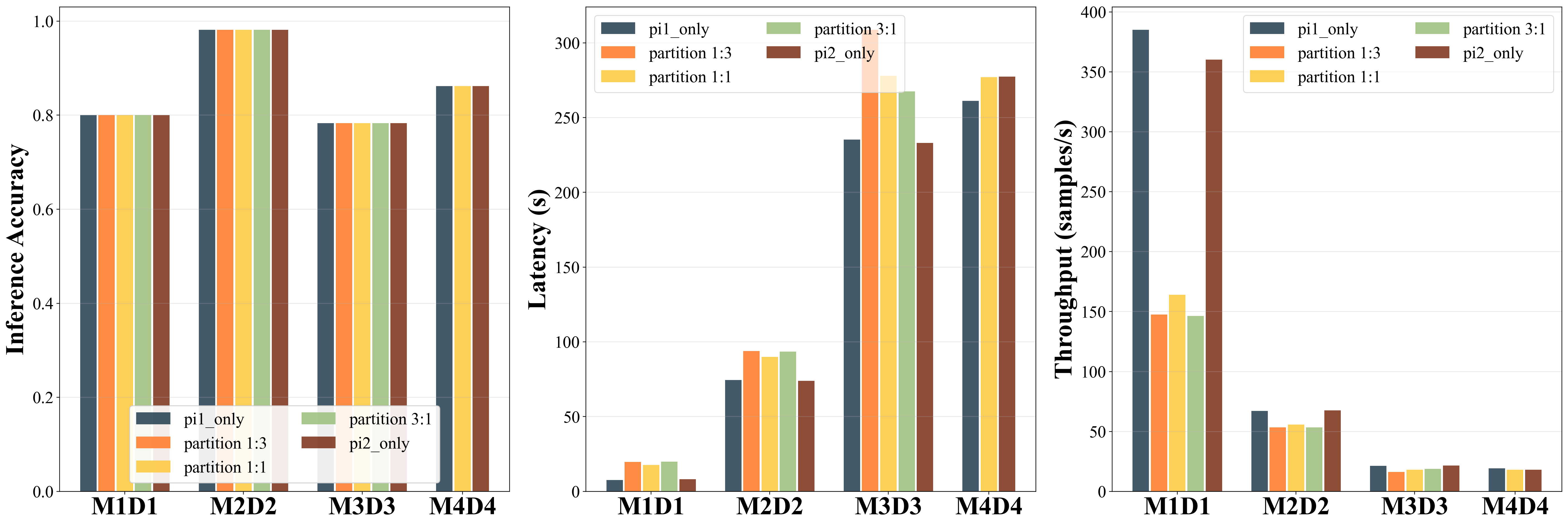}
	\caption{Model parallel inference experiments on Raspberry Pi cluster in terms of accuracy, delay(s) and throughput (sample/s), where M1D1: M5 on UrbanSound8K; M2D2: MobileNetV2 on MNIST; M3D3: ResNet18 on CIFAR-10; M4D4:TextCNN on THUCNews.}
	\label{fig:RaspberryPiclustersimulation}
\end{figure}
%Fig. 4.8 基于树莓派cluster的模型并行实验  三个柱状图

\subsection{Evaluation on Single-Node Multi-Core Parallelism}

To evaluate the local computing capability of edge nodes, we conduct multi-core inference experiments on a single Raspberry Pi by varying the number of parallel processes. The execution time, speedup, and inference accuracy under different degrees of parallelism are reported in Table~\ref{tab:multicore}, where increasing the degree of parallelism consistently reduces the inference time for all models, while the achieved speedups remain below the ideal linear scaling. Moreover, the marginal gain gradually diminishes as more cores are utilized. For example, the speedup of ResNet18 increases from 1.71 with two processes to only 2.10 with four processes. This behavior is consistent with Amdahl's law, as the achievable speedup is constrained by serial execution, scheduling overhead, and contention for shared computing and memory resources. Meanwhile, the inference accuracy remains unchanged across all degrees of parallelism, indicating that multicore execution does not affect the inference results. Overall, multi-core parallelism effectively improves inference performance, but the effective computing capability of an edge node does not scale linearly with the degree of parallelism. This observation also provides empirical support for modeling the available computing capability of ESs as resource-dependent under concurrent workloads. Further privacy leakage evaluations are provided in Appx.~\ref{privacyleakageevaluation}.

\begin{table}[t]
\centering
\caption{Multicore parallel inference performance. The accuracy remains unchanged across different degrees of parallelism: 0.8110 (M5), 0.9790 (MobileNetV2), 0.7820 (ResNet18), and 0.7690 (TextCNN).}
\label{tab:multicore}
\scriptsize
\setlength{\tabcolsep}{1pt}
\renewcommand{\arraystretch}{1}
\begin{tabular}{c|c|c|c}
\hline
\rowcolor{gray!10}
\textbf{Model} & \textbf{Parallelism} & \textbf{Time (s)} & \textbf{Speedup} \\
\hline
M5 &
1/2/3/4 &
1.8248/0.8902/0.7534/0.7025 &
1.00/2.05/2.42/2.60 \\

MobileNetV2 &
1/2/3/4 &
6.2828/3.6957/2.9826/2.7949 &
1.00/1.70/2.11/2.25 \\

ResNet18 &
1/2/3/4 &
33.5795/19.6782/16.8907/15.9593 &
1.00/1.71/1.99/2.10 \\

TextCNN &
1/2/3/4 &
13.1987/7.1296/6.1885/5.3978 &
1.00/1.85/2.13/2.45 \\
\hline
\end{tabular}
\end{table}

\section{Conclusion and Future Work}

We investigated commitment-aware model-parallel inference under
spatio-temporally evolving edge conditions. By coupling CCT-based service
commitments with predictive workload estimation and receding-horizon
scheduling, PROMISE enables current SD--ES assignments to be evaluated
against their anticipated downstream resource impact rather than only
instantaneous system states. Experiments demonstrate its
adaptability across varying system scales, workload intensities, and
prediction horizons, while the Raspberry-Pi testbed validates the
practical feasibility of model-parallel deployment and the underlying
computation, communication, and privacy characteristics. Future work will extend PROMISE toward larger hardware testbeds with
online stochastic scheduling, richer correlated workload and mobility
models, and scalable decision mechanisms for large edge systems.

%\begin{spacing}{0.98}
%	%\footnotesize
%	\bibliographystyle{ieeetr}
%	\bibliography{reference}
%\end{spacing}

% \newpage
% \clearpage
%{\footnotesize
\bibliographystyle{IEEEtran}
\bibliography{reference}

@ARTICLE{Zhou2023acce,
	author={Zhou, Huan and Li, Mingze and Wang, Ning and Min, Geyong and Wu, Jie},
	journal={IEEE Trans. Parallel Dist. Syst.}, 
	title={Accelerating Deep Learning Inference via Model Parallelism and Partial Computation Offloading}, 
	year={2023},
	volume={34},
	number={2},
	pages={475-488}}

@ARTICLE{Gao2023task,
	author={Gao, Mingjin and Shen, Rujing and Shi, Long and Qi, Wen and Li, Jun and Li, Yonghui},
	journal={IEEE Trans. Mobile Comput.}, 
	title={Task Partitioning and Offloading in DNN-Task Enabled Mobile Edge Computing Networks}, 
	year={2023},
	volume={22},
	number={4},
	pages={2435-2445}}

@ARTICLE{Zou2024scal,
	author={Zou, Xiaofeng and Chen, Cen and Lin, Peiying and Zhang, Luochuan and Xu, Yanwu and Zhang, Wenjie},
	journal={IEEE Trans. Emerg. Topics Comput. Intell.}, 
	title={Scalable Heterogeneous Scheduling Based Model Parallelism for Real-Time Inference of Large-Scale Deep Neural Networks}, 
	year={2024},
	volume={8},
	number={4},
	pages={2962-2973}}

@ARTICLE{Ren2022fine,
  author={Ren, Pei and Qiao, Xiuquan and Huang, Yakun and Liu, Ling and Pu, Calton and Dustdar, Schahram},
  journal={IEEE Trans. Serv. Comput.}, 
  title={Fine-Grained Elastic Partitioning for Distributed DNN Towards Mobile Web AR Services in the 5G Era}, 
  year={2022},
  volume={15},
  number={6},
  pages={3260-3274}}

@ARTICLE{Zheng2025opt,
  author={Zheng, Tong and Bi, Yuanguo and Han, Guangjie and Wang, Xingwei and Liu, Yuheng and Liu, Yufei and Chen, Xiangyi},
  journal={IEEE Trans. Comput.}, 
  title={Optimizing Multi-DNN Parallel Inference Performance in MEC Networks: A Resource-Aware and Dynamic DNN Deployment Scheme}, 
  year={2025},
  volume={74},
  number={11},
  pages={3938-3952}}

@ARTICLE{Wang2024fail,
  author={Wang, Li and Li, Liang and Xu, Lianming and Peng, Xian and Fei, Aiguo},
  journal={IEEE Trans. Mobile Comput.}, 
  title={Failure-Resilient Distributed Inference With Model Compression Over Heterogeneous Edge Devices}, 
  year={2024},
  volume={23},
  number={12},
  pages={12680-12692}}

@ARTICLE{iot2026static,
  author={Fang, Juan and Wang, Xinghao and Liu, Yaqi and Tang, Heng and Li, Xiaolin},
  journal={IEEE Internet of Things Journal}, 
  title={Multiagent Collaborative Inference Optimization for Large-Scale DNNs in IoT Edge Systems}, 
  year={2026},
  volume={13},
  number={11},
  pages={24938-24953}}

@ARTICLE{Han2024s2e,
  author={Han, Shujun and Zhang, Wenzhao and Xu, Xiaodong and Wang, Bizhu and Sun, Mengying and Tao, Xiaofeng and Zhang, Ping},
  journal={IEEE Internet Things J.}, 
  title={S2E-DECI: Secrecy and Energy-Efficient Dual-Aware Device-Edge Co-Inference for AIoT}, 
  year={2024},
  volume={11},
  number={24},
  pages={39142-39157}}

@ARTICLE{Qiao2025on,
  author={Qiao, Ying and Teng, Shuyang and Luo, Juan and Sun, Peng and Li, Fan and Tang, Fengxiao},
  journal={IEEE Internet Things J.}, 
  title={On-Orbit DNN Distributed Inference for Remote Sensing Images in Satellite Internet of Things}, 
  year={2025},
  volume={12},
  number={5},
  pages={5687-5703}}

@ARTICLE{Ye2025resou,
  author={Ye, Shengyuan and Ouyang, Bei and Du, Jiangsu and Zeng, Liekang and Qian, Tianyi and Ou, Wenzhong and Chu, Xiaowen and Guo, Deke and Lu, Yutong and Chen, Xu},
  journal={IEEE Trans. Mobile Comput.}, 
  title={Resource-Efficient Collaborative Edge Transformer Inference With Hybrid Model Parallelism}, 
  year={2025},
  volume={24},
  number={10},
  pages={10945-10962}}

@ARTICLE{Li2024dis,
  author={Li, Hui and Li, Xiuhua and Fan, Qilin and He, Qiang and Wang, Xiaofei and Leung, Victor C. M.},
  journal={IEEE Trans. Mobile Comput.}, 
  title={Distributed DNN Inference With Fine-Grained Model Partitioning in Mobile Edge Computing Networks}, 
  year={2024},
  volume={23},
  number={10},
  pages={9060-9074}}

@ARTICLE{Sun2025ene,
  author={Sun, Yanzan and Qiu, Jiacheng and Pan, Guangjin and Xu, Shugong and Zhang, Shunqing and Wang, Xiaoyun and Han, Shuangfeng},
  journal={IEEE Internet Things J.}, 
  title={Energy Optimization of Multitask DNN Inference in MEC-Assisted XR Devices: A Lyapunov-Guided Reinforcement Learning Approach}, 
  year={2025},
  volume={12},
  number={11},
  pages={17499-17513}}

@ARTICLE{Liu2024moei,
  author={Liu, Zhicheng and Tian, Meng and Dong, Mianxiong and Wang, Xiaofei and Qiu, Chao and Zhang, Cheng},
  journal={IEEE Trans. Mobile Comput.}, 
  title={MoEI: Mobility-Aware Edge Inference Based on Model Partition and Service Migration}, 
  year={2024},
  volume={23},
  number={10},
  pages={9437-9450}}

@ARTICLE{Liu2025ada,
  author={Liu, Sicong and Luo, Hao and Li, XiaoChen and Li, Yao and Guo, Bin and Yu, Zhiwen and Wang, YuZhan and Ma, Ke and Ding, YaSan and Yao, Yuan},
  journal={IEEE Trans. Mobile Comput.}, 
  title={AdaKnife: Flexible DNN Offloading for Inference Acceleration on Heterogeneous Mobile Devices}, 
  year={2025},
  volume={24},
  number={2},
  pages={736-748}}

@ARTICLE{Cao2024learn,
  author={Cao, Yang and Lien, Shao-Yu and Yeh, Cheng-Hao and Deng, Der-Jiunn and Liang, Ying-Chang and Niyato, Dusit},
  journal={IEEE Internet Things J.}, 
  title={Learning-Based Multitier Split Computing for Efficient Convergence of Communication and Computation}, 
  year={2024},
  volume={11},
  number={20},
  pages={33077-33096}}

@ARTICLE{Samikwa2024disnet,
  author={Samikwa, Eric and Maio, Antonio Di and Braun, Torsten},
  journal={IEEE Internet Things J.}, 
  title={DISNET: Distributed Micro-Split Deep Learning in Heterogeneous Dynamic IoT}, 
  year={2024},
  volume={11},
  number={4},
  pages={6199-6216}}

@ARTICLE{Xu2023dis,
  author={Xu, Yuzhe and Mohammed, Thaha and Di Francesco, Mario and Fischione, Carlo},
  journal={IEEE Internet Things J.}, 
  title={Distributed Assignment With Load Balancing for DNN Inference at the Edge}, 
  year={2023},
  volume={10},
  number={2},
  pages={1053-1065},
  doi={10.1109/JIOT.2022.3205410}}

@ARTICLE{Chen2025adap,
  author={Chen, Yuxuan and Li, Rongpeng and Yu, Xiaoxue and Zhao, Zhifeng and Zhang, Honggang},
  journal={Frontiers of Inf. Tech. \& Electron. Eng.}, 
  title={Adaptive Layer Splitting for Wireless Large Language Model Inference in Edge Computing: A Model-Based Reinforcement Learning Approach}, 
  year={2025},
  volume={26},
  number={2},
  pages={278-292}}

@ARTICLE{Li2023adap,
  author={Li, Pengzhen and Koyuncu, Erdem and Seferoglu, Hulya},
  journal={IEEE Open J. Commun. Soc.}, 
  title={Adaptive and Resilient Model-Distributed Inference in Edge Computing Systems}, 
  year={2023},
  volume={4},
  number={},
  pages={1263-1273}}

@ARTICLE{Shi2023auto,
  author={Shi, Hongjian and Zheng, Weichu and Liu, Zifei and Ma, Ruhui and Guan, Haibing},
  journal={IEEE J. Sel. Areas Commun.}, 
  title={Automatic Pipeline Parallelism: A Parallel Inference Framework for Deep Learning Applications in 6G Mobile Communication Systems}, 
  year={2023},
  volume={41},
  number={7},
  pages={2041-2056}}

@ARTICLE{Xue2022ddpqn,
  author={Xue, Min and Wu, Huaming and Peng, Guang and Wolter, Katinka},
  journal={IEEE Trans. Serv. Comput.}, 
  title={DDPQN: An Efficient DNN Offloading Strategy in Local-Edge-Cloud Collaborative Environments}, 
  year={2022},
  volume={15},
  number={2},
  pages={640-655}}

@INPROCEEDINGS{Li2024dnn,
  author={Li, Yuepeng and Zeng, Deze and Gut, Lin and Guo, Song and Zomaya, Albert Y.},
  booktitle={IEEE 44th Int. Con. Distrib. Comput. Syst. (ICDCS)}, 
  title={DNN Partitioning and Assignment for Distributed Inference in SGX Empowered Edge Cloud}, 
  year={2024},
  volume={},
  number={},
  pages={635-644}}

@ARTICLE{Wang2023decen,
  author={Wang, Feng and Cai, Songfu and Lau, Vincent K. N.},
  journal={IEEE J. Sel. Topics Signal Process.}, 
  title={Decentralized DNN Task Partitioning and Offloading Control in MEC Systems With Energy Harvesting Devices}, 
  year={2023},
  volume={17},
  number={1},
  pages={173-188}}

@ARTICLE{Wang2026scal,
  author={Wang, Nianfu and Wang, Wanyou and Zhong, Xiaoxiong and Liu, Jingyu and Shi, Gaotao and Li, Zhijun},
  journal={IEEE Trans. Mobile Comput.}, 
  title={ScalPipe: Scalable Collaborative Pipeline Inference for Distributed Heterogeneous Devices}, 
  year={2026},
  volume={},
  number={},
  pages={1-15}}

@ARTICLE{Xu2025cadec,
  author={Xu, Xiaolong and Hu, Yuhao and Cui, Guangming and Qi, Lianyong and Dou, Wanchun and Cai, Zhipeng},
  journal={IEEE Trans. Mobile Comput.}, 
  title={CADEC: A Combinatorial Auction for Dynamic Distributed DNN Inference Scheduling in Edge-Cloud Networks}, 
  year={2025},
  volume={24},
  number={10},
  pages={10024-10041}}

@ARTICLE{Bao2025joint,
  author={Bao, Tingting and Li, Xin and Zhao, Yongli and Lian, Meng and Jiang, Yike and Zhang, Jie},
  journal={J. Opt. Commun. Netw.}, 
  title={Joint optimization of DNN model partitioning and slice delivery for distributed edge-cloud inference over optical networks}, 
  year={2025},
  volume={17},
  number={11},
  pages={1047-1058}}

@ARTICLE{Dong2024dnn,
  author={Dong, Chongwu and Shafiq, Muhammad and Dabel, Maryam M. Al and Sun, Yanbin and Tian, Zhihong},
  journal={IEEE Trans. Consum. Electron.}, 
  title={DNN Inference Acceleration for Smart Devices in Industry 5.0 by Decentralized Deep Reinforcement Learning}, 
  year={2024},
  volume={70},
  number={1},
  pages={1519-1530}}

@ARTICLE{Xin2026load,
  author={Xin, Jingjie and Li, Xin and Kilper, Daniel and Huang, Shanguo},
  journal={IEEE Trans. Netw. Sci. Eng.}, 
  title={Load-Balance-Guaranteed DNN Distributed Inference Offloading in MEC Networks Interconnected by Metro Optical Networks}, 
  year={2026},
  volume={13},
  number={},
  pages={3391-3408}}

@ARTICLE{Lin2025top,
	author={Lin, Changyao and Chen, Zhenming and Zhang, Ziyang and Liu, Jie},
	journal={IEEE Trans. Parallel Distrib. Syst.}, 
	title={TOP: Task-Based Operator Parallelism for Asynchronous Deep Learning Inference on GPU}, 
	year={2025},
	volume={36},
	number={2},
	pages={266-281}}

@ARTICLE{Wang2026mobi,
  author={Wang, Peng and Sun, Wen and Yang, Yi and Niyato, Dusit and Wu, Dapeng Oliver},
  journal={IEEE Trans. Mobile Comput.}, 
  title={MobiSplit: Mobility-Aware Inference Partitioning and Offloading for Efficient Edge Intelligence}, 
  year={2026},
  volume={25},
  number={3},
  pages={3969-3984}}

@ARTICLE{Tang2021joint,
  author={Tang, Xin and Chen, Xu and Zeng, Liekang and Yu, Shuai and Chen, Lin},
  journal={IEEE Internet Things J.}, 
  title={Joint Multiuser DNN Partitioning and Computational Resource Allocation for Collaborative Edge Intelligence}, 
  year={2021},
  volume={8},
  number={12},
  pages={9511-9522}}

@ARTICLE{Cheng2025privacy,
  author={Cheng, Zhipeng and Xia, Xiaoyu and Wang, Hong and Liwang, Minghui and Chen, Ning and Fan, Xuwei and Wang, Xianbin},
  journal={IEEE Trans. Serv. Comput.}, 
  title={Privacy-Aware Joint DNN Model Deployment and Partitioning Optimization for Collaborative Edge Inference Services}, 
  year={2025},
  volume={18},
  number={5},
  pages={3079-3092}}

@INPROCEEDINGS{Chang2021neural,
  author={Chang, Sungkyun and Lee, Donmoon and Park, Jeongsoo and Lim, Hyungui and Lee, Kyogu and Ko, Karam and Han, Yoonchang},
  booktitle={IEEE Int. Conf. Acoust. Speech and Signal Process. (ICASSP)}, 
  title={Neural Audio Fingerprint for High-Specific Audio Retrieval Based on Contrastive Learning}, 
  year={2021},
  volume={},
  number={},
  pages={3025-3029}}

@INPROCEEDINGS{Marshall2025diff,
	author={Marshall, Ben and Li, Sirui and Meka, Shiv Akarsh and Liu, Wei},
	booktitle={Int. Joint Conf. Neural Netw. (IJCNN)}, 
	title={Differential Privacy on Large Language Models for Privacy Preserving Clinical Coding}, 
	year={2025},
	volume={},
	number={},
	pages={1-8}}

@ARTICLE{Huang2025joint,
	author={Huang, Jiale and Wu, Jigang and Wu, Yalan and Wu, Jiaxin},
	journal={IEEE Trans Netw. Science Eng.}, 
	title={Joint Request Offloading and Resource Allocation for Long-Term Utility Optimization in Collaborative Edge Inference With Time-Coupled Resources}, 
	year={2025},
	volume={12},
	number={4},
	pages={2622-2639}}

@ARTICLE{Guo2025seamless,
	author={Guo, Bingshuo and Liwang, Minghui and Xia, Xiaoyu and Li, Li and Jiao, Zhenzhen and Hosseinalipour, Seyyedali and Wang, Xianbin},
	journal={IEEE Trans. Serv. Comput.}, 
	title={Seamless Graph Task Scheduling Over Dynamic Vehicular Clouds: A Hybrid Methodology for Integrating Pilot and Instantaneous Decisions}, 
	year={2025},
	volume={18},
	number={3},
	pages={1753-1768}}

@ARTICLE{background,
  author={Wu, Minghong and Liwang, Minghui and Su, Yuhan and Li, Li and Hosseinalipour, Seyyedali and Wang, Xianbin and Dai, Huaiyu and Jiao, Zhenzhen},
  journal={IEEE Trans. Mobile Comput.}, 
  title={Toward Seamless Hierarchical Federated Learning Under Intermittent Client Participation: A Stagewise Decision-Making Methodology}, 
  year={2026},
  volume={25},
  number={4},
  pages={4715-4730}}

@ARTICLE{Taleb2017MECSurvey,
  author={Taleb, Tarik and Samdanis, Konstantinos and Mada, Badr and Flinck, Hannu and Dutta, Sunny and Sabella, Dario},
  journal={IEEE Commun. Surveys Tut.},
  title={On Multi-Access Edge Computing: A Survey of the Emerging 5G Network Edge Cloud Architecture and Orchestration},
  year={2017},
  volume={19},
  number={3},
  pages={1657-1681},
}

@TECHREPORT{ETSI_MEC003,
  author={{ETSI}},
  title={Multi-access Edge Computing (MEC); Framework and Reference Architecture},
  institution={Eur. Telecommun. Standards Inst. (ETSI)},
  type={ETSI GS MEC},
  number={003 V3.1.1},
  year={2022},
  address={Sophia Antipolis, France}
}

@ARTICLE{background1,
  author={Chen, Yan and Jia, Xiaolin and Wang, Haiquan},
  journal={IEEE Trans. Mobile Comput.}, 
  title={Hierarchical Offloading Optimization for Collaborative DNN Inference in Satellite Edge Computing Networks}, 
  year={2026},
  volume={25},
  number={8},
  pages={12128-12143},
  }

@ARTICLE{background2,
  author={Liu, Jianhua and Wang, Xin and Fei, Shichen and Tong, Weiqin and Li, Minglu and Ren, Kui},
  journal={IEEE Trans. Mobile Comput.}, 
  title={Optimal DNN Partitioning and Deployment for Dependable Inference Workflow on End Devices}, 
  year={2026},
  volume={},
  number={},
  pages={1-18},
}

@article{background3,
author = {Hao, Yuntao and Ding, Nan and Xia, Weiguo and Ge, Hongwei and Xu, Li},
title = {DNN Partitioning for Cooperative Inference in Edge Intelligence: Modeling, Solutions, Toolchains},
year = {2026},
journal={ACM Comput. Surv.},
volume = {58},
number = {8},
pages = {1-34},
}

@inproceedings{he2016deep,
    title={Deep residual learning for image recognition},
    author={He, Kaiming and Zhang, Xiangyu and Ren, Shaoqing and Sun, Jian},
    booktitle={Proc. IEEE/CVF Conf. Comput. Vis. Pattern Recognit. (CVPR)},
    pages={770--778},
    year={2016}
}

@inproceedings{sandler2018mobilenetv2,
    title={MobileNetV2: Inverted residuals and linear bottlenecks},
    author={Sandler, Mark and Howard, Andrew and Zhu, Menglong and Zhmoginov, Andrey and Chen, Liang-Chieh},
    booktitle={Proc. IEEE/CVF Conf. Comput. Vis. Pattern Recognit. (CVPR)},
    pages={4510--4520},
    year={2018}
}

@article{lecun1998gradient,
    title={Gradient-based learning applied to document recognition},
    author={LeCun, Yann and Bottou, L{\'e}on and Bengio, Yoshua and Haffner, Patrick},
    journal={Proc. IEEE},
    volume={86},
    number={11},
    pages={2278--2324},
    year={1998},
    publisher={IEEE}
}

@inproceedings{dai2017m5,
    title={M5: A lightweight deep neural network for raw audio classification},
    author={Dai, Wei and Dai, Chen and Qu, Shaoxin and Li, Jia and Das, Samarjit},
    booktitle={Proc. IEEE Int. Conf. Acoust., Speech Signal Process. (ICASSP)},
    pages={421--425},
    year={2017},
    organization={IEEE}
}

@inproceedings{salamon2014dataset,
    title={A dataset and taxonomy for urban sound research},
    author={Salamon, Justin and Jacoby, Christopher and Bello, Juan Pablo},
    booktitle={Proc. ACM Int. Conf. Multimedia (MM)},
    pages={1041--1044},
    year={2014}
}

@inproceedings{kim2014convolutional,
    title={Convolutional Neural Networks for Sentence Classification},
    author={Kim, Yoon},
    booktitle={Proc. Conf. Empirical Methods Natural Lang. Process. (EMNLP)},
    pages={1746--1751},
    year={2014}
}

@misc{sun2016thucnews,
  author={Sun, Maosong and Li, Jingyang and Guo, Zhipeng and Zhao, Yu and Zheng, Yabin and Si, Xiance and Liu, Zhiyuan},
  title={{THUCTC}: An Efficient Chinese Text Classifier},
  year={2016},
  note={THU Natural Language Processing Lab}
}

@article{tmc2025,
author={Dai, Longbao and Zeng, Fanzi and Kong, Haoran and Cai, Jianghao and Jiang, Hongbo and Li, Keqin},
  journal={IEEE Trans. Mobile Comput.}, 
  title={Throughput-Aware Cooperative Task Offloading in Dynamic Mobile Edge Computing Systems}, 
  year={2025},
  volume={24},
  number={12},
  pages={13276--13292}}

@ARTICLE{onnxformat,
  author={Liang, Daojun and Zhang, Haixia and Han, Qiaojian and Yuan, Dongfeng and Zhang, Minggao},
  journal={IEEE Trans. Ind. Inf.}, 
  title={RasPiDets: A Quasi-Real-Time Defect Detection Method With End-Edge-Cloud Collaboration}, 
  year={2025},
  volume={21},
  number={7},
  pages={5525-5535}}

@ARTICLE{amdahllaw,
  author={Pei, Songwen and Kim, Myoung-Seo and Gaudiot, Jean-Luc},
  journal={IEEE Embedded Syst. Lett.}, 
  title={Extending Amdahl’s Law for Heterogeneous Multicore Processor with Consideration of the Overhead of Data Preparation}, 
  year={2016},
  volume={8},
  number={1},
  pages={26-29}}

@ARTICLE{computespeedv1,
  author={Li, Chunlin and Zhang, Zihao and Wang, Bingxin and Lei, Mengchao and Liu, Sen and Li, Aoyong and Wan, Shaohua},
  journal={ACM Trans. Intell. Syst. Technol.}, 
  title={Joint Service Migration and Resource Allocation for DNN Tasks using SA-DDQN-DDPG in Vehicular Edge Computing}, 
  year={2025},
  volume={17},
  number={1},
  pages={1-25}}
%}

%--------------------------------------------------------------------------------------------------------------------------
%--------------------------------------------------------------------------------------------------------------------------
%--------------------------------------------------------------------------------------------------------------------------
%--------------------------------------------------------------------------------------------------------------------------
%--------------------------------------------------------------------------------------------------------------------------
\newpage
\clearpage
\appendices

\section{Detailed Literature Review}
\label{relatedwork}
Details of different perspectives on literature review.

\noindent
$\bullet$~\textit{View 1: partial dynamics vs. coupled spatio-temporal evolution.}
Existing distributed inference studies have progressively moved beyond purely fixed-resource settings, yet most characterize only a subset of the dynamic factors involved in practical end-edge inference. For example, \textit{Zhou} et al. \cite{Zhou2023acce} developed model-parallel inference with partial computation offloading that adapts task assignment to different channel conditions, while \textit{Gao} et al. \cite{Gao2023task} introduced layer-level task partitioning together with execution-delay estimation, slot-based scheduling, and dynamic pricing. \textit{Zou} et al. \cite{Zou2024scal} focused on scalable model-parallel scheduling across heterogeneous computing resources, whereas \textit{Ren} et al. \cite{Ren2022fine} employed deep reinforcement learning (DRL) to realize fine-grained elastic DNN partitioning across heterogeneous computing platforms. More recent learning-based designs have further accommodated runtime resource variations. For instance, \textit{Fang} et al. \cite{iot2026static} jointly optimized collaborative DNN inference under shared wireless resources through multiagent learning, and \textit{Zheng} et al. \cite{Zheng2025opt} investigated resource-aware dynamic DNN deployment and request scheduling for concurrent multi-DNN inference. Other works have addressed specific sources of system uncertainty: \textit{Wang} et al. \cite{Wang2024fail} studied failure-resilient distributed inference over heterogeneous devices subject to crash or timeout failures, while \textit{Han} et al. \cite{Han2024s2e} considered wireless transmission security and finite-blocklength effects in device-edge co-inference. Recent studies have incorporated richer temporal or spatial dynamics into distributed inference. \textit{Ye} et al. \cite{Ye2025resou} explicitly considered bandwidth-constrained edge environments and device-level resource dynamics, developing fault-tolerant rescheduling to accommodate unexpected stragglers. \textit{Li} et al. \cite{Li2024dis} investigated fine-grained model partitioning under heterogeneous ES/device resources and task-specific delay constraints, while \textit{Sun} et al. \cite{Sun2025ene} adopted Lyapunov-guided reinforcement learning for multitask inference energy optimization. From a mobility perspective, \textit{Liu} et al. \cite{Liu2024moei} jointly considered model partitioning and service migration to support mobility-aware edge inference. These advances demonstrate that system dynamics have received increasing attention; however, the dynamic factors are predominantly modeled individually or treated as observed runtime states. In particular, the coupled evolution among stochastic workload demand, mobility-induced communication conditions, and load-dependent service capacity remains insufficiently characterized. Such coupling is critical in our setting because current task admissions not only consume future computing resources but also reshape the effective ES capability experienced by already admitted tasks. PROMISE therefore models these factors within a unified spatio-temporal framework and further predicts their future evolution for subsequent commitment-aware scheduling.

\noindent
$\bullet$~\textit{View 2: reactive/adaptive scheduling vs. prediction-aware rolling-horizon optimization.}
Existing distributed inference studies can also be distinguished by how temporal information is incorporated into scheduling decisions. A large body of work generates partitioning or offloading decisions primarily from the currently observed task, network, and resource states. Representative examples include \textit{Zhou} et al. \cite{Zhou2023acce}, \textit{Gao} et al. \cite{Gao2023task}, \textit{Zou} et al. \cite{Zou2024scal}, \textit{Ren} et al. \cite{Ren2022fine}, \textit{Liu} et al. \cite{Liu2025ada}, and \textit{Chen} et al. \cite{Chen2025adap}. These approaches substantially improve inference efficiency through adaptive partitioning, offloading, or learning-based decision making, but do not explicitly evaluate a current scheduling decision against a predicted multi-timeslot workload trajectory. Notably, model-based or profiling-assisted methods should not be conflated with such temporal prediction: for example, the model-based reinforcement learning framework in \cite{Chen2025adap} employs a surrogate model to facilitate efficient split-point evaluation under varying network conditions, rather than explicitly forecasting future workload evolution over a rolling scheduling horizon.

Other studies broaden the decision scope to multi-task, multi-device, or more sophisticated parallel inference scenarios. \textit{Ye} et al. \cite{Ye2025resou} developed heterogeneous-resource-aware collaborative Transformer inference with dynamic rescheduling, \textit{Zheng} et al. \cite{Zheng2025opt} considered concurrent multi-DNN deployment and request scheduling, \textit{Xu} et al. \cite{Xu2023dis} formulated distributed inference offloading with load balancing as a multiple-assignment problem, and \textit{Shi} et al. \cite{Shi2023auto} developed automatic pipeline parallelism based on hardware profiling and task scheduling. More recent works further introduce learning-based or long-term control mechanisms. \textit{Li} et al. \cite{Li2024dis} employed asynchronous actor-critic learning for fine-grained DNN partitioning, while \textit{Sun} et al. \cite{Sun2025ene} used Lyapunov-guided reinforcement learning for multitask inference optimization. \textit{Liu} et al. \cite{Liu2024moei} incorporated mobility-aware service migration, \textit{Cao} et al. \cite{Cao2024learn} developed multi-tier DRL decision making with different control cycles, \textit{Samikwa} et al. \cite{Samikwa2024disnet} adapted distributed inference to dynamic computing and communication resources, and \textit{Lin} et al. \cite{Lin2025top} addressed operator scheduling under dynamically arriving asynchronous inference tasks. These studies represent important progress from static optimization toward adaptive and long-term inference control. Nevertheless, \emph{adaptivity or long-term optimization does not necessarily imply forward-looking scheduling}. Queue-based control, reinforcement learning, and runtime rescheduling can account for accumulated system states and long-term objectives without explicitly exposing the predicted future workload/resource trajectory to the current decision. Consequently, the downstream impact of a current assignment on future resource contention and commitment feasibility remains difficult to assess explicitly. PROMISE addresses this distinction by estimating future task arrivals and computational workloads over an adaptive multi-timeslot horizon and embedding these estimates into predictive ES-state rollout. The scheduler therefore evaluates current decisions according to their anticipated future consequences and repeatedly updates them in a receding-horizon manner as new states become observable.

\noindent
$\bullet$~\textit{View 3: metric-/constraint-driven QoS vs. commitment-based service provisioning.}
Existing distributed inference studies predominantly characterize service quality through performance metrics or externally specified operating constraints. Inference latency remains one of the most widely optimized metrics, as considered in \cite{Li2023adap,Wang2026scal,Bao2025joint}. Beyond latency-oriented optimization, numerous studies have incorporated additional objectives to reflect energy, cost, security, fairness, and resource utilization. For instance, \textit{Xue} et al. \cite{Xue2022ddpqn} jointly considered delay, energy consumption, and task cost in local-edge-cloud DNN offloading, while \textit{Han} et al. \cite{Han2024s2e} jointly optimized DNN partitioning and resource allocation for energy-efficient co-inference under physical-layer secrecy requirements. Other studies introduced explicit system constraints or resource-oriented objectives. Specifically, \textit{Xu} et al. \cite{Xu2023dis} formulated load-balanced DNN inference assignment by maximizing proportional fairness, \textit{Li} et al. \cite{Li2024dnn} accounted for the restricted secure memory of Intel software guard extensions (SGX) in distributed inference, \textit{Wang} et al. \cite{Wang2023decen} optimized DNN partitioning and offloading under workload and energy-queue dynamics, and \textit{Wang} et al. \cite{Wang2026mobi} considered mobility-aware inference partitioning and offloading under changing network conditions.

Other works extend the QoS formulation toward application- or market-specific objectives. \textit{Xu} et al. \cite{Xu2025cadec} formulated dynamic distributed DNN inference scheduling as a combinatorial-auction-based social-welfare optimization problem. \textit{Qiao} et al. \cite{Qiao2025on} developed on-orbit distributed DNN inference to improve the efficiency of remote-sensing inference in satellite IoT. \textit{Dong} et al. \cite{Dong2024dnn} incorporated multiple inference-performance and resource-utilization considerations through decentralized learning, while \textit{Xin} et al. \cite{Xin2026load} explicitly incorporated load imbalance into distributed inference offloading. These studies demonstrate that existing QoS designs are substantially richer than pure latency minimization, and some provide explicit resource constraints, fairness objectives, or algorithmic performance guarantees. Therefore, it is more precise to characterize them as \emph{metric- or constraint-driven QoS} rather than uniformly as ``soft QoS.'' A different service abstraction is considered in PROMISE. In the aforementioned studies, latency, energy, fairness, security, load balance, or other QoS requirements are primarily treated as performance objectives or externally imposed constraints. They do not explicitly make a platform-announced completion commitment an endogenous scheduling variable whose tightness and eventual fulfillment jointly determine service utility. PROMISE introduces the committed completion time (CCT) for this purpose: the platform proactively announces a completion commitment according to the predicted system evolution, tighter feasible commitments yield higher service rewards, and failure to fulfill the announced CCT constitutes a commitment violation. Accordingly, service reliability is no longer represented only indirectly through latency minimization or resource efficiency, but is explicitly coupled with the scheduling decision through commitment establishment and fulfillment. This motivates commitment-aware distributed inference with predictive and verifiable service provisioning under spatio-temporally evolving edge conditions.

\section{Rationale Analysis on Settings and Assumptions of PROMISE}
\label{appxration}
This section discusses the rationale behind the major settings and assumptions of PROMISE, with the goal of balancing engineering realism and analytical tractability. We first clarify how the proposed modeling captures the key characteristics of practical edge inference systems and then justify the abstractions adopted to enable efficient online optimization.

\noindent
$\bullet$~\textit{\textbf{Realism of the target problem.}}
PROMISE retains the system characteristics that directly determine inference scheduling, resource evolution, and commitment fulfillment, thereby maintaining close correspondence with practical end--edge inference services.

\noindent
\textit{(i) {Privacy-aware DNN partitioning.}}
In practical model-parallel inference, privacy-sensitive front layers are commonly executed locally, while only intermediate features are transmitted to the ES. PROMISE accordingly evaluates reconstruction-based privacy leakage using measures such as SSIM and constrains the DNN partition point according to each SD's privacy-risk tolerance. This abstraction explicitly couples privacy preservation with the amount of local computation and the residual workload offloaded to the edge.

\noindent
\textit{(ii) {Load-dependent computing capability.}}
The effective computing capability available to an inference task is generally load dependent because concurrent execution introduces CPU/GPU contention, memory-bandwidth competition, cache interference, and other shared-resource overheads. PROMISE captures this behavior through a nonlinear degradation model, where the coefficient $\beta$ characterizes the sensitivity of per-task computing capability to increasing concurrency, consistent with the performance degradation observed in practical servers \cite{Liu2024moei}. This modeling is essential to PROMISE because admitting a new task can alter not only its own execution rate but also the future completion evolution of tasks already hosted by the same ES.

\noindent
\textit{(iii) {Spatio-temporal dynamics and uncertainty.}}
Practical edge inference jointly exhibits temporal workload variations and spatially heterogeneous service conditions. Device mobility changes wireless connectivity and transmission efficiency, while task arrivals, DNN types, and input batch numbers evolve stochastically with user activities and application demands. Meanwhile, workload-dependent ES capability further couples current admissions with future service availability. PROMISE explicitly models these factors and subsequently exploits their predicted evolution in rolling-horizon scheduling.

\noindent
\textit{(iv) {Commitment-based service mechanism.}}
Practical service provisioning requires not only low latency but also predictable delivery. Failure to complete an inference request within an announced completion time can degrade service credibility and user experience. PROMISE therefore introduces CCT as an endogenous platform commitment and evaluates its fulfillment against the realized task completion time. Together with the task completion ratio, this mechanism captures the tradeoff between aggressive service commitments and reliable fulfillment, providing a scheduling-level abstraction of commitment-oriented service provisioning.

\noindent
\textit{(v) {Cross-timeslot coupling.}}
Inference tasks may occupy ES resources for multiple timeslots; hence, a current assignment reshapes future workload occupancy and computing capability. Conversely, subsequent stochastic arrivals introduce additional contention and can affect the completion feasibility of commitments made earlier. PROMISE captures this forward temporal coupling through the evolving ES state $\mathsf{sch}_n^{[\tau]}$ and explicitly incorporates predicted future demand into its rolling-horizon optimization, allowing current decisions to be evaluated according to their anticipated downstream consequences.

\noindent
$\bullet$~\textit{\textbf{Model simplifications and their justifications.}}
To maintain tractable online scheduling under dynamic and uncertain edge conditions, PROMISE abstracts several lower-level engineering details that are secondary to the targeted cross-timeslot scheduling mechanism.

\noindent
\textit{(i) {Quasi-static transmission condition within each timeslot.}}
The transmission-time coefficient $R^{[\tau],\mathsf{SD}}_{m}$ is assumed to remain constant within each scheduling timeslot while varying across timeslots. This abstraction separates scheduling-level communication dynamics from fast physical-layer fluctuations: short-term channel variations are averaged within a timeslot, whereas mobility-induced changes in effective transmission conditions are captured through the time-varying $R^{[\tau],\mathsf{SD}}_{m}$. Such slot-level modeling is sufficient for evaluating the communication delay relevant to inference scheduling without introducing unnecessary physical-layer complexity.

\noindent
\textit{(ii) {Independent task generation.}}
Task arrivals are modeled as independent Bernoulli processes, without explicitly characterizing spatial correlations among SDs or event-driven arrival bursts. This assumption provides a tractable representation of heterogeneous stochastic demand while preserving the essential uncertainty faced by the scheduler. More importantly, PROMISE does not rely on a fixed long-term arrival realization: workload predictions and scheduling decisions are continuously refreshed as new system states become observable, allowing deviations from the estimated demand to be progressively corrected. The experimental results further examine the robustness of PROMISE under this abstraction.

\noindent
\textit{(iii) {Negligible model loading and warm-up latency.}}
We omit the latency associated with loading DNN models from persistent storage and CPU/GPU warm-up. In practical edge inference systems, frequently invoked models can be maintained in memory by the serving framework, making such overhead substantially less frequent than task execution and resource scheduling. When necessary, model-loading or warm-up latency can be incorporated into the task workload or ES state without changing the proposed prediction and rolling-horizon optimization framework.

\noindent
\textit{(iv) {Negligible PT--ES control-plane latency.}}
The PT and ESs are assumed to be interconnected through reliable wired backhaul, and the latency of control-plane signaling between them is therefore neglected. This assumption is consistent with common MEC orchestration architectures, where management entities exchange lightweight control information over high-speed wired infrastructure and the resulting signaling latency is typically secondary to wireless data transmission and DNN inference execution~\cite{Taleb2017MECSurvey, ETSI_MEC003}. Importantly, this abstraction applies only to PT-ES orchestration signaling; the SD-side wireless transmission delay remains explicitly modeled in PROMISE.

\section{Detailed Derivation of $X^{[\tau']}$-Timeslot Demand Estimation}
\label{appxdemandestimation}

This appendix derives the predictive task-arrival and edge-side
computational-workload profiles used by Module A in Sec.~4.1.
A task generated at timeslot $\tau$ does not immediately become
available for edge scheduling, since it first undergoes local inference
and intermediate-feature transmission. Therefore, the future demand
should be characterized according to the scheduling timeslot at which
a task becomes ready for edge-side processing, rather than its
generation timeslot.

For SD $\bm{u}_m$, define its task-generation probability at timeslot
$\tau$ as
\begin{equation}
p_m^{[\tau]}
\triangleq
\Pr\!\left(\alpha_m^{[\tau]}=1\right)
=
\begin{cases}
p_m, & \tau=k\mathbbm{n}_m,\quad k\in\mathbb{Z}^{+},\\
0, & \text{otherwise}.
\end{cases}
\label{eq:app_generation_prob}
\end{equation}
Conditioned on task generation, let the predictive task profile include
the DNN type $l$, batch number $\mathbb{D}_m^{[\tau]}$, privacy
requirement $\rho_m^{[\tau]}$, feasible partition point
$z_{m,l}^{[\tau],\mathsf{part}}$, and transmission-time coefficient
$R_m^{[\tau],\mathsf{SD}}$. For future tasks whose exact attributes are
not yet observed, their joint statistics are characterized according to
the corresponding task and communication profiles. A realization of
$z_{m,l}^{[\tau],\mathsf{part}}$ satisfying
$\phi(z_{m,l}^{[\tau],\mathsf{part}})\leq\rho_m^{[\tau]}$ determines
the per-batch local workload
$\mathbbm{c}_m^{[\tau],\mathsf{SD}}$, edge-side workload
$\mathbbm{c}_m^{[\tau],\mathsf{ES}}$, and intermediate-feature size
$\mathbbm{i}_m^{[\tau]}$. Following Sec.~\ref{sec: time}, the total delay before edge-side scheduling is given as
\begin{align}
T_m^{[\tau],\mathsf{pre}}
&\triangleq
t_m^{[\tau],\mathsf{SDloc}}
+
t_m^{[\tau],\mathsf{SDTrans}}
\nonumber\\
&=
\frac{
\mathbb{D}_m^{[\tau]}
\mathbbm{c}_m^{[\tau],\mathsf{SD}}
}{
f_m^{\mathsf{SD}}
}
+
\mathbb{D}_m^{[\tau]}
\mathbbm{i}_m^{[\tau]}
R_m^{[\tau],\mathsf{SD}}.
\label{eq:app_preedge_delay}
\end{align}
Since the task generated at $\tau$ becomes schedulable at
$\left\lceil
\hat{\tau}_m^{[\tau],\mathsf{TransC}}
\right\rceil$, we define its scheduling offset as
\begin{equation}
\delta_m^{[\tau]}
\triangleq
\left\lceil
\frac{
T_m^{[\tau],\mathsf{pre}}
}{
\Delta\tau
}
\right\rceil,
\label{eq:app_scheduling_offset}
\end{equation}
such that
\begin{equation}
\left\lceil
\hat{\tau}_m^{[\tau],\mathsf{TransC}}
\right\rceil
=
\tau+\delta_m^{[\tau]}.
\label{eq:app_scheduling_relation}
\end{equation}
The equality in \eqref{eq:app_scheduling_relation} follows because
$\tau$ is integer-valued. Since the quantities in
\eqref{eq:app_preedge_delay} vary with the task and communication
conditions, $\delta_m^{[\tau]}$ is generally random before the
corresponding task realization is observed.

Consider an arbitrary future scheduling timeslot
\begin{equation}
q\in
\left\{
\tau'+1,\ldots,\tau'+X^{[\tau']}
\right\}.
\label{eq:app_future_slot}
\end{equation}
A task generated by $\bm{u}_m$ at $q-\delta$ becomes schedulable at
$q$ if its scheduling offset equals $\delta$. Hence, the number of
tasks becoming schedulable at $q$ is
\begin{equation}
Y^{[q]}
=
\sum_{m=1}^{|\mathcal{U}|}
\sum_{\delta\geq 0}
\alpha_m^{[q-\delta]}
\mathbbm{1}
\left\{
\delta_m^{[q-\delta]}=\delta
\right\},
\label{eq:app_task_arrival}
\end{equation}
where terms with $q-\delta\notin\mathcal{T}$ are omitted.
Taking the expectation gives
\begin{align}
\mathbb{E}\!\left(Y^{[q]}\right)
&=
\sum_{m=1}^{|\mathcal{U}|}
\sum_{\delta\geq 0}
p_m^{[q-\delta]}
\Pr\!\left(
\delta_m^{[q-\delta]}=\delta
\,\middle|\,
\alpha_m^{[q-\delta]}=1
\right).
\label{eq:app_expected_arrival}
\end{align}
From \eqref{eq:app_scheduling_offset}, the conditional offset
probability can equivalently be written as
\begin{align}
&\Pr\!\left(
\delta_m^{[\tau]}=\delta
\,\middle|\,
\alpha_m^{[\tau]}=1
\right)
\nonumber\\
&\quad=
\Pr\!\left(
(\delta-1)\Delta\tau
<
T_m^{[\tau],\mathsf{pre}}
\leq
\delta\Delta\tau
\,\middle|\,
\alpha_m^{[\tau]}=1
\right).
\label{eq:app_offset_prob}
\end{align}
Therefore, \eqref{eq:app_expected_arrival} jointly accounts for the
stochastic task-generation process and the heterogeneous local
computation and transmission delays that map task generation to
edge-side schedulability.

We next derive the corresponding edge-side computational workload.
For a task generated by $\bm{u}_m$ at $\tau$, its required edge-side
workload is
\begin{equation}
G_m^{[\tau]}
\triangleq
\mathbb{D}_m^{[\tau]}
\mathbbm{c}_m^{[\tau],\mathsf{ES}}.
\label{eq:app_edge_workload}
\end{equation}
Accordingly, the aggregate edge-side workload of tasks becoming
schedulable at $q$ is
\begin{equation}
W^{[q]}
=
\sum_{m=1}^{|\mathcal{U}|}
\sum_{\delta\geq 0}
\alpha_m^{[q-\delta]}
G_m^{[q-\delta]}
\mathbbm{1}
\left\{
\delta_m^{[q-\delta]}=\delta
\right\}.
\label{eq:app_aggregate_workload}
\end{equation}
Taking the expectation yields
\begin{align}
\mathbb{E}\!\left(W^{[q]}\right)
&=
\sum_{m=1}^{|\mathcal{U}|}
\sum_{\delta\geq 0}
p_m^{[q-\delta]}
\mathbb{E}
\Big[
G_m^{[q-\delta]}
\mathbbm{1}
\left\{
\delta_m^{[q-\delta]}=\delta
\right\}
\nonumber\\[-1mm]
&\hspace{38mm}
\bigm|
\alpha_m^{[q-\delta]}=1
\Big].
\label{eq:app_expected_workload}
\end{align}
Importantly, the workload term and scheduling-offset indicator in
\eqref{eq:app_expected_workload} are retained within the same
expectation. This is because the task attributes determining the
edge-side workload also affect the local-computation and transmission
delays, and hence the scheduling timeslot. Therefore,
$\mathbb{E}(W^{[q]})$ cannot, in general, be reduced to
$\mathbb{E}(Y^{[q]})$ multiplied by an unconditional average
per-task workload.

For completeness, the expectations in
\eqref{eq:app_expected_arrival} and
\eqref{eq:app_expected_workload} can be evaluated by marginalizing
over the task and communication variables specified in Sec.~\ref{sec: uncertain modeling}.
Let $\bm{\xi}_m^{[\tau]}$ denote a realization of these variables
conditioned on $\alpha_m^{[\tau]}=1$, and let
$\mathcal{X}_m^{[\tau]}$ denote its support. Then,
\begin{align}
\mathbb{E}\!\left(Y^{[q]}\right)
&=
\sum_{m=1}^{|\mathcal{U}|}
\sum_{\delta\geq0}
p_m^{[q-\delta]}
\int_{\mathcal{X}_m^{[q-\delta]}}
\mathbbm{1}
\left\{
\delta_m^{[q-\delta]}(\bm{\xi})=\delta
\right\}
\nonumber\\
&\hspace{29mm}\times
\mathrm{d}F_m^{[q-\delta]}(\bm{\xi}),
\label{eq:app_arrival_marginal}
\end{align}
and
\begin{align}
\mathbb{E}\!\left(W^{[q]}\right)
&=
\sum_{m=1}^{|\mathcal{U}|}
\sum_{\delta\geq0}
p_m^{[q-\delta]}
\int_{\mathcal{X}_m^{[q-\delta]}}
G_m^{[q-\delta]}(\bm{\xi})
\nonumber\\
&\hspace{16mm}\times
\mathbbm{1}
\left\{
\delta_m^{[q-\delta]}(\bm{\xi})=\delta
\right\}
\mathrm{d}F_m^{[q-\delta]}(\bm{\xi}),
\label{eq:app_workload_marginal}
\end{align}
where $F_m^{[\tau]}$ denotes the joint distribution of the relevant
task and communication variables. This general form accommodates the
discrete DNN type and batch number adopted in Sec.~\ref{sec: uncertain modeling} as well as
continuous or empirically characterized transmission conditions,
without imposing an additional distributional assumption beyond the
system model.

Finally, evaluating \eqref{eq:app_expected_arrival} and
\eqref{eq:app_expected_workload} for all
$q\in\{\tau'+1,\ldots,\tau'+X^{[\tau']}\}$ gives
\begin{equation}
\bm{Y}^{[\tau']}
=
\left[
\mathbb{E}\!\left(Y^{[\tau'+1]}\right),
\ldots,
\mathbb{E}\!\left(
Y^{[\tau'+X^{[\tau']}]}
\right)
\right],
\label{eq:app_arrival_profile}
\end{equation}
and
\begin{equation}
\bm{W}^{[\tau']}
=
\left[
\mathbb{E}\!\left(W^{[\tau'+1]}\right),
\ldots,
\mathbb{E}\!\left(
W^{[\tau'+X^{[\tau']}]}
\right)
\right].
\label{eq:app_workload_profile}
\end{equation}
Here, $\bm{Y}^{[\tau']}$ characterizes the expected number of tasks
becoming schedulable in each future scheduling timeslot, whereas
$\bm{W}^{[\tau']}$ characterizes their expected aggregate edge-side
computational workload. Both profiles are recomputed at every
scheduling timeslot and serve as certainty-equivalent future-demand
inputs to the predictive ES-state rollout in Sec.~\ref{moduleb}. They characterize
future aggregate demand only and do not prescribe future SD--ES
assignments.

\section{Experimental Settings for Sec. \ref{sec: modelparallel}}
\label{appx: modelparallel}

For the model-parallel experiments in Sec.~\ref{sec: modelparallel},
we implement the two-stage partitioned inference process using a
two-node execution setup. Specifically, each DNN is divided at a
candidate partition point into front-end and back-end submodels, which
are deployed on two Raspberry Pi nodes. The first node emulates SD-side
local inference by executing the layers preceding the partition point,
after which the resulting intermediate features are transmitted to the
second node, which emulates ES-side execution of the remaining layers.
This setup reproduces the partitioned computation and
intermediate-feature transmission behavior of the SD--ES inference
pipeline, rather than the large-scale SD--ES scheduling topology.
Large-scale stochastic workloads and dynamic resource scheduling are
evaluated separately in Sec.~\ref{simulation}. For each DNN architecture, multiple candidate partition points are configured to investigate how the partition location affects the
distribution of computation between the two execution stages and the
intermediate-feature transmission overhead. The candidate partition
points used in the experiments are summarized in
Table~\ref{tab:model_splits}.

\begin{table}[t]
\centering
\caption{Candidate partition points used in the model-parallel experiments on .}
\label{tab:model_splits}
\setlength{\tabcolsep}{4pt}
\renewcommand{\arraystretch}{1.05}
\footnotesize
\begin{tabular}{ll}
\toprule
\textbf{Model/Dataset} & \textbf{Candidate Partition Points} \\
\midrule
TextCNN/THUCNews
& \texttt{GlobalMaxPool} \\
ResNet18/CIFAR-10
& \texttt{layer1}, \texttt{layer2}, \texttt{layer3} \\
M5/UrbanSound8K
& \texttt{pool2}, \texttt{pool3}, \texttt{avgpool} \\
MobileNetV2/MNIST
& \texttt{features.5}, \texttt{features.9}, \texttt{features.13} \\
\bottomrule
\end{tabular}
\end{table}

\section{Privacy Leakage Evaluation in Edge Environments}
 \label{privacyleakageevaluation}
Our previous analysis provides a preliminary evaluation of privacy leakage
at different model partition points through intermediate feature
reconstruction, without considering practical deployment factors such as
model conversion, embedded execution, and network transmission. To further
evaluate privacy risks in a real deployment environment, we conduct feature
inversion attacks on the Raspberry Pi-based model-parallel inference
prototype. Specifically, we use ResNet18 trained on CIFAR-10 and reconstruct the original inputs from intermediate features at different partition points via gradient-based optimization. This experiment evaluates how partition depth affects privacy leakage and provides empirical support for privacy-aware model partitioning.

\noindent
$\bullet$ \textit{Feature Inversion Attack:}
Given intermediate features observed at model partition points, the attacker
aims to reconstruct the original input
$\mathbf{x}\in\mathbb{R}^{3\times32\times32}$ by optimizing a surrogate
input $\hat{\mathbf{x}}$. For an intermediate feature
$\mathbf{f}\in\mathbb{R}^{C\times H\times W}$, the basic reconstruction
objective is formulated as
\begin{equation}
\hat{\mathbf{x}}
=
\arg\min_{\tilde{\mathbf{x}}}
\mathcal{L}_{\mathsf{feat}}
\left(
F(\tilde{\mathbf{x}};\boldsymbol{\theta}),\mathbf{f}
\right)
+
\lambda_{\mathsf{tv}}
\mathcal{R}_{\mathsf{tv}}(\tilde{\mathbf{x}})
+
\lambda_{\mathsf{L}2}
\|\tilde{\mathbf{x}}\|_{2}^{2},
\label{eq:feature_inversion}
\end{equation}
where $F(\cdot;\boldsymbol{\theta})$ denotes the frozen front-end
subnetwork up to the corresponding partition point. The feature-matching
loss $\mathcal{L}_{\mathsf{feat}}$ measures the discrepancy between the
reconstructed and observed intermediate representations. For each
intercepted intermediate representation, we adopt the following
direction-based feature-matching loss:
\begin{equation}
\mathcal{L}_{\mathsf{feat}}
=
1-
\frac{
\left\langle
F(\tilde{\mathbf{x}};\boldsymbol{\theta}),\mathbf{f}
\right\rangle
}{
\left\|
F(\tilde{\mathbf{x}};\boldsymbol{\theta})
\right\|_{2}
\left\|\mathbf{f}\right\|_{2}
+\epsilon
},
\label{eq:direction_loss}
\end{equation}
where $\epsilon$ is a small constant for numerical stability. To suppress
high-frequency artifacts and encourage spatial smoothness, the total
variation (TV) regularization is defined as
\begin{equation}
\mathcal{R}_{\mathsf{tv}}(\tilde{\mathbf{x}})
=
\sum_{i,j}
\left(
\left|
\tilde{\mathbf{x}}_{i+1,j}
-
\tilde{\mathbf{x}}_{i,j}
\right|
+
\left|
\tilde{\mathbf{x}}_{i,j+1}
-
\tilde{\mathbf{x}}_{i,j}
\right|
\right).
\label{eq:tv_regularization}
\end{equation}
The coefficients $\lambda_{\mathsf{tv}}$ and
$\lambda_{\mathsf{L}2}$ control the strengths of the TV and
$\mathsf{L}_2$ regularization terms, respectively.

\noindent
$\bullet$ \textit{Optimization Strategy and Implementation Details:}
Since the input space is high-dimensional
($3\times32\times32=3072$) and the inversion objective is non-convex,
direct optimization may converge to poor local optima. We therefore
parameterize the surrogate input using an unconstrained variable
$\mathbf{z}$ and map it to the valid input range as
\begin{equation}
\tilde{\mathbf{x}}
=
\mu+\sigma\tanh(\mathbf{z}),
\label{eq:tanh_mapping}
\end{equation}
where $\mu=0.5$ and $\sigma=1.5$ control the center and range of the
reconstructed input, respectively.

The reconstruction is optimized using AdamW with AMSGrad and a weight
decay of $5\times10^{-4}$. Early stopping is applied with a patience of
30 iterations, while the learning rate is reduced by a factor of $0.5$
if the feature loss decreases by less than $10^{-5}$ over 10 consecutive
iterations. Since intermediate representations exhibit different
characteristics across network depths, layer-specific learning rates and
TV regularization coefficients are adopted, as summarized in
Table~\ref{tab:inversion_hyperparameters}. We empirically apply stronger
TV regularization to shallower layers and allow more optimization
iterations for deeper layers. All inversion experiments are performed
on the CPU. For ResNet18 at the \texttt{conv1} partition, reconstructing
one sample takes approximately $1.2$~s on average.

The implemented attack adopts \emph{joint feature inversion}, in which
the adversary jointly exploits intermediate features observed from
multiple consecutive layers. Let $\mathbb{L}'$ denote the set of
intercepted intermediate layers and $|\mathbb{L}'|$ denote the number
of such layers. Given the intercepted representations
$\{\mathbf{f}^{(l)}\}_{l\in\mathbb{L}'}$, the joint inversion objective
is formulated as
\begin{equation}
\mathcal{L}_{\mathsf{joint}}
=
\sum_{l\in\mathbb{L}'}
w_l
\mathcal{L}_{\mathsf{feat}}
\left(
F_l(\tilde{\mathbf{x}};\boldsymbol{\theta}),
\mathbf{f}^{(l)}
\right)
+
\lambda_{\mathsf{tv}}
\mathcal{R}_{\mathsf{tv}}(\tilde{\mathbf{x}})
+
\lambda_{\mathsf{L}2}
\|\tilde{\mathbf{x}}\|_2^2,
\label{eq:joint_inversion}
\end{equation}
where $F_l(\cdot;\boldsymbol{\theta})$ denotes the front-end subnetwork
up to layer $l$, and $w_l$ is the weight assigned to the corresponding
feature-matching term. Larger weights are assigned to shallower layers
to emphasize representations containing more input-specific information.
By jointly exploiting intermediate representations across multiple
network depths, this formulation provides a stronger reconstruction
setting for evaluating the privacy leakage associated with different
model partition points. Compared with single-layer inversion, joint
inversion increases the average reconstruction SSIM by $12.3\%$ on
CIFAR-10, thereby providing a more stringent basis for the subsequent
privacy evaluation.
\begin{table}[t]
\centering
\caption{Layer-specific hyperparameters for feature inversion}
\label{tab:inversion_hyperparameters}
\footnotesize
\setlength{\tabcolsep}{6pt}
\renewcommand{\arraystretch}{1.08}
\begin{tabular}{cccc}
\hline
\rowcolor{gray!10}
\textbf{Layer} &
\textbf{Learning Rate} &
$\boldsymbol{\lambda_{\mathsf{tv}}}$ &
\textbf{Max. Iter.} \\
\hline
\texttt{conv1}  & 0.05 & 0.010 & 500  \\
\texttt{layer1} & 0.03 & 0.005 & 800  \\
\texttt{layer2} & 0.02 & 0.003 & 1000 \\
\texttt{layer3} & 0.01 & 0.002 & 1200 \\
\texttt{layer4} & 0.01 & 0.001 & 1500 \\
\hline
\end{tabular}
\end{table}

Table~\ref{tab:privacy_inversion} reports the average feature-matching
loss, reconstruction MSE, and SSIM over 10 CIFAR-10 test samples at
different partition points of ResNet18. Since our interest lies in
reconstruction-based privacy leakage, we primarily use SSIM to quantify
the structural similarity between the reconstructed and original inputs. The results reveal a clear dependence of reconstruction leakage on the
partition location. The shallow \texttt{conv1} representation yields the
highest SSIM of $0.4135$, indicating that it retains substantially more
input-related structural information. After deeper feature
transformations, the SSIM drops markedly and remains mostly within
$0.15$--$0.19$, eventually reaching $0.1316$ at the output. Although
local fluctuations exist, e.g., an SSIM of $0.1904$ at
\texttt{layer3.1.conv2}, the overall trend indicates reduced structural
reconstructability with increasing representation depth. Notably, the feature-matching loss does not directly reflect input-space
reconstruction fidelity. For example, the output representation exhibits
the lowest feature loss but also the lowest SSIM. We therefore use SSIM,
rather than the optimization loss, as the primary indicator of
reconstruction-based privacy leakage. These results demonstrate the
strong dependence of privacy leakage on the DNN partition point, thereby
providing empirical support for the partition-dependent privacy characterization adopted in our privacy-aware model partitioning.

\begin{table}[t]
\centering
\caption{Feature inversion results at different partition points of ResNet18 on CIFAR-10}
\label{tab:privacy_inversion}
\footnotesize
\setlength{\tabcolsep}{5pt}
\renewcommand{\arraystretch}{1.05}
\begin{tabular}{lccc}
\rowcolor{gray!10}
\toprule
\textbf{Partition Point} &
\textbf{Feature Loss} &
\textbf{MSE} $\downarrow$ &
\textbf{SSIM} $\uparrow$ \\
\midrule
\texttt{conv1}            & 0.0832 & 0.0512 & \textbf{0.4135} \\
\texttt{layer1.0.conv1}   & 0.8555 & 0.0637 & 0.1643 \\
\texttt{layer1.1.conv2}   & 0.5452 & 0.0601 & 0.1734 \\
\texttt{layer2.0.conv1}   & 0.7709 & 0.0632 & 0.1684 \\
\texttt{layer2.1.conv2}   & 0.4449 & 0.0617 & 0.1695 \\
\texttt{layer3.0.conv1}   & 0.7864 & 0.0631 & 0.1672 \\
\texttt{layer3.1.conv2}   & 0.4812 & 0.0628 & 0.1904 \\
\texttt{layer4.1.conv2}   & 0.8403 & 0.0653 & 0.1454 \\
\texttt{output}           & 0.0009 & 0.0644 & 0.1316 \\
\bottomrule
\end{tabular}
\end{table}

\end{document}